\RequirePackage{lineno}
\documentclass[twocolumn,letterpaper,aps,rmp,superscriptaddress,showpacs,floatfix]{revtex4-1}

\usepackage{graphicx}	\usepackage{xspace}     \usepackage{amsmath}
\usepackage{hyperref}\usepackage{color}\usepackage{multirow}\usepackage{array}
\usepackage{amssymb}

\newcolumntype{C}[1]{>{\centering\let\newline\\\arraybackslash\hspace{0pt}}m{#1}}
\newcommand{\jhcs}{jet-hadron correlations\xspace}
\newcommand{\dhcs}{dihadron correlations\xspace}
\newcommand{\Dhcs}{Dihadron correlations\xspace}

\newcommand{\dhc}{dihadron correlation\xspace}
\newcommand{\jlc}{jet-like correlation\xspace}
\newcommand{\jlcs}{jet-like correlations\xspace}
\newcommand{\pT}{$p_T$\xspace}
\newcommand{\ET}{$E_T$\xspace}
\newcommand{\meanpT}{$\langle p_T \rangle$\xspace}
\newcommand{\meanET}{$\langle E_T \rangle$\xspace}

\newcommand{\highpT}{high-$p_T$\xspace}
\newcommand{\kT}{$k_T$\xspace}
\newcommand{\akT}{anti-$k_T$\xspace}

\newcommand{\vtwo}{$v_2$\xspace}
\newcommand{\vthree}{$v_3$\xspace}
\newcommand{\qhat}{$\hat{q}$\xspace}
\newcommand{\Aj}{$A_{J}$\xspace}
\newcommand{\RdAu}{$R_{dAu}$\xspace}
\newcommand{\RpPb}{$R_{pPb}$\xspace}
\newcommand{\RAA}{$R_{AA}$\xspace}
\newcommand{\RCP}{$R_{CP}$\xspace}
\newcommand{\IAA}{$I_{AA}$\xspace}
\newcommand{\Npart}{$N_{part}$\xspace}
\newcommand{\Nch}{$N_{ch}$\xspace}
\newcommand{\Nbin}{$N_{bin}$\xspace}
\newcommand{\zerodeg}{0$^{\circ}$\xspace}
\newcommand{\oneeightydeg}{180$^{\circ}$\xspace}
\newcommand{\dphi}{$\Delta\phi$\xspace}
\newcommand{\deta}{$\Delta\eta$\xspace}
\newcommand{\GeV}{GeV/$c$\xspace}
\newcommand{\MeV}{MeV/$c$\xspace}

\newcommand{\psiR}{$\psi_{R}$\xspace}
\newcommand{\vn}{$v_{n}$\xspace}
\newcommand{\vnum}[1]{$v_{#1}$\xspace}

\newcommand{\vntrigger}{$v_{n}^{\mathrm{t}}$\xspace}
\newcommand{\vnassoc}{$v_{n}^{\mathrm{a}}$\xspace}

\newcommand{\eref}[1]{equation~\ref{#1}}
\newcommand{\Sref}[1]{Section~\ref{#1}}
\newcommand{\Fref}[1]{Figure~\ref{#1}}
\newcommand{\Tref}[1]{Table~\ref{#1}}
\newcommand{\Eref}[1]{Equation~\ref{#1}}
\newcommand{\dNdphi}{$\frac{dN}{d\phi}$}
\newcommand{\pp}{$p$+$p$\xspace}
\newcommand{\ee}{$e^+$+$e^-$\xspace}
\newcommand{\Cu}{Cu+Cu\xspace}
\newcommand{\Au}{Au+Au\xspace}
\newcommand{\UU}{U+U\xspace}
\newcommand{\Pb}{Pb+Pb\xspace}
\newcommand{\dAu}{$d$+Au\xspace}
\newcommand{\dA}{$d$+A\xspace}
\newcommand{\pPb}{$p$+Pb\xspace}
\newcommand{\pA}{$p$+A\xspace}
\newcommand{\AplusA}{$A$+$A$\xspace}
\newcommand{\sNN}{$\sqrt{s_{\mathrm{NN}}}$\xspace}
\newcommand{\sqrts}{$\sqrt{s}$\xspace}
\newcommand{\ns}{near-side\xspace}
\newcommand{\as}{away-side\xspace}
\newcommand{\ptassoc}{$p_T^{\mathrm{a}}$\xspace}
\newcommand{\pttrig}{$p_T^{\mathrm{t}}$\xspace}
\newcommand{\pttrigrange}[2]{#1 $< p_T^{\mathrm{t}} <$ #2~GeV/$c$\xspace}
\newcommand{\ptassocrange}[2]{#1 $< p_T^{\mathrm{a}} < $ #2~GeV/$c$\xspace}

\newcommand{\etarange}[1]{$|\eta|<$#1\xspace}

\newcommand{\FJ}{$FastJet$\xspace}

\begin{document}

\title{ Review of Jet Measurements in Heavy Ion Collisions}

\author{Megan Connors}\affiliation{Georgia State University, Atlanta, GA, USA-30302.}\affiliation{RIKEN BNL Research Center, Upton, NY 11973-5000}
\author{Christine Nattrass}\affiliation{University of Tennessee, Knoxville, TN, USA-37996.}
\author{Rosi Reed}\affiliation{Lehigh University, Bethlehem, PA, USA-18015.}
\author{Sevil Salur}\affiliation{Rutgers University, Piscataway, NJ USA-08854.\\
All authors contributed equally to this manuscript. Authors are listed alphabetically.}

\date{\today}

\begin{abstract} 

A hot, dense medium called a Quark Gluon Plasma (QGP) is created in ultrarelativistic heavy ion collisions.
Early in the collision, hard parton scatterings generate high momentum partons that traverse the medium, which then fragment into sprays of particle called jets. Understanding how these partons interact with the QGP  and fragment into final state particles provides critical insight into quantum chromodynamics. Experimental measurements from high momentum hadrons, two particle correlations, and full jet reconstruction at the Relativistic Heavy Ion Collider (RHIC) and the Large Hadron Collider (LHC) continue to improve our understanding of energy loss in the QGP.  Run 2 at the LHC recently began and there is a jet detector at RHIC under development.  Now is the perfect time to reflect on what the experimental measurements have taught us so far, the limitations of the techniques used for studying jets, how the techniques can be improved, and how to move forward with the wealth of experimental data such that a complete description of energy loss in the QGP can be achieved. 

Measurements of jets to date clearly indicate that hard partons lose energy.  Detailed comparisons of the nuclear modification factor between data and model calculations led to quantitative constraints on the opacity of the medium to hard probes.  However, while there is substantial evidence for softening and broadening jets through medium interactions, the difficulties comparing measurements to theoretical calculations limit further quantitative constraints on energy loss mechanisms.  Since jets are algorithmic descriptions of the initial parton, the same jet definitions must be used,  including the treatment of the underlying heavy ion background, when making data and theory comparisons. 
We call for an agreement between theorists and experimentalists on the appropriate treatment of the background, Monte Carlo generators that enable experimental algorithms to be applied to theoretical calculations, and a clear understanding of which observables are most sensitive to the properties of the medium, even in the presence of background. This will enable us to determine the best strategy for the field to improve quantitative constraints on properties of the medium in the face of these challenges.

\end{abstract}

\maketitle

\tableofcontents

l\section{Introduction} \label{Sec:Intro}

In ultrarelativistic heavy ion collisions, the temperature is so high that the nuclei melt, forming a hot, dense liquid of quarks and gluons called the Quark Gluon Plasma (QGP).  
Hard quark and gluon scatterings occur early in the collision, prior to the formation of the QGP.  These quarks and gluons, known as partons, traverse the medium and then fragment into collimated sprays of particles called jets.  These partons lose energy to the medium and the jets they produce are thus modified.  This process, called jet quenching, is studied with experimental measurements of high momentum hadrons, two particle correlations, and jet reconstruction at the Relativistic Heavy Ion Collider (RHIC) and the Large Hadron Collider (LHC).  
After nearly two decades of experimental measurements have taught us so far, we reflect on the limitations of the techniques used for studying jets, how the techniques can be improved, and how to move forward with the wealth of experimental data such that a complete description of energy loss in the QGP can be achieved. 

Our goal in the following sections is to provide an overview of what we have learned from jet measurements and what the field needs to do in order to improve our quantitative understanding of jet quenching and the properties of the medium from RHIC energies (\sNN = 7.7--200 GeV) to LHC energies (\sNN = 2.76--5.02 TeV).  We will discuss measurements using the ALICE, ATLAS, and CMS detectors at the LHC, and the BRAHMS, PHENIX, Phobos, and STAR detectors at RHIC.  The main goal of this paper is to review experimental techniques and measurements. While we discuss some models and their interpretation, a full review of the theory of partonic interactions with the medium is outside the scope of this paper.  In this section, we provide an overview of the formation of the QGP and other processes which impact the measurement of jets and their interaction with the medium.  One key factor in measuring jets in heavy ion collisions is accounting for the effect of the fluctuating background on different observables. \Sref{Sec:ExpMethods} discusses the various measurement techniques and approaches to background subtraction and suppression and how these techniques may impact the results and their interpretation.  We include measurements of nuclear modification factors, dihadron and multi-hadron correlations, and reconstructed jets.  We follow this with a discussion of results in \Sref{Sec:Results} organized by what they tell us about the medium.  Do jets lose energy in the medium?  Is fragmentation modified in the medium?  Do jets modify the medium?  Are there cold nuclear matter effects?  We show that there is substantial evidence for both partonic energy loss and modified fragmentation.  The evidence for modification of the medium by jets is considerably more scant.  Our understanding of cold nuclear matter effects is rapidly evolving, but currently there do not appear to be substantial cold nuclear matter effects for jets. 

We conclude with a discussion of what we have learned and the way forward for the field in \Sref{Sec:Discussion}.
There are extensive detailed measurements of jets, benefited by improved detector technologies, high cross sections, and higher luminosities, and there have been dramatic improvements in our theoretical understanding and capabilities.  
However, experimental techniques and the bias they may impose are frequently neglected, and it is not currently possible to apply experimental algorithms to most models.
The current status of comparisons between models and data motivates our call for an agreement between theorists and experimentalists on the appropriate treatment of the background, Monte Carlo generators that enable experimental algorithms to be applied to theoretical calculations, and a clear understanding of which observables are most sensitive to the properties of the medium, even in the presence of background. 
This will enable us to quantitatively constrain properties of the medium.

\subsection{Formation and evolution of the Quark Gluon Plasma}
Quarks and gluons become deconfined under extremely high energy and density conditions.  This deconfined state became known as the QGP~\cite{Shuryak:1980tp}. 
With the advancements in accelerator physics, it can be created and studied in high energy heavy ion collisions.
 
The formation of the QGP requires energy densities above 0.2-1 GeV/fm$^3$~\cite{Karch:2002abc,Bazavov:2014pvz}.  These energy densities can  currently be reached in high energy heavy ion collisions at RHIC located at Brookhaven National Laboratory in Upton, NY and the LHC located at CERN in Geneva, Switzerland.  Estimates of the energy density indicate that central heavy ion collisions with an incoming energy per nucleon pair as low as \sNN = 7.7 GeV, the lower boundary of collision energies accessible at RHIC, can reach energy densities above 1 GeV/fm$^3$~\cite{Adare:2015bua} and that collisions at 2.76 TeV, accessible at the LHC, reach energy densities as high as 12 GeV/fm$^3$~\cite{Adam:2016thv,Chatrchyan:2012mb}. Contrary to initial na\"ive expectations of a gas-like QGP, the QGP formed in these collisions was shown to behave like a liquid of quarks and gluons ~\cite{Adcox:2004mh,Adams:2005dq,Back:2004je,Arsene:2004fa,Heinz:2013th}.

\begin{figure}
\begin{center}

\rotatebox{0}{\resizebox{8cm}{!}{
	\includegraphics{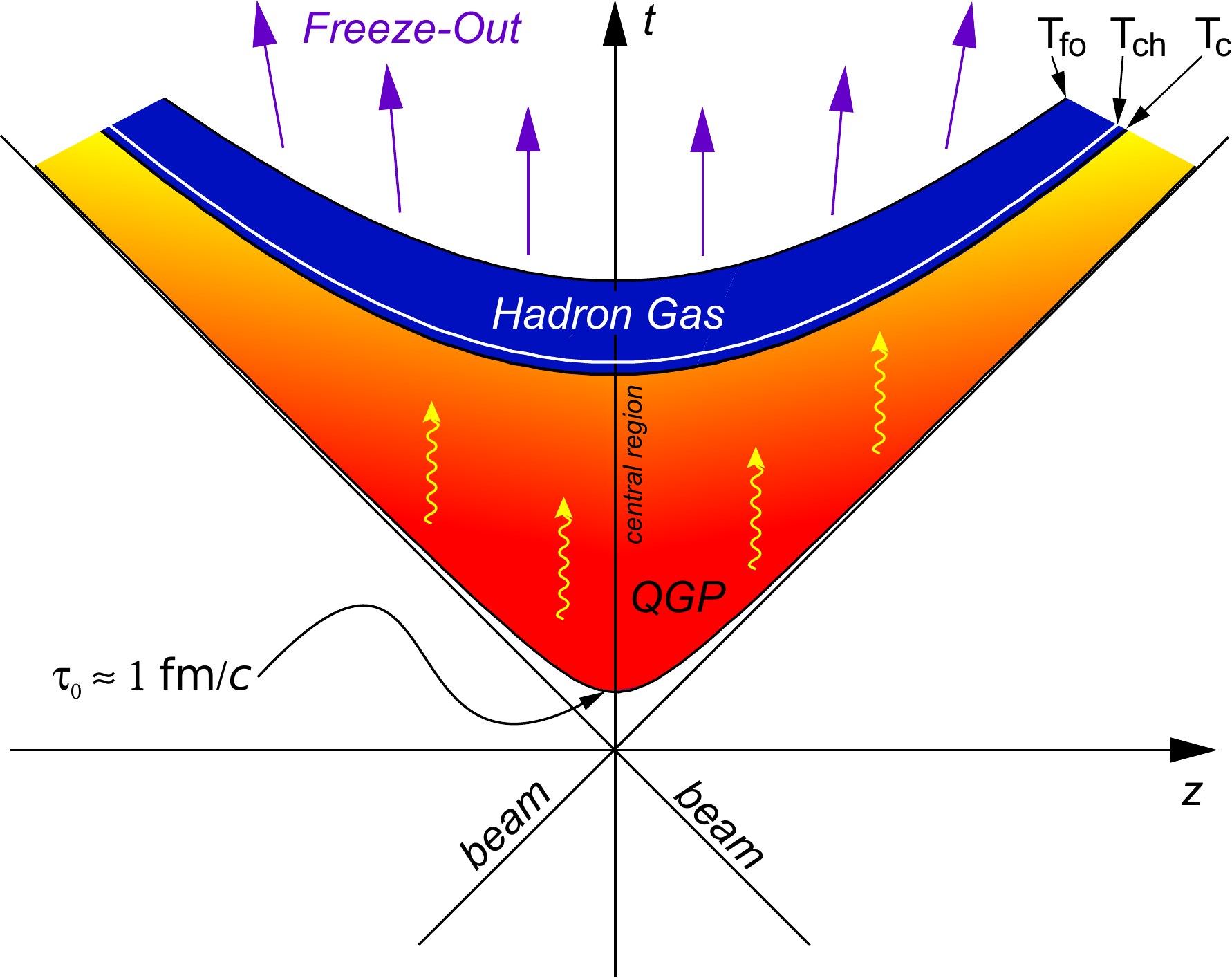}
}}\caption{
A light cone diagram showing the stages of a heavy ion collision.  The abbreviation  $T_{\mathrm fo}$ is for the thermal freeze-out temperature, $T_{\mathrm ch}$ is for the chemical freeze-out temperature, and $T_{\mathrm c}$ is for the critical temperature where the phase transition between a hadron gas and a QGP occurs.  $\tau_0$ is the formation time of the QGP. Figure courtesy of Thomas Ullrich.  
}\label{Fig:lightcone}
\end{center}
\end{figure}

\begin{figure*}
\begin{center}

\rotatebox{0}{\resizebox{15cm}{!}{
	\includegraphics{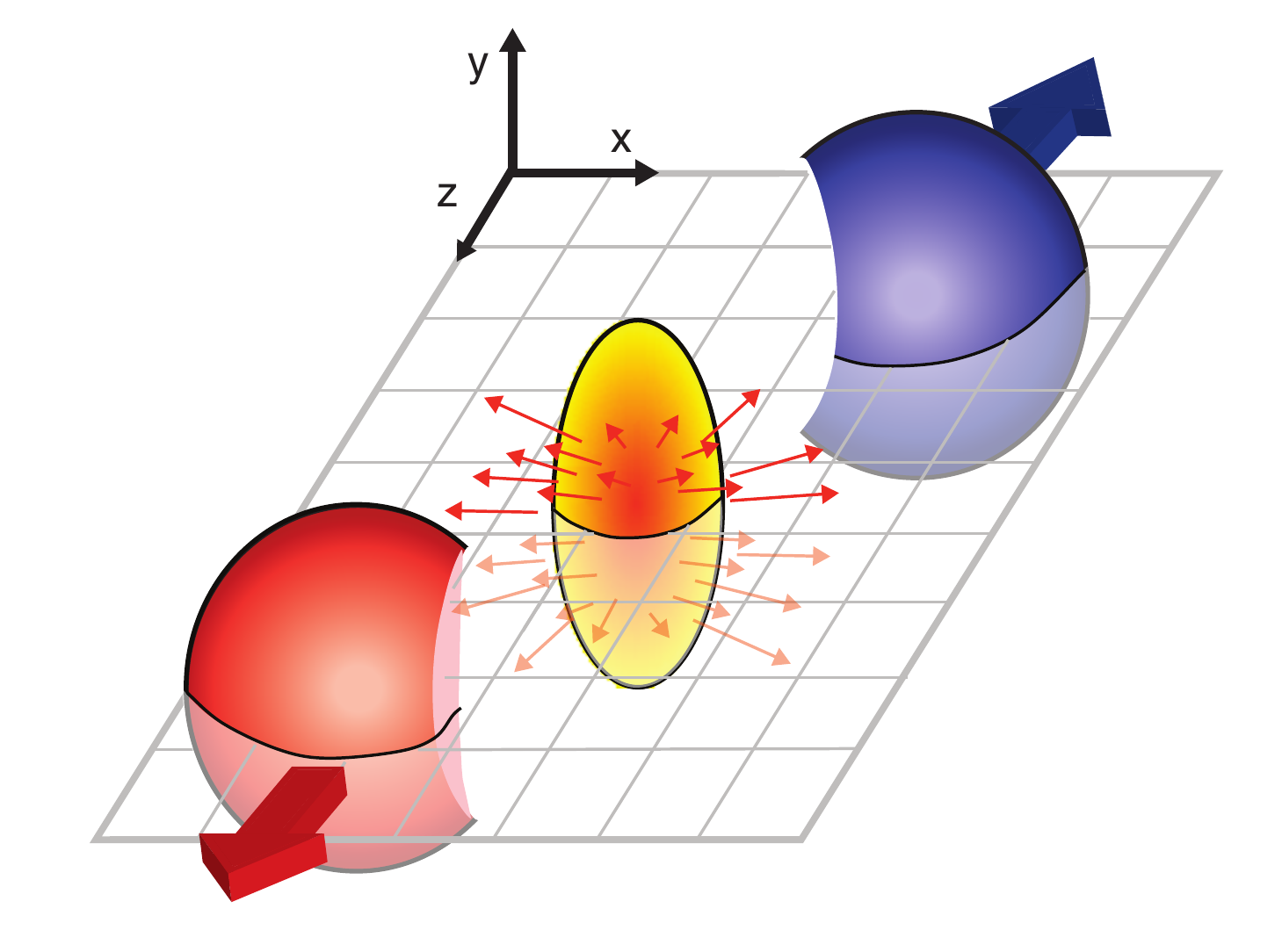}
	\includegraphics{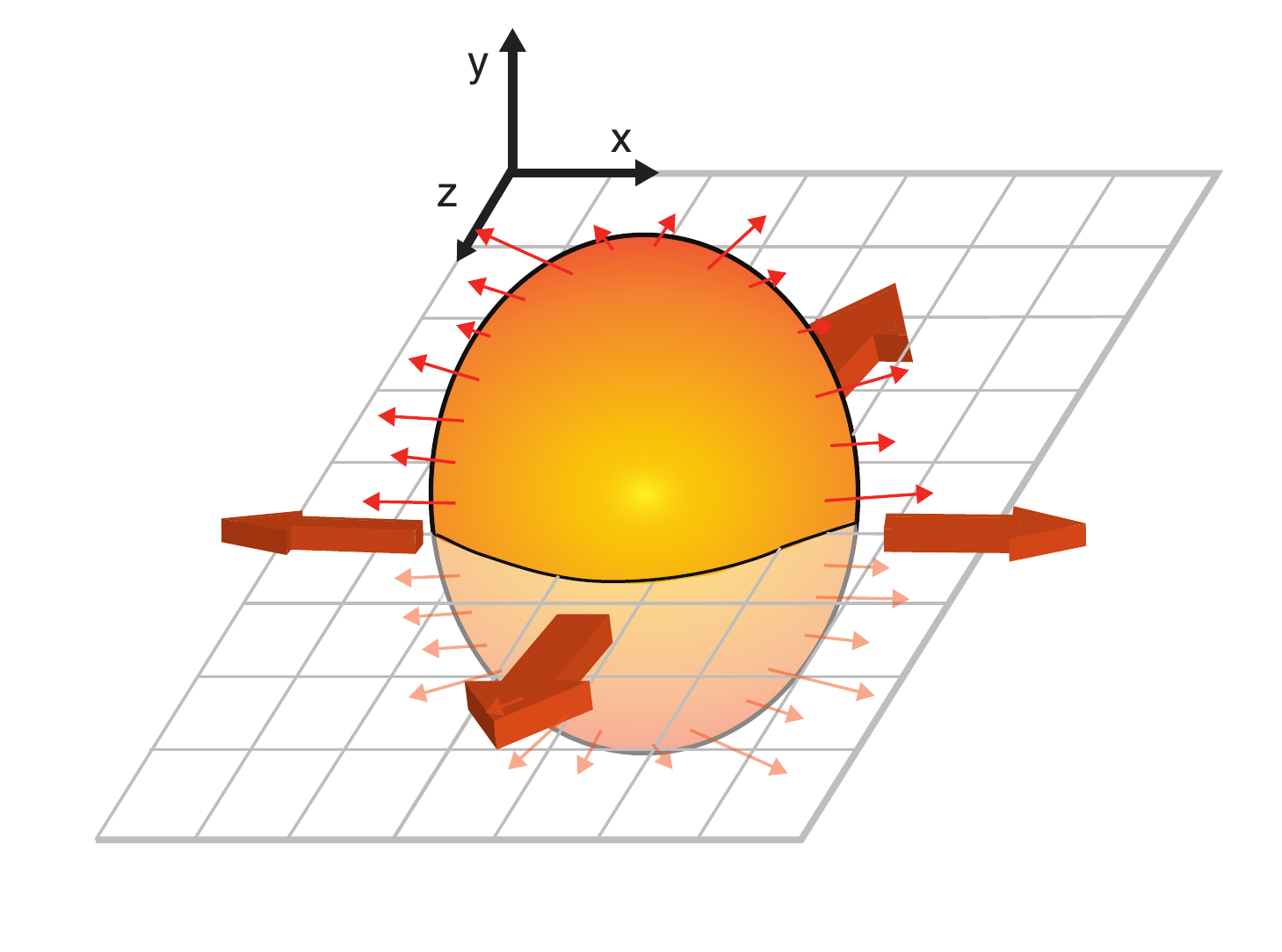}
}}\caption{
Schematic diagrams showing the initial overlap region (left) and the spatial anisotropy generated by this anisotropic overlap region.  This anisotropy can be quantified using the Fourier coefficients of the momentum anisotropy.
Figure courtesy of Boris Hippolyte.
}\label{Fig:flowschematic}
\end{center}
\end{figure*}

The heavy ion collision and the evolution of the fireball, as depicted in Figure~\ref{Fig:lightcone}, has several stages, and the measurement of the final state particles can be affected by one or all of these stages depending on the production mechanism and interaction time within the medium.  The initial state of the incoming nuclei is not precisely known, but its properties impact the production of final state particles.  The incoming nuclei are often modeled as either an independent collection of nucleons called a Glauber initial state~\cite{Miller:2007ri}, or a wall of coherent gluons called a Color Glass Condensate~\cite{Iancu:2000hn}.  In either initial state model, both the impact parameter of the nuclei and fluctuations in the positions of the incoming quarks or gluons, called partons, lead to an asymmetric nuclear overlap region.  This asymmetric overlap is shown schematically in \Fref{Fig:flowschematic}.  The description of the initial state most consistent with the data is between these extremes~\cite{Moreland:2014oya}.  The proposed electron ion collider is expected to resolve ambiguities in the initial state of heavy ion collisions~\cite{Geesaman:2015fha}.  

In all but the most central collisions, some fraction of the incoming nucleons do not participate in the collision and escape unscathed.  These nucleons, called spectators, can be observed directly and used to measure the impact parameter of the collision.  Before the formation of the QGP, partons in the nuclei may scatter off of each other just as occurs in \pp collisions.  An interaction with a large momentum transfer (${Q}$) is called a hard scattering, a process which is, in principle, calculable with perturbative quantum chromodynamics (pQCD).  The majority of these hard scatterings are 2$\rightarrow$2, which result in high momentum partons traveling 180$^{\circ}$ apart in the plane transverse to the beam as they travel through the evolving medium.  These hard parton scatterings are the focus of this paper.  

As the medium evolves, it forms a liquid of quarks and gluons.  
The liquid reaches local equilibrium, with temperature fluctuations in different regions of the medium.  The liquid QGP phase is expected to live for 1-10 fm/$c$, depending on the collision energy \cite{Harris:1996zx}.  As the medium expands and cools, it reaches a density and temperature where partonic interactions cease, a hadron gas is formed, and the hadron fractions are fixed.  This point in the collision evolution is called chemical freeze-out~\cite{Fodor:2004nz,Adams:2005dq,Adam:2015vda}.  As the medium expands and cools further, collisions between hadrons cease and hadrons reach their final energies and momenta.  This stage of the collision, thermal freeze-out, occurs at a somewhat lower temperature than the chemical freeze-out.

Thermal photons, in a manner analogous to black body radiation, reveal that the QGP may reach temperatures of 300--600 MeV in central collisions at both 200 GeV~\cite{Adare:2008ab} and 2.76 TeV~\cite{Adam:2015lda}.  The temperature can also be inferred from the sequential melting of bound states of a bottom quark and antiquark~\cite{Chatrchyan:2012np}.    The ratios of final state hadrons are used to determine that the chemical freeze-out temperature is around 160 MeV~\cite{Fodor:2004nz,Adams:2005dq,Adam:2015vda} and that the thermal freeze out occurs at about 100--150 MeV, depending on the collision energy and centrality~\cite{Abelev:2013vea,Adcox:2003nr,Arsene:2005mr,Back:2006tt}.  

The properties of the medium are determined from the final state particles that are measured.  The initial gluon density can be related to the final state hadron multiplicity through the concept of hadron-parton duality~\cite{VanHove:1987zv}, leading to estimates of gluon densities of around 700 per unit pseudorapidity at the top RHIC energy of \sNN = 200 GeV~\cite{Adler:2004zn} and 2000 per unit pseudorapidity at the top LHC energy of \sNN = 5.02 TeV~\cite{Adam:2015ptt,Aamodt:2010pb,Chatrchyan:2011pb,Aad:2016zif,ATLAS:2011ag}.  

The azimuthal anisotropy in the momentum distribution of final state hadrons is the result of the initial state anisotropy.  The survival of these anisotropies provides evidence that the medium flows in response to pressure gradients~\cite{Adam:2016izf,Chatrchyan:2013kba,Aad:2014vba,Adler:2003kt,Alver:2006wh,Adler:2001nb}.  This asymmetry is illustrated schematically in \Fref{Fig:flowschematic}.  The shape and magnitude of these anisotropies can be used to constrain the viscosity to entropy ratio, revealing that the QGP has the lowest viscosity to entropy ratio ever observed~\cite{Adcox:2004mh,Adams:2005dq,Back:2004je,Arsene:2004fa}.  Hadrons containing strange quarks are enhanced in heavy ion collisions above expectations from \pp collisions~\cite{ABELEV:2013zaa,Khachatryan:2016yru,Abelev:2013xaa}.  This is due to a combination of the suppression of strangeness in \pp collisions due to the limited phase space for the production of strange quarks, and the higher energy density available for the production of strange quarks in heavy ion collisions.  Correlations between particles may provide evidence for increased production of strangeness due to the decreased strange quark mass in the medium~\cite{Abelev:2009ac,Adam:2015vje}.  Baryon production is enhanced for both light~\cite{Adler:2003cb,Arsene:2009nr,Abelev:2006jr} and strange quarks~\cite{ABELEV:2013zaa,Khachatryan:2016yru,Abelev:2013xaa,Abelev:2007xp}, an observation generally interpreted as evidence for the direct production of baryons through the recombination of quarks in the medium~\cite{Fries:2003vb,Greco:2003xt,Dover:1991zn,Hwa:2002tu}.

Hard parton scattering occurs early in the collision evolution, prior to the formation of the QGP, so that their interactions with the QGP probe the entire medium evolution. Therefore, they can be used to reveal the properties of the medium, such as its stopping power and transport coefficients.
Since the differential production cross section of these hard parton scatterings is calculable in pQCD, and these calculations have been validated over many orders of magnitude in proton-proton collisions, in principle they form a well calibrated probe.  The initial production must scale by the number of nucleon collisions, which means that their interactions with the medium would cause deviations from this scaling. Since the majority of these hard partons are produced in pairs, they can be used both as a probe and a control.     
Particle jets of this nature are formed in ${e}^{+}{e}^{-}$ and proton-proton (\pp) collisions as well and are
observed to fragment similarly in ${e}^{+}{e}^{-}$ and \pp collisions.

In a heavy ion collision, where a QGP is formed, the hard scattered quarks and gluons are expected to interact strongly with the hot QCD medium due to their color charges, and lose energy, either through collisions with medium partons, or through gluon bremsstrahlung.  The energy loss of high momentum partons due to strong interactions is a process called jet quenching, and results in modification of the properties of the resulting jets in heavy ion collisions compared to expectations from proton-proton collisions ~\cite{Bjorken:1982tu, Gyulassy:1990ye,Baier:1994bd}.  This energy loss was first observed in the suppression of high momentum hadrons produced in heavy ion collisions at RHIC~\cite{Adams:2003kv,Adler:2003qi,Back:2004bq} and later also observed at the LHC~\cite{Aamodt:2010jd,CMS:2012aa}.  The modification can be observed through measurements of jet shapes, particle composition, fragmentation, splitting functions and many other observables.
Detailed studies of jets to characterize how and why partons lose energy in the QGP require an understanding of how evidence for energy loss may be manifested in the different observables, and the effect of the large and complicated background from other processes in the collision.

Early studies of the QGP focused on particles produced through soft processes, measuring the bulk properties of the medium.  With the higher cross sections for hard processes with increasing collision energy, higher luminosity delivered by colliders, and detectors better suited for jet measurements, studies of jets are enabling higher precision measurements of the properties of the QGP~\cite{Akiba:2015jwa}.  The 2015 nuclear physics Long Range Plan (LRP)~\cite{Geesaman:2015fha} highlighted the particular need to improve our quantitative understanding of jets in heavy ion collisions.  Here we assess our current understanding of jet production in heavy ion collisions in order to inform what shape future studies should take in order to optimize the use of our precision detectors.

\subsection{Jet definition}
\begin{figure}
\begin{center}
\rotatebox{0}{\resizebox{8cm}{!}{
	\includegraphics[trim={0 6cm 0 6cm},clip]{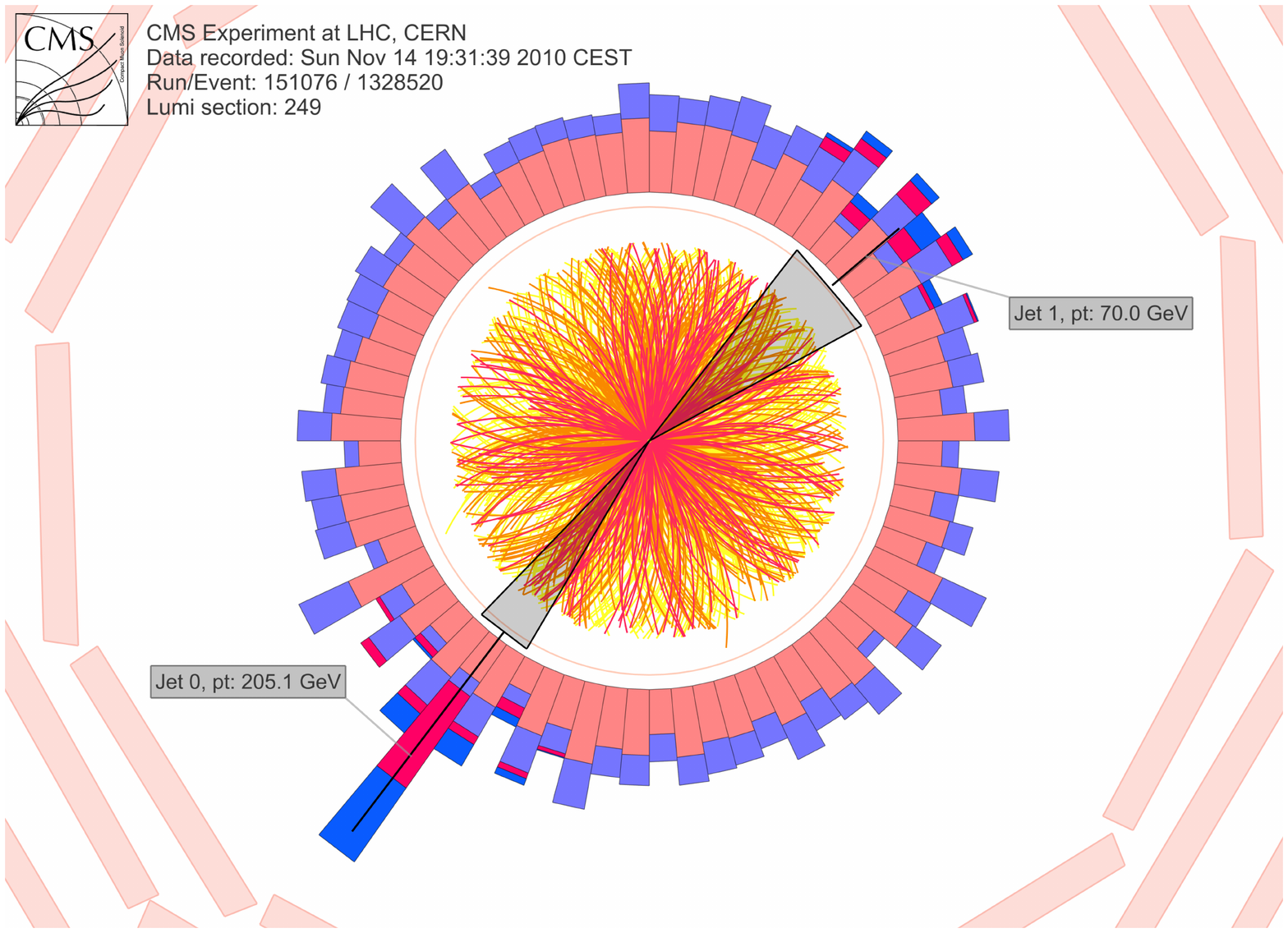}
}}\caption{
Event display showing a dijet event in a \Pb collision at \sNN = 2.76 TeV~\cite{CMS:1309898}.  This shows the large background for jet measurements in heavy ion collisions.
}\label{Fig:jeteventdisplay}
\end{center}
\end{figure}

In principle, using a jet finding algorithm to cluster all of the daughter particles of a given parton will give access to the full energy and momentum of the parent parton.  However, even in \ee collisions, the definition of a jet is ambiguous, even on the partonic level.  For instance, in $e^+ e^- \rightarrow q \bar{q}$, the quark may emit a gluon.  If this gluon is emitted at small angles relative to the quark, it is usually considered part of the jet, whereas if it is emitted at large angles relative to the parent parton, it may be considered a third jet.  This ambiguity led to the Snowmass Accord, which stated that in order to be comparable, experimental and theoretical measurements had to use the same definition of a jet and that the definition should be theoretically robust~\cite{Huth:1990mi}.  

The choice of which final state particles should be included in the jet is also somewhat arbitrary and more difficult in \AplusA collisions than in \pp collisions.  \Fref{Fig:jeteventdisplay} shows an event display from a \Pb collision at \sNN = 2.76 TeV, showing the large background in the event.  If a hard parton emits a soft gluon and that gluon thermalizes with the medium, are the particles from the hadronization of that soft gluon part of the jet or part of the medium?  Any interaction between daughters of the parton and medium particles complicates the definition of what should belong to the jet and what should not.  This ambiguity in the definition of the observable itself makes studies of jets qualitatively different from, e.g., measurements of particle yields.  These aspects of jet physics need to be taken into account in the choice of a jet finding algorithm and background subtraction methods in order to be able to interpret the resulting measurements.  

One of the main motivations for studies of jets in heavy ion collisions was to provide measurements of observables with a production cross-section that can be calculated using pQCD, which yields a well calibrated probe.  In certain limits, this is feasible, although it is worth noting that many observables are sensitive to non-perturbative effects.  One such non-perturbative effect is hadronization, which can affect even the measurements of relatively simple observables such as the jet momentum spectra. 

In addition to the ambiguities inherent in the definition of what is and is not a jet, there is the question of how to deal with the large background in heavy ion collisions.  For example, measurements of reconstructed jets usually have a minimum momentum threshold for constituents in order to suppress the background contribution.  If the corrections for these analysis techniques are insensitive to assumptions about the background and hadronization, the results may still be perturbatively calculable.  However, these techniques for dealing with the background may also bias the measured jet sample, for instance by selecting gluon jets at a higher rate than quark jets.  In the context of jets in a heavy ion collision, these analysis cuts are part of the definition of the jet and can not be ignored.  

The interpretation of the measurement of any observable cannot be fully separated from the techniques used to measure it because both measurements and theoretical calculations of jet observables must use the same definition of a jet.  As we review the literature, we discuss how the jet definitions and techniques used in experiment may influence the interpretation of the results.  Even though our goal is an understanding of partonic interactions within the medium, a detailed understanding of soft particle production is necessary to understand the methods for suppressing and subtracting the contribution of these particles to jet observables.

\subsection{Interactions with the medium}
There are several models used to describe interactions between hard partons and the medium, however, a full review of theoretical calculations is beyond the scope of this paper.  We briefly summarize theoretical frameworks for interactions of hard partons with the medium here and refer readers to~\cite{Qin:2015srf,Burke:2013yra} and the references therein for details.
The production of final state particles in nuclear collisions is described by assuming that these processes can be factorized~\cite{Majumder:2007iu,Majumder:2010qh}. The nuclear parton distribution functions $x_a f^A_a(x_a)$ and $x_b f^B_b(x_b)$ describe the probability of finding partons with momentum fraction $x_a$ and $x_b$, respectively. The differential cross sections for partons $a$ and $b$ interacting with each other to produce a parton $c$ with a momentum $p$ can be described using pQCD.  The production of a final state hadron h is then given by fragmentation function $D^h_c(z)$ where $z = p^h/p$ is the fraction of the parton's momentum carried by the final state hadron.  The differential cross section for the production of hadrons as a function of their transverse momenta \pT and rapidity $y$ at leading order is then given by
{\small
\begin{equation}
 \frac{d^3\sigma^{h}}{dy d^{2}p_T} = \frac{1}{\pi} \int dx_a \int dx_b f^A_a(x_a) f^B_b(x_b) \frac{d\sigma_{ab\rightarrow cX}}{d\hat{t}} \frac{D^h_c(z)}{z}.\label{Eq:Factorization}
\end{equation}
}
\noindent where $\hat{t}=(\hat{p}-x_a P)^2$, $\hat{p}$ is the four-momentum of parton, $c$, and $P$ is the average momentum of a nucleon in nucleus A. The nuclear parton distribution functions and the fragmentation functions cannot be calculated perturbatively.  The parton distribution functions describe the initial state of the incoming nuclei.  
Any differences between the nuclear and proton parton distribution functions, which describe the distribution of partons in a nucleon, are considered cold nuclear matter effects.  
Cold nuclear matter effects may include coherent multiple scattering within the nucleus~\cite{Qiu:2004da}, gluon shadowing and saturation~\cite{Gelis:2010nm}, or partonic energy loss within the nucleus~\cite{Wang:2001ifa,Bertocchi:1976bq,Vitev:2007ve}.  Most models for interactions of partons with a QGP factorize this process and only modify the fragmentation functions~\cite{Majumder:2007iu}.  
One goal of studies of high momentum particles in heavy ion collisions is to study the modification of these fragmentation functions, which will allow us to understand how and why partons lose energy within the QGP and to determine the microscopic structure of the medium.  We note that the theoretical definition in \Eref{Eq:Factorization} associates the production of a final state hadron with a particular parton.  This is not possible experimentally, so the experimentally measured quantity also referred to as a fragmentation function is not the same as $D^h_c(z)$ in \Eref{Eq:Factorization}. 

Medium-induced gluon radiation (bremsstrahlung) and collisions with partons in the medium cause the partons to lose energy to the medium, often described as a modification of the fragmentation functions in \Eref{Eq:Factorization}.
There are four major approaches to describing these interactions.
The GLV model~\cite{Vitev:2002pf,Djordjevic:2003zk,Djordjevic:2004nq,Wicks:2005gt,Djordjevic:2008iz} and its CUJET implementation~\cite{Buzzatti:2011vt} assumes that the scattering centers in the medium are nearly static and that the mean free path of a parton is much larger than the color screening length in the medium.  This assumption is valid for a thinner medium.

The Higher Twist~\cite{Majumder:2009ge} framework assumes medium modified splitting functions during fragmentation calculated by including higher twist corrections to the differential cross sections for deep inelastic scattering off of nuclei.  These corrections are enhanced by the length of the medium.
The higher twist model has also been adapted to include multiple gluon emissions~\cite{Majumder:2010qh,Majumder:2009ge,Collins:1985ue}.  

In the BDMPS~\cite{Baier:1996kr,Baier:1998yf,Baier:2000mf} approach and its equivalents~\cite{Wiedemann:2000ez,Wiedemann:2000tf,Albacete:2004gw,Eskola:2004cr,Armesto:2011ht,Zakharov:1996fv} the effect of multiple parton scatterings is evaluated using a path integral over a path ordered Wilson line~\cite{Wiedemann:2000ez,Wiedemann:2000za}.  This assumes infinite coherence of the radiated gluons and a thick medium.
YaJEM~\cite{Renk:2013rla,Renk:2008pp} and JEWEL~\cite{Zapp:2013vla,Zapp:2013zya} are Monte Carlo implementations of the BDMPS framework.

The energy loss mechanism in the AMY model is similar to BDMPS but the rate equations for partonic energy loss are solved numerically and convoluted with differential pQCD cross sections and fragmentation functions to determine the final state  differential hadronic cross sections~\cite{Qin:2007rn,Qin:2009bk,Jeon:2003gi,Arnold:2002ja}.  This is applied in a realistic hydrodynamical environment~\cite{Song:2007fn,Song:2007ux,Qiu:2011hf,Qiu:2012uy}.  
The MARTINI model~\cite{Qin:2007rn,Schenke:2010rr} is a Monte Carlo model implementation of the AMY formalism which uses PYTHIA~\cite{Sjostrand:2006za} to describe the hard scattering and a Glauber initial state~\cite{Miller:2007ri}.  Partonic energy loss occurs in the medium, taking temperature and hydrodynamical flow into account~\cite{Nonaka:2006yn,Schenke:2010nt,Schenke:2010rr}.

There are additional approaches, including embedding jets into a hydrodynamical fluid~\cite{Tachibana:2017syd} and using the correspondence between Anti-deSitter space and conformal field theories~\cite{Gubser:2006qh}.
There is a new description of jet quenching in which coherent parton branching plays a central role to the jet-medium interactions \cite{CasalderreySolana:2012ef,Mehtar-Tani:2014yea}. In this work it is assumed that the hierarchy of scales governing jet evolution allow the jet to be separated int a hard core, which interacts with the medium as a single coherent antenna, and softer structures that will interact in a color decoherent fashion.  In order for this to be valid, there must be a large separation of the intrinsic jet scale and the characteristic momentum scale of the medium.  While this certainly is valid for the highest momentum jets at the LHC, it is not clear at which scales in collision energy and jet energy this assumption breaks down. 
We refer readers to a recent theoretical review for a more complete picture of theoretical descriptions of partonic energy loss in the QGP~\cite{Qin:2015srf}.

Medium-induced bremsstrahlung occurs when the medium exchanges energy, color,
and longitudinal momentum with the jet.  Since both the energy and longitudinal momentum of the hard partons exceeds that of the medium partons, these exchanges cause the parton as a whole to lose energy.  Additionally, since the hard partons have much higher transverse momentum than the medium partons, any collision will reduce the momentum of the jet as a whole.  Both of these effects will broaden the resulting jet and soften the average final state particles produced from the jet.  Collisional energy loss similarly broadens and softens the jet.
Partonic energy loss in the medium is quantified by the jet transport coefficients $\hat{q} = Q^2/L$, where $Q$ is the transverse momentum lost to the medium and $L$ is the path-length traversed; $\hat{e}$, the longitudinal momentum lost per unit length; and $\hat{e_2}$, the fluctuation in the longitudinal momentum per unit length~\cite{Muller:2012hr,Majumder:2012ti}.

The JET collaboration systematically compared each of these models to data to determine how well the transport properties of partons in the medium can be constrained~\cite{Burke:2013yra}.  This substantially improved our quantitative understanding of partonic energy loss in the medium, but only used a small fraction of the available data.  The Jetscape collaboration~\cite{Jetscape} has formed to develop a Monte Carlo framework which enables combinations of different models of the initial state, the hydrodynamical evolution of medium, and partonic energy loss to be used within the same framework.  The goal is a Bayesian analysis comparing models to data to quantitatively determine properties of the medium, similar to~\cite{Novak:2013bqa,Bernhard:2016tnd}.  Jetscape will incorporate many of the available jet observables into this Bayesian analysis.  Part of the motivation for this paper is to evaluate which experimental observables might provide effective input for this effort and what factors need to be considered for these comparisons.

In light of the ambiguities in the jet definition discussed above, we note that whether or not the energy is lost depends on this definition.  The functional experimental definition of lost energy is any energy which no longer retains short-range correlations with the parent parton, meaning that it is further than about half a unit in pseudorapidity and azimuth.  Energy which retains short-range correlations with the parent parton is still considered part of the jet and any short-range modifications are considered modifications of the fragmentation function.

\subsection{Separating the signal from the background}\label{Sec:Intro:Background}

Hard partons traverse  a medium which is flowing and expanding, with fluctuations in the density and temperature. 
Since the mean transverse momentum of unidentified hadrons in \Pb collisions at \sNN = 2.76 TeV is ~680 \MeV~\cite{Abelev:2013bla}, sufficiently high \pT hadrons are expected to be produced dominantly in jets and production from soft processes is expected to be negligible.  It is unclear precisely at which momentum the particle yield is dominated by jet production rather than medium production.  Moreover, most particles produced in jets are at low momenta even though the jet momentum itself is dominated by the contribution of a few high \pT~particles.  Particularly if jets are modified by processes such as recombination, strangeness enhancement, or hydrodynamical flow, these low momentum particles produced in jets may carry critical information about their parent partons' interactions with the medium.    Methods employed to suppress and subtract background from jet measurements are dependent on assumptions about the background contribution and can change the sensitivity of measurements to possible medium modifications.  The resulting biases in the measurements can be used as a tool rather than treated as a weakness in the measurement; however, they must be first understood.

The largest source of correlated background is due to collective flow.  
The azimuthal distribution of particles created in a heavy ion collision can be written as
\begin{equation}
 \frac{dN}{d(\phi-\psi_{R})} \propto 1 + \sum_{n=1}^{\infty} 2 v_{n} \cos(n(\phi - \psi_{R}))\label{Eqtn:FlowFourierDecomposition}
\end{equation}
\noindent where $N$ is the number of particles, $\phi$ is the angle of a particle's momentum in azimuth in detector coordinates and \psiR is the angle of the reaction plane in detector coordinates~\cite{Poskanzer:1998yz}.  The Fourier coefficients \vn are thought to be dominantly from collective flow at low momenta~\cite{Adcox:2004mh,Adams:2005dq,Back:2004je,Arsene:2004fa}, although \eref{Eqtn:FlowFourierDecomposition} is valid for any correlation because any distribution can be written as its Fourier decomposition.  The magnitude of the Fourier coefficients \vn decreases with increasing order.  The sign of the flow contribution to the first order coefficient $v_{1}$ is dependent on the incoming direction of the nuclei and changes sign when going from positive to negative pseudorapidities.  For most measurements, which average over the direction of the incoming nuclei, $v_{1}$ due to flow is zero, although we note that there may be contributions to $v_{1}$ from global momentum conservation.

The even \vn arise mainly from anisotropies 
 in the average overlap region of the incoming nuclei, considering the nucleons to be smoothly distributed in the nucleus with the density depending only on the radius.  
 The odd \vn for $n>1$ are generally understood to arise from the fluctuations in the positions of the nucleons within the nucleus.  These fluctuations also contribute to the even \vn, though these coefficients are dominated by the overall geometry.  
Jets themselves can lead to non-zero \vn through jet quenching, complicating background subtraction for jet studies.   At high momenta (\pT~$ \gtrsim $~5-10 \GeV) the \vn are thought to be dominated by jet production.   Furthermore, the \vn fluctuate event-by-event even for a given centrality class. This means that independent measurements, which differ in their sensitivity to jets, averaged over several events cannot be used blindly to subtract the correlated background due to flow.

To measure jets, experimentalists have to make some assumptions about the interplay between hard and soft particles and about the form of the background. 
Without such assumptions, experimental measurements are nearly impossible.  Some observables are more robust to assumptions about the background than others, however, these measurements are not always the most sensitive to energy loss mechanisms or interactions of jets with the medium.  An understanding of data requires an understanding of the measurement techniques and assumptions about the background.  We therefore discuss the measurement techniques and their consequences in great detail in \Sref{Sec:ExpMethods} before discussing the measurements themselves in \Sref{Sec:Results}.

\section{Experimental methods}\label{Sec:ExpMethods}

This section focuses on different methods for probing jet physics including inclusive hadron measurements, dihadron correlations, jet reconstruction algorithms and jet-particle correlations and a brief description of relevant detectors. In addition to explaining the measurement details and how the effect of the background on the observable is handled for each, this section highlights strengths and weaknesses of these different methods which are important for interpreting the results.  We emphasize background subtraction and suppression techniques because of potential biases they introduce.

\subsection{Detectors}
Measurements of heavy ion collisions often focus on midrapidity, with precision, particle identification, and tracking in a high multiplicity environment.  Some measurements, such as those of single particles, are not significantly impacted by a limited acceptance, while the acceptance corrections for reconstructed jets are more complicated when the acceptance is limited.  We briefly summarize the colliders, RHIC and the LHC, and the most important features of each of their detectors for measurements of jets, referring readers to other publications for details.

The properties of the medium are slightly different at RHIC and the LHC, with the LHC reaching the highest temperatures and energy densities and RHIC providing the widest range of collision energies and systems.  The relevant properties of each collider are summarized in \Tref{Tab:Colliders}.  Some properties of each detector are summarized in \Tref{Tab:Detectors}.

{ \small
\begin{table}
\begin{center}
\caption{Collision systems, collision energies (\sqrts) for \pp collisions, collision energies per nucleon (\sNN) for \AplusA collisions, charged particle multiplicities ($dN/d\eta$) for central collisions, energy densities for central collisions, and the temperature compared to the critical temperature for formation of the QGP $T/T_{c}$ for both RHIC and the LHC.}
\label{Tab:Colliders}
\begin{tabular}{|c | p{3.5cm} | p{3.5cm}|}
\hline
Collider        & RHIC & LHC \\ \hline
Collisions        & \pp, \dAu, \Cu, \Au, \UU & \pp, \pPb, \Pb \\ 
\sqrts & 62--500 GeV & 0.9--14 TeV \\ 
\sNN       & 7.7--500 GeV & 2.76--5.02 TeV \\ 
$dN/d\eta$       & 192.4$\pm$16.9 -- 687.4$\pm$36.6~\cite{Adare:2015bua} & 1584$\pm$76~\cite{Aamodt:2010pb}, 1943$\pm$54~\cite{Adam:2015ptt} \\
$\epsilon$        & 1.36$\pm$0.14 GeV/fm$^3$~\cite{Adare:2015bua} -- 4.9$\pm$0.3 GeV/fm$^3$~\cite{Adams:2004cb} &  12.3$\pm$1.0 GeV/fm$^3$~\cite{Adam:2016thv} \\
$T/T_{c}$\footnote{Calculated using $T=196$ MeV at \sNN = 200 GeV, $T=280$ MeV at \sNN = 2.76 TeV, and $T=292$ MeV at \sNN = 5.02 TeV from~\cite{Srivastava:2016hwr} assuming that $T_c$ = 155 MeV from the extrapolation of the chemical freeze-out temperature using comparisons of data to statistical models in~\cite{Floris:2014pta}.}        & 1.3 & 1.8--1.9 \\
\hline
\end{tabular}
\end{center}
\end{table}
}

\begin{table*}
\begin{center}
\caption{Summary of acceptance of detectors at RHIC and the LHC and when detectors took data.  When not otherwise listed, azimuthal acceptance is 2$\pi$.}
\label{Tab:Detectors}
\begin{tabular}{|c | c | c| c |c | c |}
\hline
Collider        & Detector &EMCal & HCal & Tracking & Taking data\\ \hline
\multirow{ 5}{*}{RHIC} & BRAHMS & N/A & N/A & $0<\eta<4$ & 2000--2006\\ 
                       & PHENIX & $|\eta|<0.35$ & N/A & $|\eta|<0.35$, $2\times \Delta\phi = 90^{\circ}$& 2000--2016\\ 
                       & PHOBOS & N/A & N/A & $0<|\eta|<2$, $2\times \Delta\phi = 11^{\circ}$ & 2000--2005\\ 
                       & STAR   & $|\eta|<1.0$ & N/A & $|\eta|<1.0$ & 2000--\\ 
                       & sPHENIX& $|\eta|<1.0$ & $|\eta|<1.0$ & $|\eta|<1.0$ & future \\ \hline
\multirow{ 4}{*}{LHC}  & ALICE  & $|\eta|<0.7$, $\Delta\phi = 107^{\circ}$ and $\Delta\phi = 60^{\circ}$ & N/A & $|\eta|<0.9$ & 2009--\\ 
                       & ATLAS  & $|\eta|<4.9$ & $|\eta|<4.9$ & $|\eta|<2.5$ & 2009--\\ 
                       & CMS    & $|\eta| < 3.0$ & $|\eta| < 5.2$ & $|\eta| < 2.5$ & 2009--\\ 
                       & LHCb   & N/A & N/A & $|\eta|<0.35$ & 2009--\\ 
\hline
\end{tabular}
\end{center}
\end{table*}

The BRAHMS~\cite{Adamczyk:2003sq}, PHENIX~\cite{Adcox:2003zm}, and PHOBOS~\cite{Back:2003sr} experiments are experiments which have completed their taking data at RHIC.  The STAR~\cite{Ackermann:2002ad} experiment is taking data at RHIC and sPHENIX~\cite{Adare:2015kwa} is a proposed upgrade at RHIC to be built in the existing PHENIX hall.
STAR has full azimuthal acceptance and nominally covers pseudorapidities $|\eta|<1$ with a silicon inner tracker and a time projection chamber (TPC), surrounded by an electromagnetic calorimeter~\cite{Ackermann:2002ad}.  An inner silicon detector was installed before the 2014 run.  Particle identification is possible both through energy loss in the TPC and a time of flight (TOF) detector.  STAR also has forward tracking and calorimetry.  The PHENIX central arms cover $|\eta|<0.35$ and are split into two 90$^{\circ}$ azimuthal regions~\cite{Adcox:2003zm}.  They consist of drift and pad chambers for tracking, a TOF for particle identification, and precision electromagnetic calorimeters.  There are both midrapidity and forward silicon for precision tracking and forward electromagnetic calorimeters.  PHENIX also has two muon arms at forward rapidities ($-1.15<|\eta|<-2.25$ and $1.15<|\eta|<-2.44$) with full azimuthal coverage.  The PHOBOS detector consists of a large acceptance scintillator with wide acceptance for multiplicity measurements ($|\eta|<3.2$) and two spectrometer arms capable of both particle identification and tracking covering $0<|\eta|<2$ and split into two 11$^{\circ}$ azimuthal regions~\cite{Back:2003sr}.  The BRAHMS detector has a spectrometer arm capable of particle identification with wide rapidity coverage ($0\lesssim y\lesssim 4$)~\cite{Adamczyk:2003sq}.  sPHENIX will have full azimuthal acceptance and acceptance in pseudorapidity of approximately $|\eta|<1$ with a TPC combined with precision silicon tracking and both electromagnetic and hadronic calorimeters~\cite{Adare:2015kwa}.  sPHENIX is optimized for measurements of jets and heavy flavor at RHIC.

The LHC has four main detectors, ALICE, ATLAS, CMS, and LHCb.  ALICE, which is primarily devoted to studying heavy ion collisions at the LHC, has a TPC, silicon inner tracker, and TOF covering $|\eta|<0.9$ and full azimuth~\cite{Aamodt:2008zz}.  It has an electromagnetic calorimeter (EMCal) covering $|\eta|<0.7$ with two azimuthal regions covering $107^{\circ}$ and $60^{\circ}$ in azimuth and a forward muon arm.  Both ATLAS and CMS are multipurpose detectors designed to precisely measure jets, leptons and
photons produced in pp and heavy ion collisions. The ATLAS detector's precision tracking is performed by a high-granularity silicon pixel detector, followed by the silicon microstrip tracker and complemented by the transition radiation tracker  for the $|\eta|<2.5$ region. The hadronic and electromagnetic calorimeters
provide hermetic azimuthal coverage in the $|\eta|<4.9$ range.  The muon spectrometer surrounds the calorimeters covering $|\eta|<2.7$ with full azimuthal coverage~\cite{Aad:2008zzm}. 
The main CMS detectors are silicon trackers which measure charged particles within the pseudorapidity range $|\eta| < 2.5$, an electromagnetic calorimeter partitioned into a barrel region ($|\eta| < 1.48$) and two endcaps ($| \eta| < 3.0$), and hadronic calorimeters covering the range $|\eta| < 5.2$. 
All CMS detectors listed here have full azimuthal coverage~\cite{Chatrchyan:2008aa}.
LHCb focuses on measurements of charm and beauty at forward rapidities. The LHCb detector consists of a single spectrometer covering $1.6<|\eta|<4.9$ and full azimuth~\cite{Alves:2008zz}.  This spectrometer arm is capable of tracking and particle identification, however, tracking is limited to low multiplicity collisions.

\subsection{Centrality determination}
The impact parameter $b$, defined as the transverse distance between the centers of the two colliding nuclei,
cannot be measured directly.  Glancing interactions with a large impact parameter generally produce fewer particles while collisions with a small impact parameter generally produce more particles, with the number of final state particles increasing monotonically with the overlap volume between the nuclei.  This correlation can be used to define the collision centrality as a fraction of the total cross section.  High multiplicity events have a low average $b$ and low multiplicity events have a large average $b$.  The former are called central collisions and the latter are called peripheral collisions. In large collision systems, the variations in the number of particles produced due to fluctuations in the energy production by individual soft nucleon-nucleon collisions is small compared to the variations due to the impact parameter.  The charged particle multiplicity, \Nch, can then be used to constrain the impact parameter.

Usually the correlation between the impact parameter and the multiplicity is determined using a Glauber model~\cite{Miller:2007ri}.  The distribution of nucleons in the nucleus is usually approximated as a Fermi distribution in a Woods-Saxon potential and the multiplicity is assumed to be a function of the number of participating nucleons (\Npart) and the binary number of interactions between nucleons (\Nbin).  The experimentally observed multiplicity is fit to determine a parametric description of the data and the data are binned by the fraction of events.  For example, the 10\% of all events with the highest multiplicity are referred to as 0-10\% central.  There are a few variations in technique which generally lead to consistent results~\cite{Abelev:2013qoq}.  
Centralities determined assuming that the distribution of impact parameters at a fixed multiplicity is Gaussian are consistent with those using a Glauber model~\cite{Das:2017ned}.

The largest source of uncertainty from centrality determination in heavy ion collisions is due to the normalization of the multiplicity distribution at low multiplicities.  In general an experiment identifies an anchor point in the distribution, such as identifying the \Nch where 90\% of all collisions produce at least that multiplicity.  Because the efficiency for detecting events with low multiplicity is low, the distribution is not measured well for low \Nch, so identification of this anchor point is model dependent.  This inefficiency does not directly impact measurements of jets in 0-80\% central collisions because these events are typically high multiplicity, however, it can lead to a significant uncertainty in the correct centrality.  This uncertainty is largest at low multiplicities, corresponding to more peripheral collisions.

As the phenomena observed in heavy ion collisions have been observed in increasingly smaller systems, this approach to determining centrality has been applied to these smaller systems as well.  While the term ``centrality" is still used, this is perhaps better understood as event activity, since the correlation between multiplicity and impact parameter is weaker in these systems and other effects may become relevant~\cite{Bzdak:2014rca,Armesto:2015kwa,Alvioli:2014eda,Coleman-Smith:2013rla,Alvioli:2014sba,Alvioli:2013vk}.  The interpretation of the ``centrality" dependence in small systems should therefore be done carefully.

\subsection{Inclusive hadron measurements}

Single particle spectra at high momenta, which are dominated by particles resulting from hard scatterings, can be used to study jets.  To quantify any modifications to the hadron spectra in nucleus-nucleus (\AplusA) collisions, the nuclear modification factor was introduced.  The nuclear modification factor in \AplusA collisions is defined as 
\begin{equation}
 R_{AA} = \frac{\sigma_{NN}}{\langle N_{bin}\rangle} \frac{d^2N_{AA}/dp_{T}d\eta}{d^2\sigma_{pp}/dp_{T}d\eta}
\end{equation} where $\eta$ is the pseudorapidity, \pT is the transverse momentum, $\langle N_{bin}\rangle$ is the average number of binary nucleon-nucleon collisions for a given range of impact parameter,  and $\sigma_{NN}$ is the integrated nucleon-nucleon cross section.  $N_{AA}$ and  $\sigma_{pp}$ in this context are the yield in $AA$ collision and cross section in \pp collisions for a particular observable.  If nucleus-nucleus collisions were simply a superposition of nucleon-nucleon collisions, the high \pT~particle cross-section should scale with the number of binary collisions and therefore \RAA~=~1.  An \RAA~$<$~1 indicates suppression and an \RAA~$>$~1 indicates enhancement.  \RAA is often measured as a function of \pT and centrality class.  Measurements of inclusive hadron \RAA are relatively straightforward as they only require measuring the single particle spectra and a calculation of the number of binary collisions for each centrality class based on a Glauber model~\cite{Miller:2007ri}.  Theoretically, hadron \RAA can be difficult to interpret, particularly at low momenta, because different physical processes that are not calculable in pQCD, such as hadronization, can change the interpretation of the result. 
Interpretation of \RAA usually focuses on high \pT, where calculations from perturbative QCD (pQCD) are possible.  An alternative to \RAA is \RCP, where peripheral heavy ion collisions are used as the reference instead of \pp collisions
\begin{equation}
 R_{CP} = \frac{\langle N_{bin}^{peri}\rangle}{\langle N_{bin}^{cent}\rangle} \frac{d^2N_{AA}^{cent}/dp_{T}d\eta}{d^2N_{AA}^{peri}/dp_{T}d\eta}
\end{equation}
where $cent$ and $peri$ denote the values of $\langle N_{bin}\rangle$ and $N_{AA}$ for central and peripheral collisions, respectively.
This is typically done either when there is no \pp reference available or the \pp reference has much larger uncertainties than the \AplusA reference.  It does have the advantage that other nuclear effects could be present in the \RCP~cross-section and cancel in the ratio, and that these collisions are recorded at the same time and thus have the same detector conditions.  However, there can be QGP effects in peripheral collisions so this can make the interpretation difficult.  The pQCD calculations used to interpret these results are sensitive in principle to hadronization effects, however, if the \RAA of hard partons does not have a strong dependence on \pT, the \RAA of the final state hadrons will not have a strong dependence on \pT.  \RAA will therefore be relatively insensitive to the effects of hadronization and more theoretically robust.

\subsection{Dihadron correlations}\label{Sec:dihadroncorrmethod}

\begin{figure}
\begin{center}

\rotatebox{0}{\resizebox{5cm}{!}{
        \includegraphics{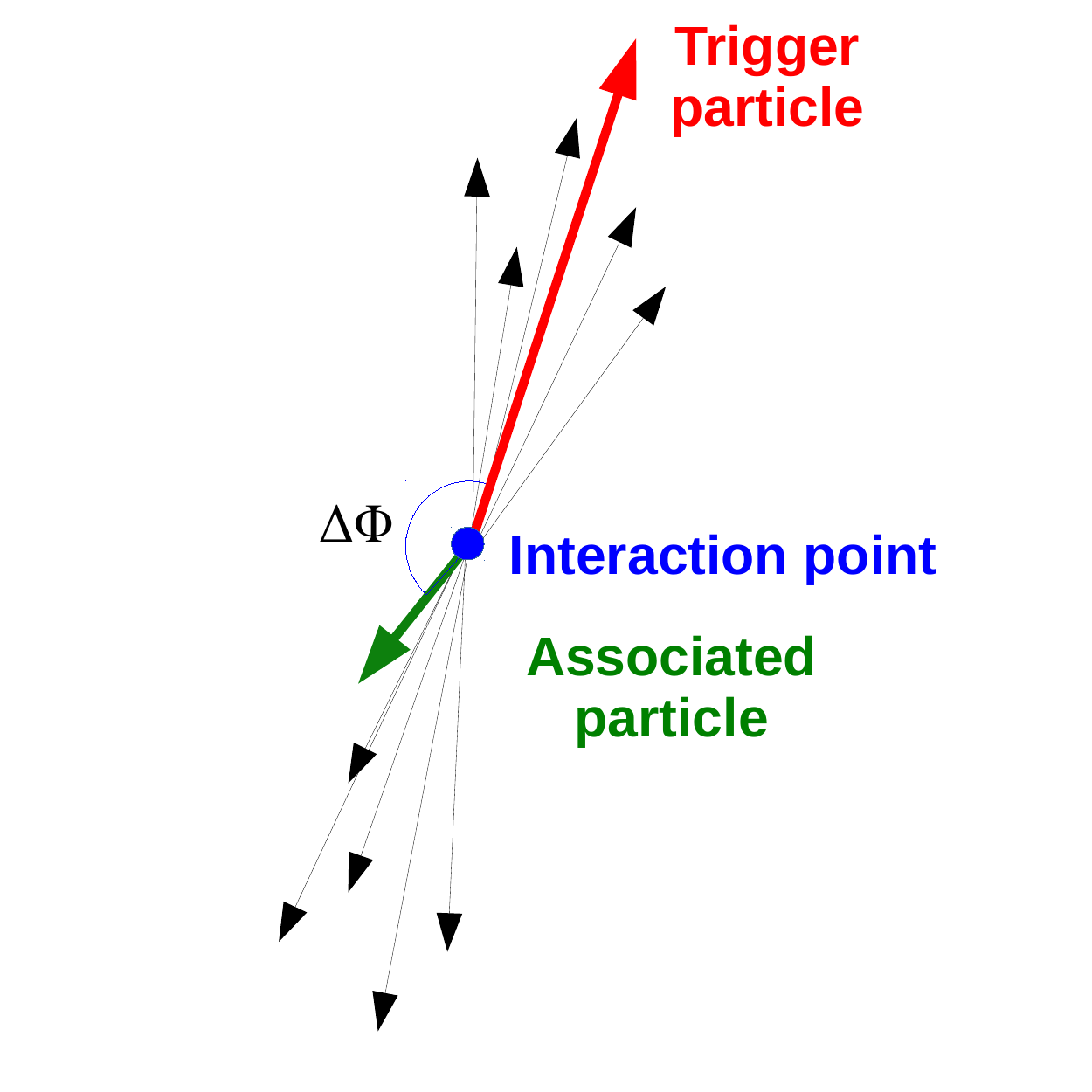}
}}\caption{
Schematic diagram showing the identification of a \highpT hadron in a \pp collision and its use to define a coordinate system for dihadron correlations.
}\label{Fig:DihadronMethodSchematic}
\end{center}
\end{figure}

A hard parton scattering usually produces two partons that are separated by \oneeightydeg in the transverse plane (commonly stated as back-to-back). In a typical \dhc study~\cite{Adler:2002tq,Adler:2005ad,Abelev:2009af,Aamodt:2011vg,Alver:2009id}, a \highpT hadron is identified and used to define the coordinate system because its momentum is assumed to be a good proxy for the jet axis of the parton it arose from.  This hadron is called the trigger particle.  The azimuthal angle of other hadrons' momenta in the event is calculated relative to the momentum of this trigger particle.  These hadrons are commonly called the associated particles.  This is illustrated schematically in \Fref{Fig:DihadronMethodSchematic}.  The associated particle is typically restricted to a fixed momentum range, also typically higher than the \meanpT of tracks in the event and lower than the momenta of trigger particles.  The distribution of associated particles relative to the trigger particle can be measured in azimuth (\dphi), pseudorapidity (\deta), or both.

\begin{figure*}
\begin{center}
\rotatebox{0}{\resizebox{16cm}{!}{
        \includegraphics{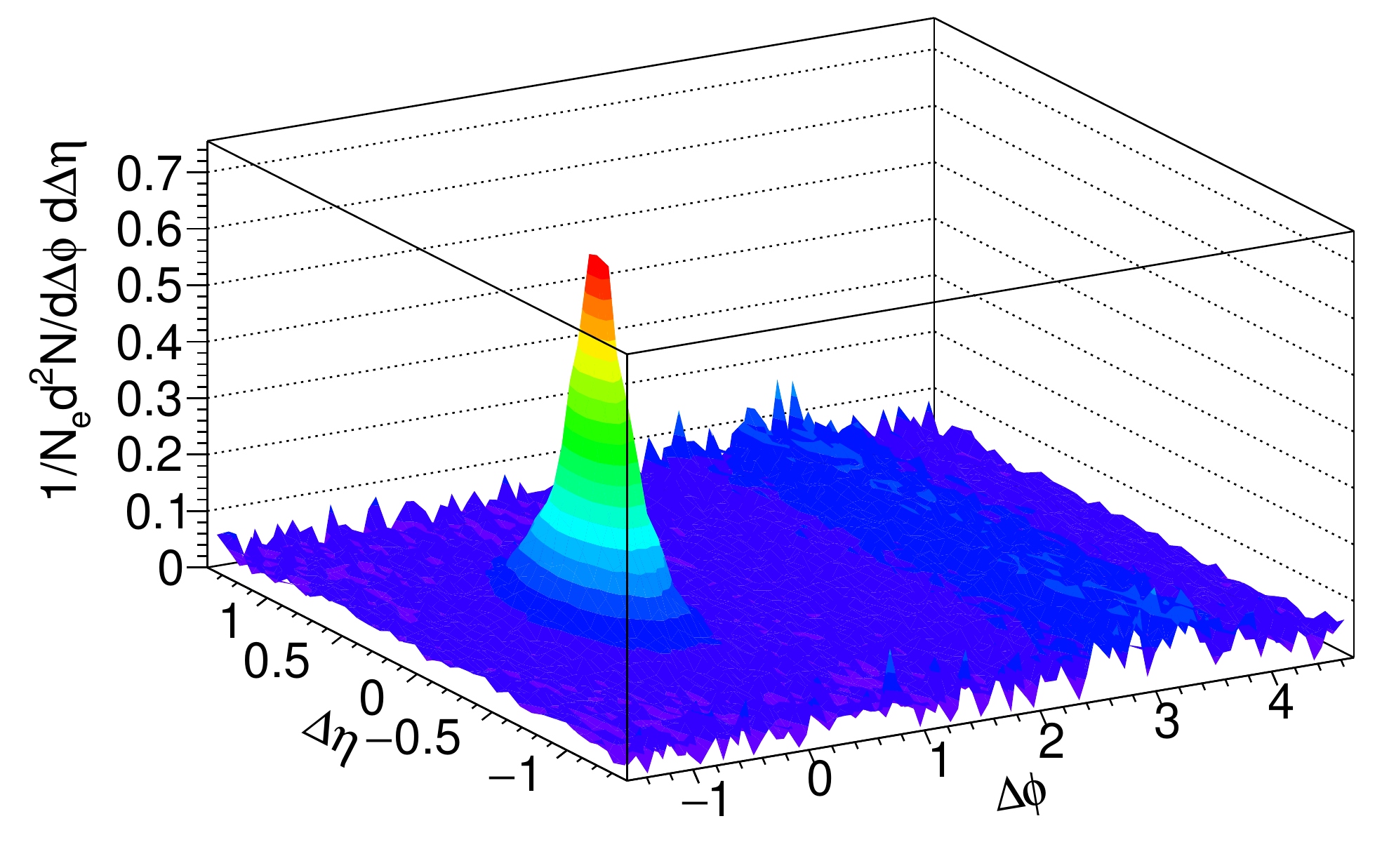}
        \includegraphics{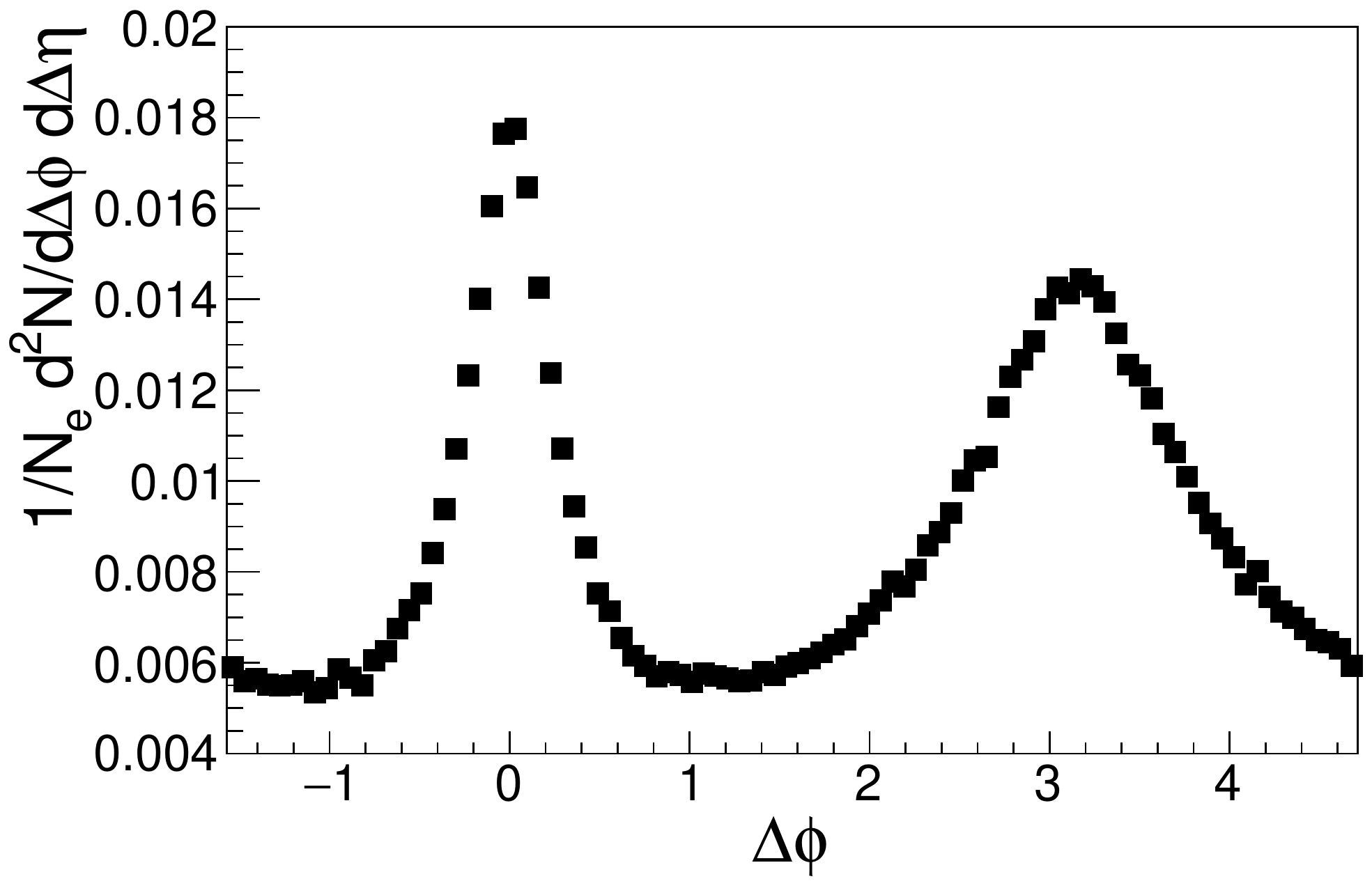}
}}\caption{\Dhcs for trigger momenta \pttrigrange{10}{15} and \ptassocrange{1.0}{2.0} within pseudorapidities \etarange{0.5} and associated particles within \etarange{0.9} in \pp collisions at \sqrts = 2.76 TeV in PYTHIA~\cite{Sjostrand:2006za}.  The signal is normalized by the number of equivalent \Pb collisions.  Left: Correlation function as a function of \dphi and \deta.  Right: Projection onto \dphi.}\label{Fig:Signal}
\end{center}
\end{figure*}

\Fref{Fig:Signal} shows a sample \dhc in \dphi and \deta and its projection onto \dphi for trigger momenta \pttrigrange{10}{15} within pseudorapidities \etarange{0.5} and associated particles within \etarange{0.9} with momenta and \ptassocrange{1.0}{2.0} in \pp collisions at \sqrts = 2.76 TeV in PYTHIA~\cite{Sjostrand:2006za}.  The peak  near \zerodeg, called the \ns, is narrow in both \dphi and \deta and results from associated particles from the same parton as the trigger particle.  The peak near \oneeightydeg, called the \as, is narrow only in \dphi and is roughly independent of pseudorapidity.  This peak arises from associated particles produced by the parton opposing the one which generated the trigger particle. The partons are back-to-back in the frame of the partons, but the rest frame of the partons is not necessarily the same as the rest frame of the incoming nuclei because the incoming partons may not carry the same fraction of the parent nucleons' momentum, x. Since most of the momenta of both the partons and the nucleons are in the direction of the beam (which is universally taken to be the $z$ axis), a difference in pseudorapidity is observed, while the influence on the azimuthal position is negligible.  This causes the \as to be broad in \deta without requiring modified fragmentation or interaction with the medium, as evident in~\Fref{Fig:Signal}.

\subsubsection{Background subtraction methods}

\Dhcs typically have a low signal to background ratio, often less than 1:25.  The raw signal in \dhcs is typically assumed to arise from only two sources, particles from jets and particles from the underlying event, which are correlated with each other due to flow. The production mechanisms of the signal and the background are assumed to be independent so they can be factorized.  These assumptions are called the two source model~\cite{Adler:2005ee}.  The correlation of two particles in the background due to flow is given by~\cite{Adler:2002tq,Bielcikova:2003ku}
\begin{equation}
 \frac{dN}{\pi d\Delta\phi} = B( 1 + \sum_{n=1}^{\infty} 2 v_{n}^{\mathrm{t}} v_{n}^{\mathrm{a}} cos(n\Delta\phi))\label{Eqtn:DHCorrelationBkgd}
\end{equation}
\noindent where $B$ is a constant which depends on the normalization and the multiplicity of trigger and associated particles in an event, the \vntrigger are the \vn for the trigger particle, the \vnassoc are the \vn for the associated particle, and \dphi is the difference in azimuthal angle between the associated particle and the trigger.  The \vn for the trigger particle may arise either from flow, if the trigger particle is not actually from a jet, or from jet quenching, since the path length dependence of partonic energy loss leads to a suppression of jets out-of-plane.
Because \dhcs are typically measured by averaging over positive and negative pseudorapidities, the average $v_1$ due to flow is zero and the $n$~=~1 term is usually omitted.  Global momentum conservation also leads to a $v_1$ signal which is approximately inversely proportional to the particle multiplicity~\cite{Borghini:2000cm}.  The momentum conservation term is typically assumed to be negligible, which may be valid for higher multiplicity events.
The pseudorapidity range for both trigger and associated particles is typically restricted to a region where the \vn do not change dramatically so that the pseudorapidity dependence of  \dNdphi is negligible.  
The azimuthal dependence of any additional sources of long range correlations could be expanded in terms of their Fourier coefficients without loss of generality.

There are two further assumptions commonly used in order to subtract this background: that the appropriate \vn are the same as the \vn measured in other analyses and that there is a region in \dphi near \dphi $\approx$ 1 where the signal is zero.  The latter assumption is called the Zero-Yield-At-Minimum (ZYAM) method~\cite{Adams:2005ph}. Early studies of \dhcs fit the data near \dphi $\approx$ 1 to determine the background level~\cite{Adler:2002tq,Adams:2004wz,Adare:2006nr,Adler:2006sc,Adare:2006nr}.  Later studies typically use a few points around the minimum~\cite{Adler:2005ee,Aggarwal:2010rf,Agakishiev:2010ur}. An alternative to ZYAM for determining the background level, $B$ in Equation \ref{Eqtn:DHCorrelationBkgd}, is the absolute normalization method \cite{Sickles:2009ka}. This method makes no assumption about the background level based on the shape of the underlying background but rather estimates the level of combinatorial pairs from the mean number of trigger and mean number of associated particles in all events as a function of event multiplicity.

It has been suggested that Hanbury-Brown-Twiss (HBT) correlations~\cite{Lisa:2005dd,Lisa:2008gf}, quantum correlations between identical particles from the same source, may contribute to the \ns peak in some momentum regions.  If the momenta of the trigger and associated particles are sufficiently different, these contributions are expected to be negligible.  Distinguishing resonances from jet-like correlations is more difficult.  A high momentum resonance can itself be considered a jet or part of a jet.  The appropriate classification for lower momentum resonances is less clear, but functionally any short range correlations are considered part of the signal in \dhcs.  

The background is then dominated by contributions from flow.  However, this does not mean that the \vn measured in other analyses are necessarily the Fourier coefficients of the background for \dhcs.  
Methods for measuring \vn have varying sensitivities to non-flow (such as jets) and fluctuations~\cite{Voloshin:2008dg}.  Fluctuations in \vn may either increase or decrease the effective \vn, depending on their physical origin and its correlation with jet production.
The correct \vn in \eref{Eqtn:DHCorrelationBkgd} is also complicated by proposed decorrelations between the reaction planes for soft and hard processes, which would change the effective \vn~\cite{Aad:2014fla,Jia:2012ez}.  A recent method uses the reaction plane dependence of the background in \eref{Eqtn:DHCorrelationBkgd} to extract the background level and shape from the correlation itself~\cite{Sharma:2015qra}.

The majority of measurements of \dhcs in heavy ion collisions in the literature omit odd \vn since these studies were done before the odd \vn were observed and understood to arise due to collective flow.  The first direct observation of the odd \vn was in \highpT \dhcs, where subtraction of only the even \vn led to two structures called the ridge (on the \ns)~\cite{Abelev:2009af,Alver:2009id} and the shoulder or Mach cone (on the \as)~\cite{Abelev:2009af,Adare:2007vu,Adare:2008ae,Adare:2008ae,Afanasiev:2007wi,Agakishiev:2010ur}.  This means that the majority of studies of \dhcs at low and intermediate momenta (\pT~$\lesssim$~3~\GeV) do not take the odd \vn into account and therefore include distortions due to flow.  Exceptions are studies which used the \deta dependence on the \ns to subtract the ridge and focused on the \jlc~\cite{Abelev:2016dqe,Agakishiev:2011st,Abelev:2009af,Abelev:2009ah}.  An understanding of the low momentum jet components is important because many of medium modifications of the jet manifest as differences in distributions at low momenta.  While some of the iconic RHIC results showing jet quenching did not include odd \vn~\cite{Adams:2004wz} and the complex structures at low and intermediate momenta are now understood to arise due to flow rather than jets~\cite{Nattrass:2016cln}, some of the broad conclusions of these studies are robust, and studies at sufficiently high momenta (\pT~$\gtrsim$~3~\GeV)  are still valid because the impact of the higher order \vn is negligible.  \Sref{Sec:Results} focuses on results robust to the omission of the odd \vn and more recent results.

\subsection{Reconstructed jets}\label{Sec:JetMethod}

A jet is defined by the algorithm used to group final state particles into jet candidates.  In QCD any parton may fragment into two partons, each carrying roughly half of the energy and moving in approximately the same direction.  This is a difficult process to quantify theoretically and leads to divergencies in theoretical calculations.  A robust jet finding algorithm would find the same jet with the same \pT regardless of the details of the fragmentation and would thus be {\it collinear safe}.  Additionally, QCD allows for an infinite number of very soft partons to be produced during the fragmentation of the parent parton.  All experiments have low momentum thresholds for their acceptance so these particles cannot generally be observed and the production of soft partons leads to theoretical divergencies as well.  A robust jet finding algorithm will find the same jets, even in the presence of a large number of soft partons and would thus be {\it infrared safe}.  In order for the jet definition to be robust, the jet-finding algorithm must be both infrared and collinear safe~\cite{Salam:2009jx}.

Jet finding algorithms are generally characterized by a resolution parameter.  In the case of a conical jet, this is the radius of the jets
\begin{equation}
 R = \sqrt{\Delta\phi^2 + \Delta\eta^2}
\end{equation}
where \dphi is the distance from the jet axis in azimuth and \deta is the distance from the jet axis in pseudorapidity.  A conical jet is symmetric in \dphi and \deta, although it is not theoretically necessary for jets to be symmetric.  We will focus the discussion on conical jets, since they are the most intuitive to understand.  The most common jet-finding algorithm in heavy ion collisions, \akT, usually reconstructs conical jets.  The majority of jet measurements include corrections up to the energy of all particles in the jet, whether or not they are observed directly.  The ALICE experiment also measures charged jets, which are corrected only up to the energy contained in charged constituents.

We emphasize that a measurement of a jet is not a direct measurement of a parton.  A jet is a composite object comprising several final state hadrons.  If the jet reconstruction algorithm applied to theoretical calculations and data is the same, experimental measurements of jets can be comparable to theoretical calculations of jets.  However, even theoretically, it is unclear which final state particles should be counted as belonging to one parton.  What the original parton's energy and momentum were before it fragmented is therefore an ill-posed question.  The only valid comparisons between theory and experiment are between jets comprised of final state hadrons and reconstructed with the same algorithm.  This understanding was the conclusion of the Snowmass Accord~\cite{Huth:1990mi}.  Ideally both the jet reconstruction algorithms and the treatment of the combinatorial background in heavy ion collisions would also be the same for theory and experiment.

\subsubsection{Jet-finding algorithms}

Infrared and collinear safe sequential recombination algorithms such as the \kT, \akT and Cambridge/Aachen (CAMB) are encoded in \FJ~\cite{Salam:2009jx,Cacciari:2011ma, Cacciari:2008gn, bib_antikt,Cacciari:2010te}.  The \FJ~\cite{Cacciari:2011ma} framework takes advantage of advanced computing algorithms in order to decrease computational times for jet-finding.  This is essential for jet reconstruction in heavy ion collisions due to the large combinatorial background.
Due to the ubiquity of the \akT jet-finding algorithm in studies of jets in heavy ion collisions, it is worth describing this algorithm in detail.  The \akT algorithm is a sequential recombination algorithm, which means that a series of stpdf for grouping particles into jet candidates is repeated until all particles in an event are included in a jet candidate.  The steps are:
\begin{enumerate} 
 \item Calculate
\begin{equation}
 d_{ij} = {\mathrm {min}}(1/p_{T,i}^2,1/p_{T,j}^2)\frac{(\eta_i-\eta_j)^2+(\phi_i-\phi_j)^2}{R^2}
\end{equation}
and 
\begin{equation}
 d_{i} = 1/p_{T,i}^2
\end{equation}
for every pair of particles where $p_{T,i}$ and $p_{T,j}$ are the momenta of the particles, $\eta_i$ and $\eta_j$ are the pseudorapidities of the particles, and $\phi_i$ and $\phi_j$ are the azimuthal angles of the particles.  
\item Find the minimum of the $d_{ij}$ and $d_{i}$.  If this minimum is a $d_{ij}$, combine these particles into one jet candidate, adding their energies and momenta, and return to the first step.  
\item If the minimum is a $d_{i}$, this is a final state jet candidate.  Remove it from the list and return to the first step.  Iterate until no particles remain.  
\end{enumerate}
The original implementation of the \akT used rapidity rather than pseudorapidity~\cite{bib_antikt}, however, in practice most experiments cannot identify particles to high momenta and the difference is negligible at high momenta so pseudorapidity is used in practice.

The \akT algorithm has a few notable features for jet reconstruction in heavy ion collisions.  Since $d_{ij}$ is smallest for pairs of \highpT particles, the \akT algorithm starts clustering \highpT particles into jets first and forms a jet around these particles.  The \akT algorithm creates jets which are approximately symmetric in azimuth and pseudorapidity, at least for the highest energy jets.  Particularly in heavy ion collisions, it must be recognized that the ``jets'' from a jet-finding algorithm are not necessarily generated by hard processes.  Since all final state particles are grouped into jet candidates, some jet candidates will comprise only particles whose production was not correlated because they were created in the same hard process but which randomly happen to be in the same region in azimuth and pseudorapidity.  These jet candidates are called fake or combinatorial jets.  Particles that are correlated through a hard process will be grouped into jet candidates, which will also contain background particles.  Care must therefore be used when interpreting the results of a jet-finding algorithm as it is possible to have jet candidates in an analysis that come from processes that may not be included in the calculation used to interpret the results.

There are two important additional points to be made with regard to jet-finding algorithms as applied to heavy ion collisions.  While jet-finding algorithms have been optimized for measurements in small systems such as \ee and \pp collisions, these algorithms are computationally efficient and well-defined both theoretically and experimentally.  Although we may want to consider how we {\it use} these algorithms, there is no need for further development of jet-finding algorithms for use in heavy ion collisions.  However, there is a difference between jet-finding in principle and in practice.  While these jet-finding algorithms are infrared and collinear safe {\it if all particles are input into the jet-finding algorithm}, most experimental measurements restrict the momenta and energies of the tracks and calorimeter clusters input into the jet-finding algorithms.  Some apply other selection criteria to the population of jets, such as requiring a high momentum track, which are not infrared or collinear safe.  These techniques are not necessarily avoidable, especially in the high background environment of heavy ion collisions, however, they must be considered when interpreting the results.

\subsubsection{Dealing with the background}
Combinatorial jets and distortions in the reconstructed jet energy due to background need to be taken into account in order to interpret a measured observable.  This can be done either in the measurement, or in theoretical calculations that are compared to the measurement.  The latter is particularly difficult in a heavy ion environment because the background has contributions from all particle production processes.   
\begin{figure*}
\begin{center}

\rotatebox{0}{\resizebox{!}{5cm}{
	\includegraphics{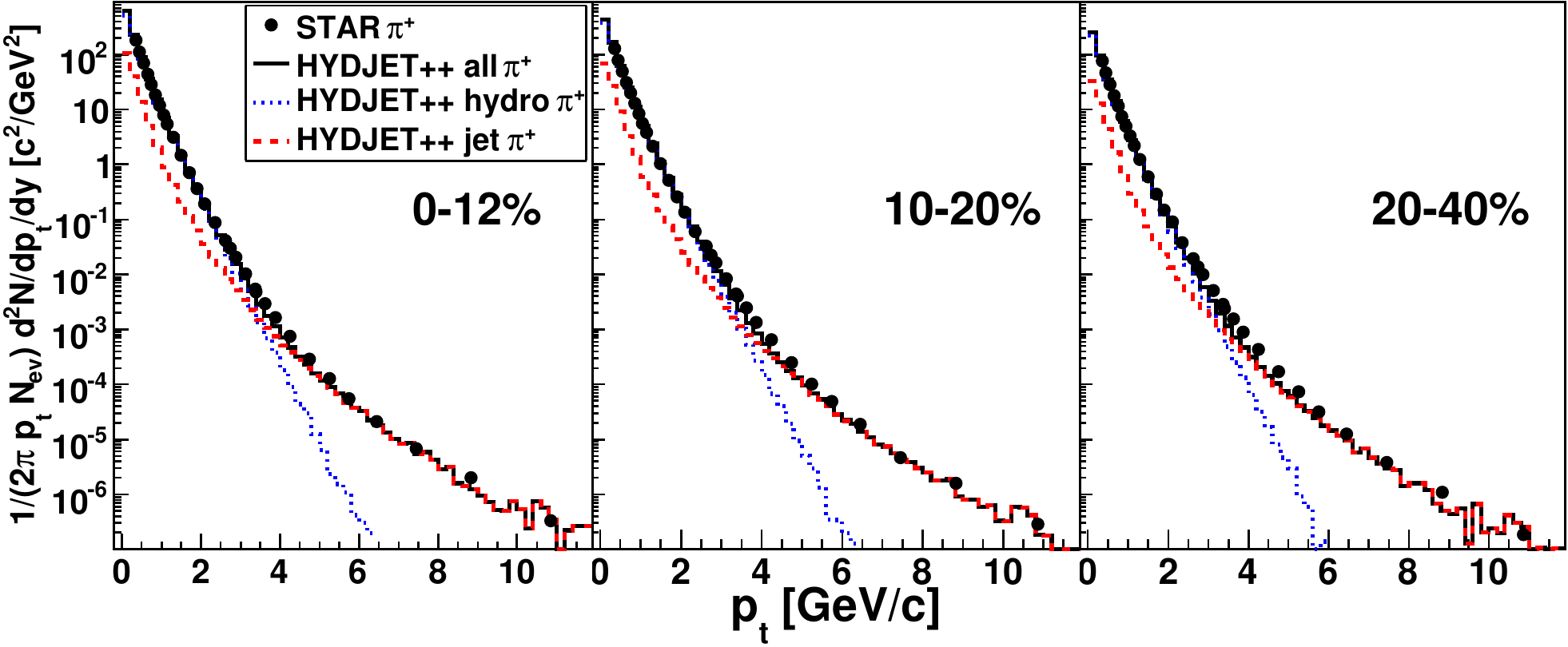}
}}\caption{
Figure from~\cite{Lokhtin:2009be} comparing HYDJET~\cite{Lokhtin:2008xi} calculations to STAR data~\cite{Abelev:2006jr}.  Particle production in HYDJET is separated into those from hard and soft processes.  This shows that at sufficiently high momenta, particle production is dominated by hard processes.
}\label{Fig:HYDJETpTRange}
\end{center}
\end{figure*}

While it is impossible to know  which particles in a jet candidate come from hard processes and which come from the background, and indeed it is even ambiguous to make this distinction on theoretical level, differences between particles in the signal and the background on average can be used to reduce the impact of particles from the background and calculate the impact of the remaining background on an ensemble of jet candidates.  As mentioned in~\Sref{Sec:Intro}, the average momentum of particles in the background is much lower than that of those in the signal.  \Fref{Fig:HYDJETpTRange} shows a comparison of HYDJET to STAR data~\cite{Lokhtin:2009be} and the particles produced by hard and soft processes in HYDJET. At sufficiently high \pT, particle production is dominated by hard processes.  HYDJET has been tuned to match fluctuations and \vn~from heavy ion collisions, so this qualitative conclusion should be robust.  Jets themselves can contribute to background for the measurement of other jets, however, the probability of multiple jets overlapping spatially and fragmenting into several high momentum particles is low.  Therefore, introducing a minimum momentum for particles to be used in jet-finding reduces the number of background particles in the jet candidates.  This also reduces the number of combinatorial jets, since there are very few high momentum particles which were not created from a hard process.
While this selection criterion reduces the background contribution, it is not collinear safe.  Additionally, as most of the modification of the jet fragmentation function is observed for constituents with ${p}_{T} < 3$ GeV, this could remove the modification signature for particular observables.

The effect of the  background can also be reduced by focusing on smaller jets or higher energy jets.  For a conical jet, the jet area is $A_{jet} = \pi R^2$.  The average number of background particles in the jet candidate is proportional to the area.  The background energy scales with the area of the jet, but is independent of the jet energy (assuming that the signal and background are independent), so the fractional change in the reconstructed jet energy due to background is smaller for higher energy jets as the majority of the jet energy is focused in the core of the jet.  Furthermore, in elementary collisions, the distribution of final state particles in the jet as a function of the fraction of the jet energy carried by the particle is approximately independent of the jet energy.  This means that the difference in the average momentum for signal particles versus background particles is larger for high energy jets.  
Since jets that interact with the medium are expected to lose energy and become broader, studies of high momentum, narrow jets alone cannot give a complete picture of partonic energy loss in the QGP.  Furthermore, even in \pp collisions, theoretical calculations are more difficult for jets with smaller cone sizes because they are sensitive to the details of the hadronization~\cite{Abelev:2013fn}.

The fraction of combinatorial jet candidates can also be reduced by requiring additional evidence of a hard process, such as requiring that the candidate jet has at least one particle above a minimum threshold, requiring that the jet candidate have a hard core, or identifying a heavy flavor component within the jet candidate.  We note that the distinction between fake jets and the background contribution in jets from hard processes is ambiguous, particularly for low momentum jets, however, the corrections for these effects are generally handled separately.
Below we review methods for addressing the impact of background particles on the jet energy and corresponding methods for dealing with any remaining combinatorial jets.  Each of these methods have strengths and weaknesses, and may lead to biases in the surviving jet population.

There are five classes of methods for background subtraction in the four experiments which have published jet measurements in heavy ion collisions.  ALICE and STAR use measurements of the average background energy/momentum density in the event to subtract the background contribution from jet candidates.  ATLAS uses an iterative procedure, first finding jet candidates, then omitting them from the calculation of the background energy distribution, and then using this background distribution to find new jet candidates.  CMS subtracts background before jet finding, omitting jet candidates from the background subtraction.  In addition, an event mixing method was recently applied to STAR data to estimate the average contribution from the background to both the jet energy and combinatorial jets.  Constituent subtraction refers to corrections to account for background before jet finding.  Each of these are described in greater detail below.

\paragraph{ALICE/STAR}  In this method the background contribution to a jet candidate is assumed to be proportional to the area of that candidate.  The area of each jet is estimated by filling an event with many very soft, small area particles (ghost particles), rerunning the jet-finder, and then counting how many are clustered into a given jet.  The background energy/momentum density per unit area ($\rho$) is measured by either using randomly oriented jet cones or the \kT jet-finding algorithm and calculating the momentum over the area of the cone or \kT~jet.  The median of the energy per unit area of the collection is used to reduce the impact from real jets in the event on the determination of the background density.  
The two highest energy jets in the event are omitted from the distribution of jets used to determine the background energy density. Since the background has a \pT~modulation that is correlated with the reaction plane, an event plane dependent $\rho$~can be determined as well~\cite{Adam:2015mda}.

This method was proposed in~\cite{Cacciari:2008gn} for measurements in \pp collisions under conditions with high pile up and its feasibility in heavy ion collisions demonstrated in~\cite{Abelev:2012ej}.  The strength of this method is that it can be used even with jets clustered with low momentum constituents.  However, the energy of individual jets is not known precisely since only the average background contribution is subtracted, but the background itself could fluctuate which smears the measurement of the jet energy and momentum.  Additionally measurements of the background energy density can include some contribution from real jets.  Subtracting the average contribution to a jet candidate due to the background may not fully take into account the tendency of jet-finding algorithms to form combinatorial jets around hot spots in the background.

\paragraph{ATLAS}  
We outline the approach in~\cite{Aad:2012vca}.  We note that the details of the analysis technique are optimized for each observable.  ATLAS measures both calorimeter and track jets.  Track jets are reconstructed using charged tracks with \pT~$>$~4~\GeV.  The high momentum constituent cut strongly suppresses combinatorial jets, and ATLAS estimates that a maximum of only 4\% of all R~=~0.4 \akT track jet candidates in 0-10\% central \Pb collisions contain a 4 \GeV~background track.  For calorimeter jet measurements, ATLAS estimates the average background energy per unit area and the \vtwo using an iterative procedure~\cite{Aad:2012vca}.  In the first step, jet candidates with R~=~0.2 are reconstructed.  The background energy is estimated using the average energy modulated by the \vtwo calculated in the calorimeters, excluding jet candidates with at least one tower with \ET~$>$~\meanET.  Jets from this step with \ET~$>$~25 GeV and track jets with \pT~$>$~10 \GeV are used to calculate a new estimate of the background and a new estimate of \vtwo, excluding all clusters within $\Delta$R~$<$~0.4 of these jets.  This new background modulated by the new \vtwo and jets with \ET~$>$~20~GeV were considered for subsequent analysis.

Combinatorial jets are further suppressed by an additional requirement that they match a track jet with high momentum (e.g. \pT~$>$~7~\GeV~\cite{Aad:2012vca}) or a high energy cluster (e.g. \ET~$>$~7~GeV~\cite{Aad:2012vca}) in the electromagnetic calorimeter.  These requirements strongly suppress the combinatorial background, however, they may lead to fragmentation biases and may suppress the contribution from jets which have lost a considerable fraction of their energy in the medium.  These biases are likely small for the high energy jets which have been the focus of ATLAS studies, however, the bias is stronger near the 20 GeV lower momentum threshold of ATLAS studies.

\paragraph{CMS}  In measurements by CMS the background is subtracted from the event before the jet-finding algorithm is run.  The average energy and its dispersion is calculated as a function of $\eta$.  Tower energies are recalculated by subtracting the mean energy plus the mean dispersion.  Negative energies after this step are set to zero.  These tower energies are input into a jet-finding algorithm and the background is recalculated, omitting towers contained in the jets.  The tower energies are again calculated by subtracting the mean energy plus the dispersion and setting negative values to zero.

\paragraph{Event Mixing}
The goal of event mixing is to generate the combinatorial background -- in the case of jet studies, fake jets.  In STAR, the fraction of combinatorial jets in an event class is generated by creating a mixed event where every track comes from a different event~\cite{Adamczyk:2017yhe}.  The data are binned in classes of multiplicity, reconstructed event plane, and z-vertex position so that the mixed event accurately reflects the distribution of particles in the background.  Jet candidates are reconstructed using this algorithm in order to calculate the contribution from combinatorial jets, which can then be subtracted from the ensemble.  This is a very promising method, particularly for low momentum jets, but we note that it is sensitive to the details of the normalization at low momenta.  It is also computationally intensive, which may make it impractical, and it is unclear how to apply it to all observables.

\paragraph{Constituent Subtraction}
The constituent background subtraction method was first developed to remove pile-up contamination from LHC based experiments, where it is not unusual to have contributions from multiple collisions in a single event.  Unlike the area based subtraction methods described above, the constituent method subtracts the background constituent-by-constituent.  The intention is to correct the 4-momentum of the particles, and thus correct the 4-momentum of the jet \cite{Berta:2014eza}.  It is necessary to consider the jet 4-momentum for some of the new jet observables that will be described in this paper, such as jet mass.  The process is an iterative scheme that utilizes the ghost particles, which are nearly zero momentum particles with a very small area on the order of 0.005 which are embedded into the event by many jet finding algorithms.  The jet finder is then run on the event, and the area is determined by counting the number of ghost particles contained within the jet.  Essentially the local background density is determined and then subtracted from the constituents, which are thrown out if they reach zero momentum.  The effect of this background scheme on the applicable observables is under study and it is not clear as of yet what its effect is compared to the more traditional area based background subtraction schemes.

\subsection{Particle Flow}
 The particle flow algorithm was developed in order to use the information from all available sub-detectors in creating the objects that are then clustered with a jet-finding algorithm.  Many particles will leave signals in multiple sub-detectors. For instance a charged pion will leave a track in a tracker and shower in a hadronic calorimeter.  If information from both detectors is used, this would double count the particle.  However, excluding a particular sub-detector would remove information about the energy flow in the collision as well.  Tracking detectors generally provide better position information while hadronic calorimeters are sensitive to more particles but whose positions are altered by the high magnetic field necessary for tracking.  The goal is to use the best information available to determine a particle's energy and position simultaneously.

The particle flow algorithm operates by creating stable particles from the available detectors.  Tracks from the tracker are extrapolated to the calorimeters -- in the case of CMS, an electromagnetic calorimeter and a hadronic calorimeter \cite{CMS-PAS-PFT-09-001}.  If there is a cluster in the associated calorimeter, it is linked to the track in question.  Only the closest cluster to the track is kept as a charged particle should only have a single track.  The energy and momentum of the cluster and track are compared.  If the energy is low enough compared to the momentum, only a single hadron with momentum equal to a weighted average of the track and calorimeter is created.  The exact threshold should depend on the details of the detector and its energy resolution.  If the energy is above a certain threshold, neutral particles are then created out of the excess energy.  If that excess is only in an electromagnetic calorimeter, the neutral particle is assumed to be a photon.  If the excess is in a hadronic calorimeter, the neutral particle is assumed to be a hadron.  If there is some combination, multiple neutral particles may be created, with the photon given preference in terms of "using up" the excess energy.

By grouping the information into individual particles, the particle flow algorithm reduces the
sensitivity of the measurement of the jet energy to the jet fragmentation pattern.  This is a correction that can be done prior to unfolding, which is described below.  The particle flow algorithm can be a powerful tool, however, it depends on the details of the sub-detectors that are available, their energy resolution, and their granularity.  For example, the ALICE detector has precision tracking detectors and an electromagnetic calorimeter but no hadronic calorimeter.  The optimal particle flow algorithm for the ALICE detector is to use the tracking information when available and only use information from the electromagnetic calorimeter if there is no information from the tracking detectors.  Additionally, the magnetic field strength plays a role, as this will dictate how much the charge particle paths diverge from one another before reaching the calorimeter and how far charged particles are deflected before reaching the calorimeters.  To fully utilize this algorithm, the energy resolution of all calorimeters must be known precisely, and the distribution of charged and neutral particles must be known. 
 
\subsection{Unfolding}
\begin{figure*}
\begin{center}
\rotatebox{0}{\resizebox{!}{6cm}{
	\includegraphics{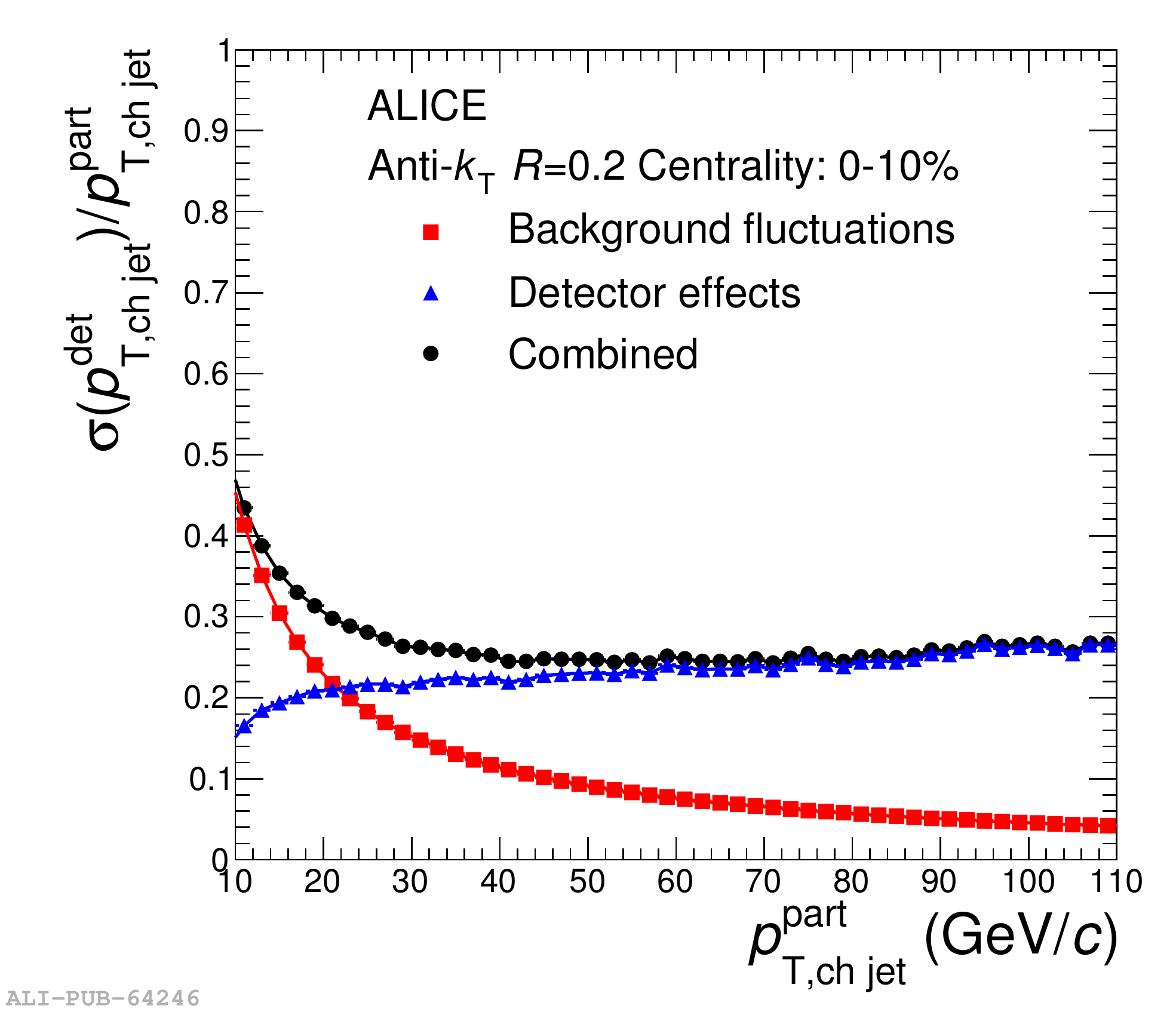}
	\includegraphics{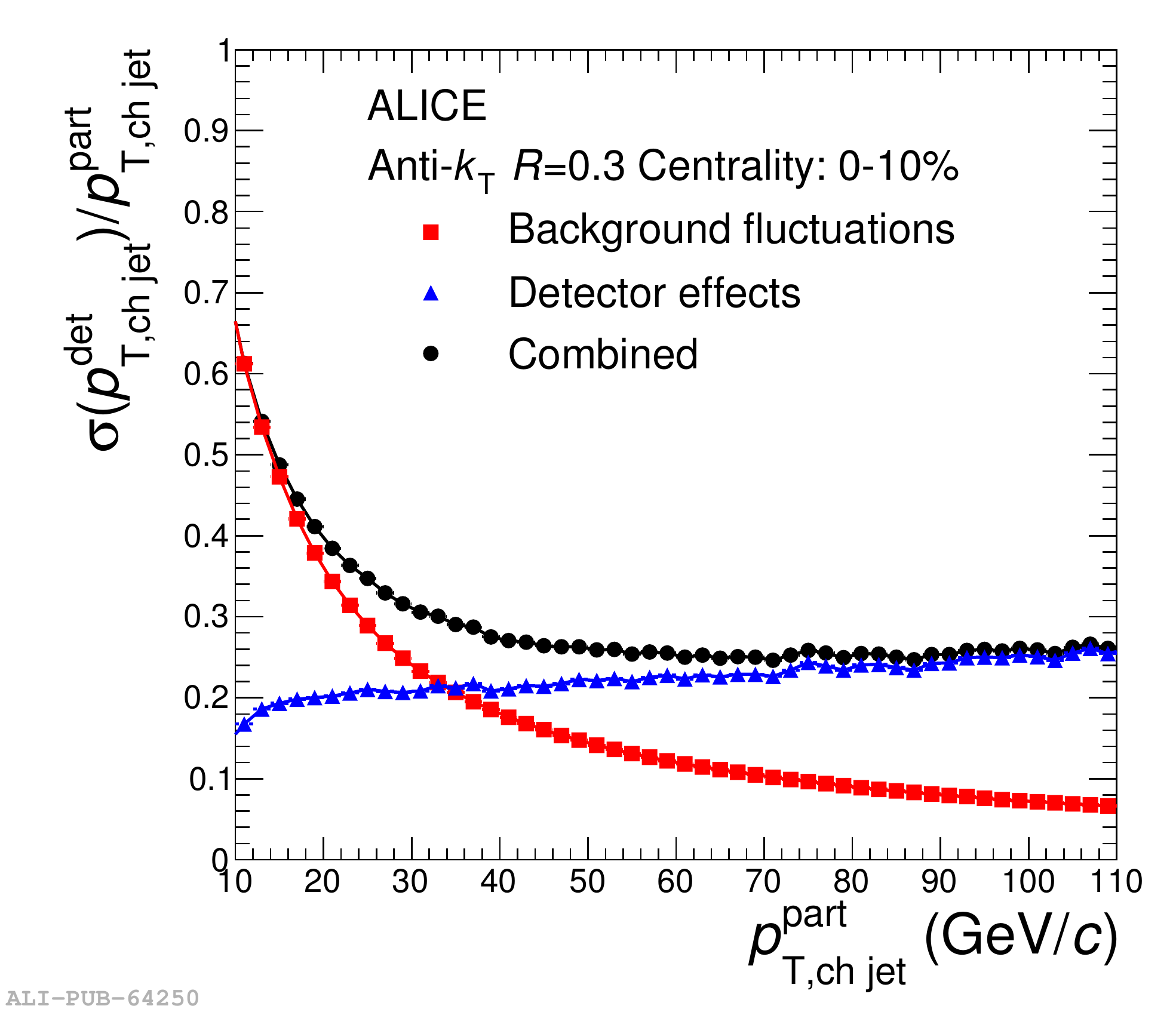}
}}\caption{
Figure from ALICE~\cite{Abelev:2013kqa}.  On the left is the standard deviation of the combined jet response (black circles) for R=0.2 \akT jets , including background fluctuations (red squares) and detector effects (blue triangles) for 0-10\% central \Pb events.  On the right is the standard deviation of the combined jet response (black circles) for R=0.3 \akT jets , including background fluctuations (blue triangles) and detector effects (red squares) for 0-10\% central Pb-Pb events.  The background effects increase the jet energy resolution more for larger jets, as can be seen from the difference in the background distributions in both plots.  For high momentum jets, where the momentum of the jet is much larger than background fluctuations, the jet energy resolution will be dominated by detector effects.
}\label{Fig:JetResolution}
\end{center}
\end{figure*}

Before comparing measurements to theoretical calculations or other measurements, they must be corrected for both detector effects and smearing due to background fluctuations.
Both the jet energy scale (JES) and the jet energy resolution (JER) need to be considered in any correction procedure.  The jet energy scale is a correction to the jet to recover the true 4-vector of the original jet (and not of the parton that created it).  The background subtraction methods described above are examples of corrections to the jet energy scale due to the addition of energy from the underlying background.  Precision measurements of the energy scale, as done by the ATLAS collaboration~\cite{ATLAS-CONF-2015-016}, are an important step in understanding the detector response and necessary to reduce the systematic uncertainties.  The jet energy resolution is a measure of the width of the jet response distribution. An example from the ALICE experiment can be seen in \Fref{Fig:JetResolution}.   In heavy-ion collisions there are two components, the increase in the distribution due to the fluctuating background that will be clustered into the jet, and due to detector effects.

In most measurements of reconstructed jets, the jet energy resolution is on the order of 10-20\% for the high momentum jets, where detector effects dominate.  This can be understood because even a hadronic calorimeter is not equally efficient at observing all particles.  In particular, the measurement of neutrons, antineutrons, and the $K_L^0$ is difficult.  The high magnetic field necessary for measuring charged particle momentum leads to a lower threshold on the momenta of reconstructed particles and can sweep charged particles in or out of the jet.  As a result, even an ideal detector has a limited accuracy for measuring jets.  The large fluctuations in the measured jet energy due to these effects distort the measured spectrum.  This is qualitatively different from measurements of single particle observables, where the momentum resolution is typically 1\% or better, often negligible compared to other uncertainties.  This means that measurements of jet observables must be corrected for fluctuations due to the finite detector resolution if they will be compared to theoretical calculations or to measurements of the same observable in a different detector, or even from the same detector with different running conditions.  Fluctuations in the background in \AplusA collisions lead to further distortions in the reconstructed jet energy.  Correcting for these effects is generally referred to as unfolding in high energy physics, although it is called unsmearing or deconvolution in other fields.

Here we summarize unfolding methods, based on the discussion in~\cite{Adye:2011gm,Cowan:2002in}.  If the true value of an observable in a bin $i$ is given by $y_i^{true}$, then the observed value in bin j, $y_j^{reco}$, is given by
\begin{equation}
 y_j^{reco} = \sum_{i=0}^{N} R_{ij} y_i^{true}\label{Eq:MatrixInversion}
\end{equation}
where $R_{ij}$ is the response matrix relating the true and reconstructed values.  

The response matrix is generally determined using Monte Carlo models including particle production, propagation of those particles through the detector material and simulation of its response, and application of the measurement algorithm, although sometimes data-driven corrections are incorporated into the response matrix.  As an example, we consider the analysis of jet spectra.  The truth result ($y_i^{true}$) is usually generated by an event generator such as PYTHIA~\cite{Sjostrand:2006za} or DPMJET~\cite{Ranft:1999qe}.  The jet finding algorithm to be used in the analysis is run on this truth event, which generates the particle level jets comprising $y_i^{true}$.  The truth event is then run through a simulation of the detector response.  It is common to include a simulated background from a generator such as HIJING~\cite{hijing}, but not required.  This creates the reconstructed event, and as before, the jet finding algorithm used in the analysis is run on this event to create the detector level jets that make up $y_j^{reco}$.  Next, the particle level jets must be matched to detector level jets to build the response matrix, with unmatched jets determining the reconstruction efficiency.  There are several ambiguities in this method.  The first is that it comes with an assumption of the spectra shape and fragmentation pattern of the jets from the simulation.  The second is that there is not always a one-to-one correspondence between the truth and detector level jets.  The detector response may cause the energy of a particular truth jet to be split into two detector level jets.  However, the response matrix requires a one-to-one correspondence, which necessitates a choice. 

If one could simply invert the response matrix,it would be possible to determine $y_i^{true} = \sum_{i=0}^{N} R_{ij}^{-1} y_j^{reco}$.  However, response matrices for jet observables are generally ill-conditioned and not invertible.  The further the jet response matrix is from a diagonal matrix, the more difficult the correction procedure is.  This is one reason the background subtraction methods outlined in the preceding section are employed.  By correcting the jet energy scale on a jet-by-jet basis, the response matrix is much closer to a diagonal matrix, however this is not a sufficient correction.   The process of unfolding is thus required to determine $y_i^{true}$ given the information in Equation \ref{Eq:MatrixInversion}.

One of the main challenges in unfolding is that it is an ill-posed statistical inverse problem which means that even though the mapping of $y_i^{true}$ to $y_j^{reco}$ is well-behaved, the inverse mapping of  $y_j^{reco}$ to $y_i^{true}$ is unstable with respect to statistical fluctuations in the smeared observations.  This is a problem even if the the response matrix is known with precision.  The issue is that  within the statistical uncertainties, the smeared data can be explained by the actual physical solution, but also by a large family of wildly oscillating unphysical solutions.   The smeared observations alone cannot distinguish among these alternatives, so additional a priori information about physically plausible solutions needs to be included.  This method of imposing physically plausible solutions is called
regularization, and it essentially is a method to reduce the variance of the unfolded truth points by introducing a bias.  The bias generally comes in the form of an assumption about the smoothness of the observable, however, this assumption always results in a loss of information.

If an observable is described well by models, it may be possible to correct the measurement using the ratio of the observed to the true value in Monte Carlo:
\begin{equation}
 \gamma_j^{true} =  \frac{\gamma_j^{true,MC}}{y_j^{reco,MC}}  y_j^{reco}
\end{equation}
where $\gamma_j^{true}$ is the estimate of the true value, $\gamma_j^{true,MC}$ is the true value in the Monte Carlo model, and $y_j^{reco,MC}$ is the measurement predicted by the model.  This approach is called a bin-by-bin correction.  It is also satisfactory when the response matrix is nearly diagonal which is generally true when the bin width is wider than the resolution in the bin.  In this circumstance, the inversion of the response matrix is generally stable and the measurement is not affected significantly by statistical fluctuations in the measurement or the response matrix.  For example, bin-by-bin efficiency corrections to measurements of single particle spectra may be adequate as long as the momentum resolution is fairly good and the input spectra have roughly the same shape as the true spectra.  This approach can work for measurements of reconstructed jets in systems such as \pp collisions [e.g. fragmentation function measurements].  Unfortunately, for typical jet measurements, the desired binning is significantly narrower than the jet energy resolution, and fluctuations in the response matrix then lead to instabilities if the response matrix is inverted.   Additionally, the high background environment of heavy ion collisions leads to lower energy resolution, and Monte Carlo models generally do not describe the data well.  Bin-by-bin corrections are therefore usually inadequate for measurements in heavy ion collisions.

Several algorithms have been developed to solve \eref{Eq:MatrixInversion}.  The two most commonly used algorithms are Single Value Decomposition (SVD)~\cite{Hocker:1995kb} and Bayesian Unfolding~\cite{DAgostini:1994zf}.  Bayesian unfolding uses a guess, which is called the prior of the true distribution, usually from a Monte Carlo model, as the start of an iterative procedure.  This method is regularized by choosing how many iterations to use, where choosing an early iteration will result in a distribution that is closer to the prior, and thus more regularized.  As the number of iterations increase there is a positive feedback which is driven by fluctuations in the response matrix and spectra, that makes the asymptotically unfolded spectrum diverge sharply from reality. The SVD formalism is a way by which to factorize a matrix into a set of matrices. This is used to write the 'unfolding' equation as a set of linear equations, with the assumption that the response matrix $R$ can be decomposed into three matrices such that $R = U S V^T$ where $U$ and $V$ are orthogonal and $S$ is diagonal.  The regularization method for using SVD formalism in unfolding uses a dampened least squares method to couple all the linear equations that come out of the process and solve them.  One then chooses a parameter, $k$, which corresponds to the ${k}^{\textrm{th}}$ singular value of the decomposed matrix, and suppresses the oscillatory divergences in the solution. 

It is worth noting that for any approach, there is a trade off between potential bias imposed on the results by the input from the Monte Carlo and the uncertainty in the final result.  In practice, different methods and different training for Bayesian unfolding are compared for determination of the systematic uncertainties.  
For measurements where models describe the data well or where the resolution leads to minimal bin-to-bin smearing, bin-by-bin corrections are often preferred, both because of the potential bias and because of the difficulty of unfolding.

In order to confirm whether a particular algorithm used in unfolding is valid, it is necessary to perform closure tests, demonstrations that the method leads to the correct value when applied to a Monte Carlo model.  The most simple tests are to convolute the Monte Carlo truth distribution with the response matrix to form a simulated detector distribution.  This distribution can then be unfolded and compared to the original truth distribution.  For this test, one should use roughly the same statistical precision as will be available in the data given how strongly the unfolding procedure is driven by statistics.  However, this does not test the validity of the response matrix, or of the choice of spectral shape for the input distribution, or of the effect of combinatorial jets that will appear in the measured data.  A more rigorous closure test can be done by embedding the detector level jets into minimally biased data, and performing the background and unfolding procedures on the embedded data to compare with the truth distribution.  

Another approach is to ``fold'' the reference to take detector effects into account.  For example, the initial measurements of the dijet asymmetry did not correct for the effect of background or detector resolution in \Pb but instead embedded \pp jets in a \Pb background in order to smear the \pp by an equivalent amount~\cite{Chatrchyan:2011sx,Aad:2010bu}.  This may lead to a better comparison between data and a particular theory, but since the response matrix is generally not made available outside of the collaboration, it can only be done by experimentalists at the time of the publication.  However, this would be an important cross-check for any model as it removes the mathematical uncertainty due to the ill posed inverse problem.

\subsection{Comparing different types of measurements}

The ultimate goal of measurements of jets in heavy ion collisions is not to learn about jets but to learn about the QGP.  Measurements of jets in \ee and \pp collisions are already complicated and the addition of a large combinatorial background in heavy ion collisions imposes greater experimental challenges.  Suppressing and subtracting the background imposes biases on the resultant jet collections.  Additionally, selection criteria applied to the collection of jet candidates in order to remove the combinatorial contribution will also impose a bias.  The exact bias imposed by these assumptions cannot be known without a complete understanding of the QGP, which is what we are trying to gain by studying jets.  
Occasionally various methods are claimed to be ``unbiased'', but is unclear what this means precisely since every measurement is biased towards a subset of the population of jets created in heavy ion collisions.   Any particular measurement may have several types of bias.  We discuss a few types of bias below.

\paragraph*{Survivor bias}
As jets interact with the medium and lose energy to the medium, they may begin to look more like the medium.  There are fluctuations in how much energy each individual parton will lose in the medium, and selecting jets which look like jets in a vacuum may skew our measurements towards partons which have lost less energy in the medium.

\paragraph*{Fragmentation bias}
Many measurement techniques select jets which have hard fragments, which may lead to a survivor bias since interactions with the medium are expected to soften the fragmentation function.  Some measurements may preferentially select jets which fragment into a particular particle, such as a neutral pion or a proton.  This in turn can bias the jet population towards quark or gluon jets.  If fragmentation is modified in the medium, it could also bias the population towards jets which either have or have not interacted with the medium.

\paragraph*{Quark bias}
Even in \ee collisions, quark and gluon jets have different structures on average, with gluon jets fragmenting into more, softer particles at larger radii~\cite{OPAL:1995ab,Abreu:1995hp}.  A bias may also be imposed by the jet-finding algorithm.  OPAL found that gluon jets reconstructed with the \kT jet finding algorithm generally contained more particles than those reconstructed with the cone algorithm in~\cite{Abe:1991ui} and that gluon jets contain more baryons~\cite{Ackerstaff:1998ev}.

The measurement techniques described above generally focus on higher momentum jets which fragment into harder constituents and have narrower cone radii.  This surely induces a bias towards quark jets.  Since gluon jets are expected to outnumber quark jets significantly~\cite{Pumplin:2002vw}, this may not be quantitatively significant overall, depending on the measurement and the collision energy.  In some measurements, survivor bias is used as a tool.  For instance measurements of hadron-jet correlations select a less modified jet by identifying a hard hadron and then look for its partner jet on the \as~\cite{Adam:2015doa}.  Correlations requiring a trigger on both the near and away sides select jets biased to be near the surface of the medium~\cite{Agakishiev:2011nb}.  These biases are inherently unavoidable and they must be understood in order to properly interpret data.  However, once they are well understood, the biases can be engineered to purposefully select particular populations of jets, for instance to select jets biased towards the surface in order to increase the probability that the away side jet has traversed the maximum possible medium.

As our experience with the \vn modulated background in dihadron correlations shows, the issue is not merely which measurements are most sensitive to the properties of the medium but the possibility that our current understanding of the background may be incomplete.  However, the potential error introduced varies widely by the measurement -- single particle spectra, dihadron correlations, and reconstructed jets all have completely different biases and assumptions about the background.  Our certainty in the interpretation of the results is therefore enhanced if the same conclusions can be drawn from measurements of multiple observables.  We therefore discuss a variety of different measurements in \Sref{Sec:Results} and demonstrate that they all lead to the same conclusions -- partons lose energy in the medium and their constituents are broadened and softened in the process.
 \section{Overview of experimental results}\label{Sec:Results}
RHIC and the LHC have provided a wealth of data which enhance our understanding of the properties of the QGP. This section of the article reviews experimental results available at of the time of publication, and is organized according to the physics addressed by the measurement rather than according to observable to focus on the implications of the measurements. Therefore the same observable may appear in multiple subsections. The questions that jet studies attempt to answer to understand the QGP are: Are there cold nuclear matter effects which must be taken into consideration in order to interpret results in heavy ion collisions?  Do partons lose energy in the medium and how much?  How do partons fragment in the medium?  Is fragmentation the same as in vacuum or is it modified?  Where does the lost energy go and how does it influence the medium?  Finally, in the next section we will discuss how well these questions have been answered and the questions that remain.

\subsection{Cold nuclear matter effects}\label{Sec:ResultsColdNuclearMatter}
Cold nuclear matter effects refer to observed differences between \pp and \pA or \dA collisions where a hot medium is not expected, but the presence of a nucleus in the initial state could influence the production of the final observable.  These effects may result from coherent multiple scattering within the nucleus~\cite{Qiu:2004da}, gluon shadowing~\cite{Gelis:2010nm}, or partonic energy loss within the nucleus~\cite{Wang:2001ifa,Bertocchi:1976bq,Vitev:2007ve}.  While such effects are interesting in their own right, if present, they would need to be taken into account in order to interpret heavy ion collisions correctly.
Studies of open heavy flavor at forward rapidities through spectra~\cite{Adare:2012yxa} and correlations~\cite{Adare:2013xlp} of leptons from heavy flavor decays indicate that heavy flavor is suppressed in cold nuclear matter.  The J/$\psi$ is also suppressed at forward rapidities~\cite{Adare:2013ezl}.  Recent studies have also indicated that there may be collective effects for light hadrons in \pA collisions~\cite{Adam:2015bka,Khachatryan:2015waa,Aad:2014lta} and even high multiplicity \pp events~\cite{Khachatryan:2016txc,Aad:2015gqa}.  Studies of jet production in \pA or \dA collisions are necessary to quantify the cold nuclear matter effects and decouple which effects observed in \AplusA data come from interactions with the medium. 
\begin{figure}
\begin{center}

\subsubsection{Inclusive charged hadrons}
\rotatebox{0}{\resizebox{8cm}{!}{
	\includegraphics{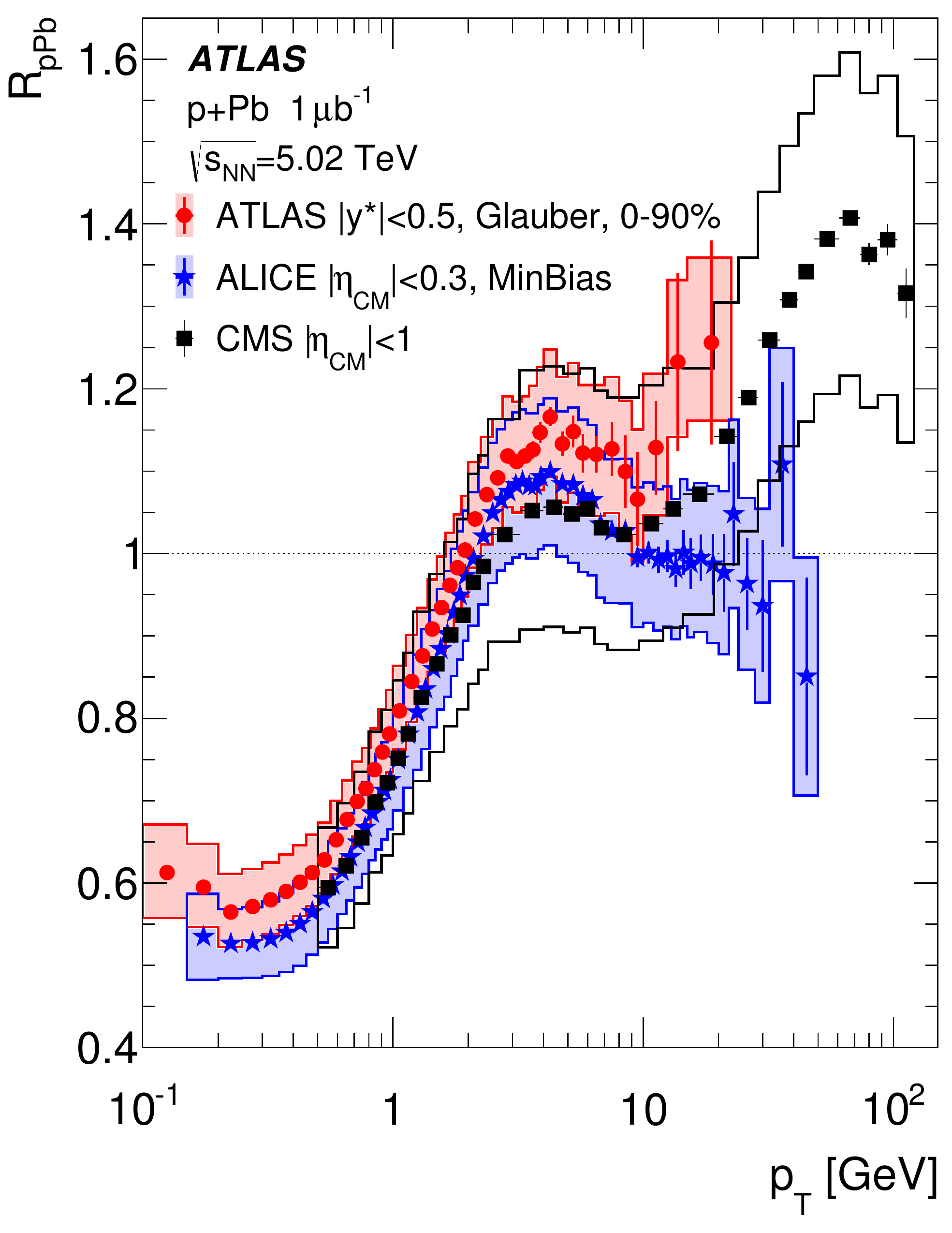}
}}\caption{ 
Figure from ATLAS~\cite{Aad:2016zif}.
The nuclear modification factor of charged hadrons in \pPb collisions at \sNN = 5.02 TeV measured by the ALICE~\cite{ALICE:2012mj}, ATLAS~\cite{Aad:2016zif}, and CMS~\cite{Khachatryan:2015xaa} experiments.  The data in this figure used an extrapolation of \pp data from \sNN = 2.76 TeV and 7 TeV as there was not a \pp reference at the same energy available at this time.  This shows that \RpPb is consistent with one within uncertainties for high \pT hadrons.
}\label{Fig:ATLASRppb}
\end{center}
\end{figure}

Measurements of inclusive hadron \RdAu at \sNN = 200 GeV~\cite{Adler:2006wg,Abelev:2009hx} and \RpPb at \sNN = 5.02 TeV ~\cite{ALICE:2012mj,Aad:2016zif,ATLAS-CONF-2016-108,Khachatryan:2015xaa,Khachatryan:2016odn} are consistent with one within the systematic uncertainties of these measurements, indicating that the large hadron suppression observed in \AplusA collisions can not  be due to cold nuclear matter effects.  This is shown in \Fref{Fig:ATLASRppb}.  We note here that the CMS results shown here were updated with a \pp reference measured at \sNN = 5.02 TeV~\cite{Khachatryan:2016odn}, which is also consistent with an \RpPb~of one.

\subsubsection{Reconstructed jets}

\begin{figure*}
\begin{center}

\rotatebox{0}{\resizebox{15cm}{!}{
	\includegraphics{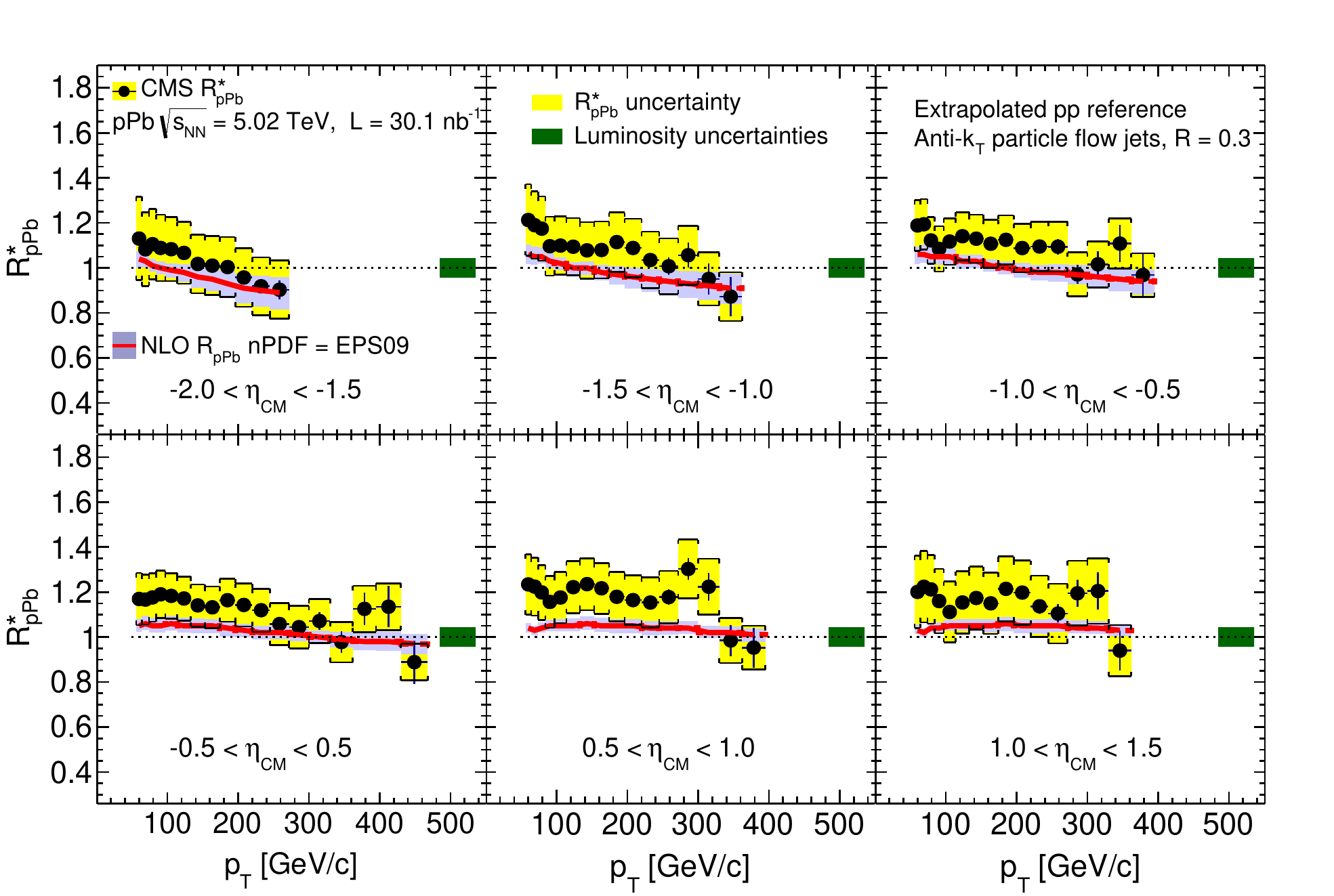}
}}\caption{ Figure from CMS~\cite{Khachatryan:2016xdg}.  The nuclear modification factor of jets in \pPb collisions measured by the CMS experiment in various rapidity bins. This shows that cold nuclear matter effects are small for jets.
}\label{Fig:CMSRppb}
\end{center}
\end{figure*}

Measurements of reconstructed jets in \dAu collisions at \sNN = 200 GeV and \pPb collisions at 5.02 TeV indicate that the minimum bias \RdAu~\cite{Adare:2015gla} and \RpPb~\cite{ATLAS:2014cpa,Adam:2016jfp}, respectively, are also consistent with one. Figure~\ref{Fig:CMSRppb} shows \RpPb measured by the CMS experiment and compared with NLO calculations including cold nuclear matter effects. The theoretical predictions and the experimental measurements in \Fref{Fig:CMSRppb} show that cold nuclear matter effects are small for jets for all \pT and pseudorapidity measured at the LHC.   A centrality dependence at midrapidity in 200 GeV \dAu and 5.02 TeV \pPb collisions which cannot be fully explained by the biases in the centrality determination as studied in~\cite{Adare:2013nff,Aad:2015ziq} is observed.  It has been proposed that the forward multiplicities used to determine centrality are anti-correlated with hard processes at midrapidity~\cite{Bzdak:2014rca,Armesto:2015kwa} or 
that the rare high-$x$ parton configurations of the proton which produce high-energy jets have a smaller cross-section for inelastic interactions with nucleons in the nucleus~\cite{Alvioli:2014eda,Coleman-Smith:2013rla,Alvioli:2014sba,Alvioli:2013vk}.  The latter suggests that high \pT jets may be used to select proton configurations with varying sizes due to quantum fluctuations. 
While this is interesting in its own right and there may be initial state effects, there are currently no indications of large partonic energy loss in small systems, thus scaling the production in \pp with the number of binary nucleon-nucleon collisions as a reference appears to valid for comparison to larger systems.

\subsubsection{Dihadron correlations}

Detailed studies of the jet structure in \dAu and comparisons to both PYTHIA and \pp collisions using \dhcs at \sNN = 200 GeV found no evidence for modification of the jet structure at midrapidity in cold nuclear matter~\cite{Adler:2005ad}.  Studies of correlations between particles at forward rapidities (1.4$< \eta <$ 2.0 and -2.0$< \eta <$ -1.4) in order to search for fragmentation effects at low $x$ also found no evidence for modified jets in cold nuclear matter~\cite{Adler:2006hi}.  
However, \jlcs with particles at higher rapidities (3.0$< \eta <$ 3.8) indicated modifications of the correlation functions in \dAu collisions at \sNN = 200 GeV~\cite{Adare:2011sc}.  This indicates that nuclear effects may have a strong dependence on $x$ and that studies of cold nuclear matter effects for each observable are important in order to demonstrate the validity of the baseline for studies in hot nuclear matter.  While there is little evidence for effects at midrapidity, observables at forward rapidities may be influenced by effects already present in cold nuclear matter.  Searches for acoplanarity in jets in \pPb collisions observed no difference between jets in \pPb and \pp collisions~\cite{Adam:2015xea}.

\subsubsection{Summary of cold nuclear matter effects for jets}

Based on current evidence from \pPb and \dAu collisions, \pp collisions are an appropriate reference for jets, however, since numerous cold nuclear matter effects have been documented, each observable should be measured in cold nuclear matter in order to properly interpret data in hot nuclear matter.
We therefore conclude that, based on the current evidence, \pPb and \dAu collisions are appropriate reference systems for hard processes in \AplusA collisions, although caution is needed, particularly at at large rapidities and high multiplicities, and future studies in small systems may lead to different conclusions.

 \subsection{Partonic energy loss in the medium}\label{Sec:EnergyLoss}

\begin{figure}
\begin{center}

\rotatebox{0}{\resizebox{8cm}{!}{
        \includegraphics{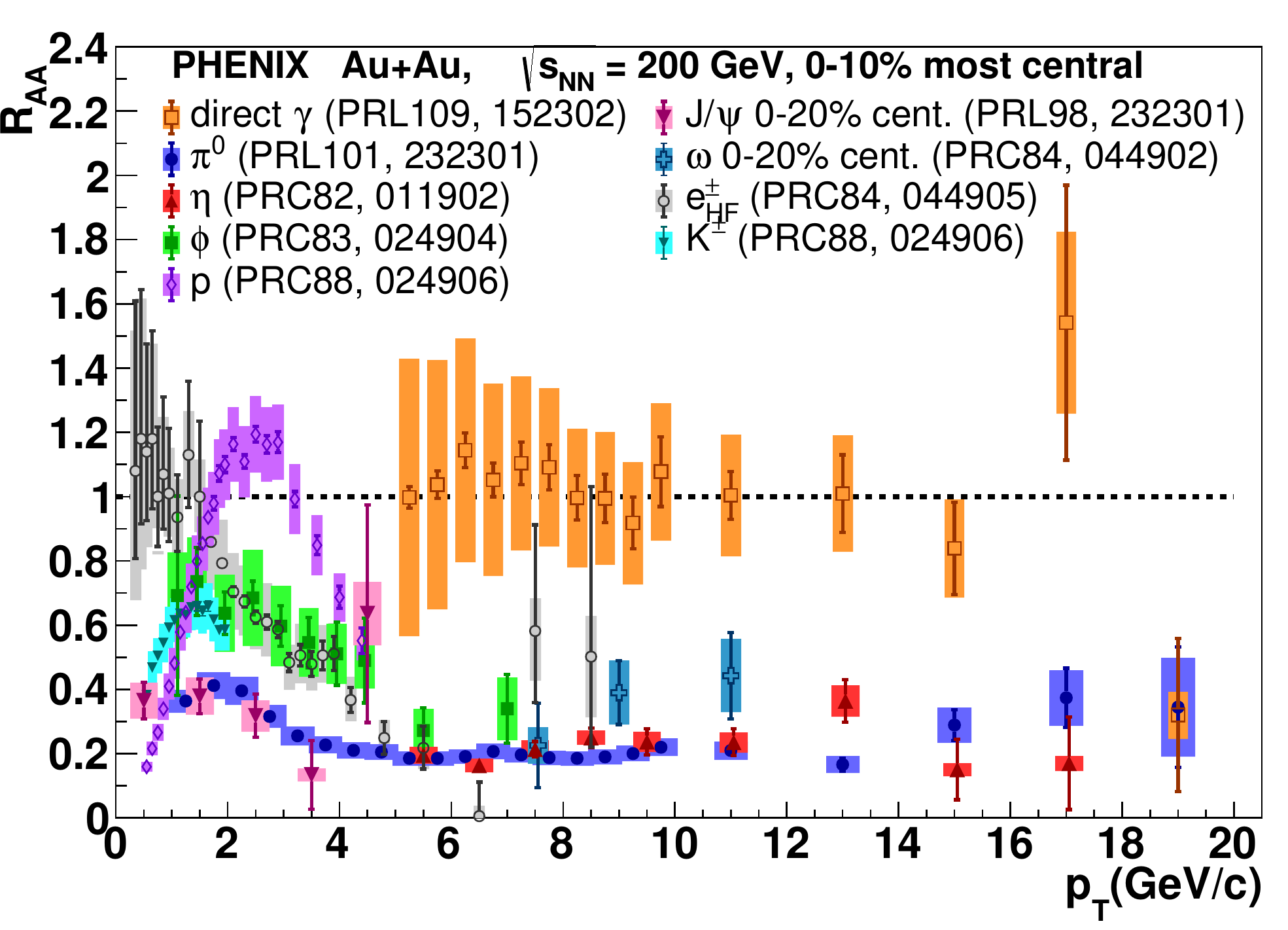}
}}\caption{
\RAA from PHENIX for direct photons~\cite{Afanasiev:2012dg}, $\pi^0$~\cite{Adare:2008qa}, $\eta$~\cite{Adare:2010dc}, $\phi$~\cite{Adare:2015ema}, $p$~\cite{Adare:2013esx}, J/$\psi$~\cite{Adare:2006ns}, $\omega$~\cite{Adare:2011ht}, $e^{\pm}$ from heavy flavor decays~\cite{Adare:2010de}, and $K^{\pm}$~\cite{Adare:2013esx}.  This demonstrates that colored probes (\highpT final state hadrons) are suppressed while electroweak probes (direct photons) are not at RHIC.
}\label{Fig:PHENIXtshirt}
\end{center}
\end{figure}

\begin{figure}
\begin{center}

\rotatebox{0}{\resizebox{8cm}{!}{
        \includegraphics{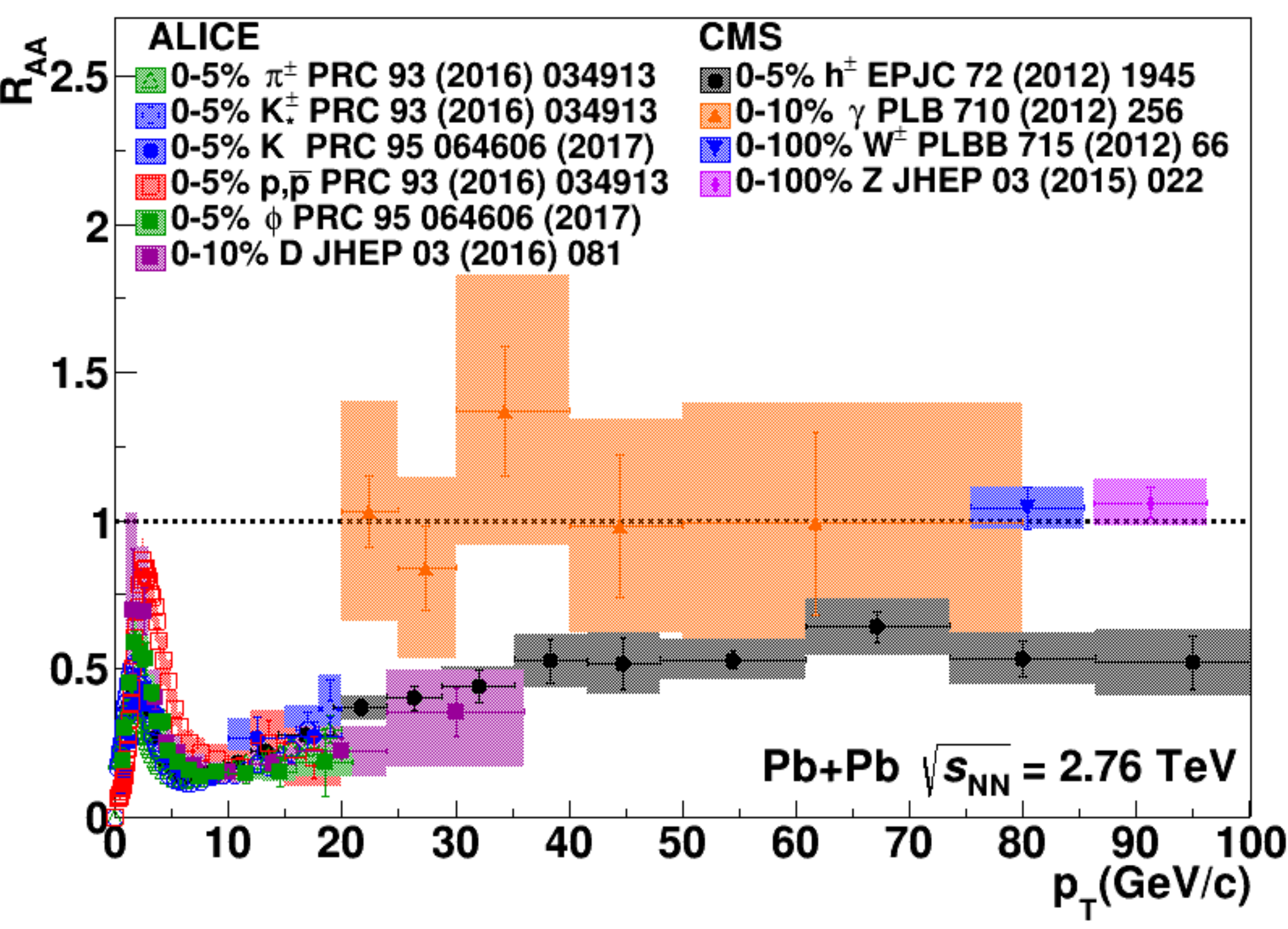}
}}\caption{
\RAA from ALICE for identified $\pi^{\pm}$, $K^{\pm}$, and $p$~\cite{Adam:2015kca} and D mesons~\cite{Adam:2015sza} and CMS for charged hadrons ($h^{\pm}$)~\cite{CMS:2012aa}, direct photons~\cite{Chatrchyan:2012vq}, W bosons~\cite{Chatrchyan:2012nt}, and Z bosons~\cite{Chatrchyan:2011ua}.  The W and Z bosons are shown at their rest mass and identified through their leptonic decay channel.
This demonstrates that colored probes (\highpT final state hadrons) are suppressed while electroweak probes (direct photons, W, Z) are not at the LHC.
}\label{Fig:LHCtshirt}
\end{center}
\end{figure}

Electroweak probes such as direct photons, which do not interact via the strong force, are expected to escape the QGP unscathed while probes which interact strongly lose energy in the medium and are suppressed at high momenta.  \Fref{Fig:PHENIXtshirt} shows a compilation of results from PHENIX demonstrating that colored probes (\highpT final state hadrons) are suppressed while electroweak probes (direct photons) are not at RHIC energies.  \Fref{Fig:LHCtshirt} shows a similar compilation of results from the LHC, demonstrating that this is also true at higher energies.  This  observed suppression in charged hadron spectra was the first indication of jet quenching in heavy ion collisions.  The lowest value of the nuclear modification factor \RAA for light hadrons is about 0.2 in collisions at \sNN~=~200~GeV~\cite{Adams:2003kv,Adler:2003qi,Back:2004bq} and about 0.1 in \Pb collisions at LHC for \sNN~=~2.76 TeV and \sNN~=~5.02 TeV~\cite{Aamodt:2010jd,CMS:2012aa,CMS-PAS-HIN-15-015}.  The \RAA of the charged hadron spectra appears to reach unity at \pT~$\approx$~100 \GeV~\cite{CMS-PAS-HIN-15-015}.  This is expected from all QCD-inspired energy loss models that at some point \RAA must reach one, because at leading order the differential cross section for interactions with the medium is proportional to $1/Q^2$~\cite{Levai:2001dc}.  
    Studies of \RCP as a function of collision energy indicate that suppression sets in somewhere between \sNN~=~27 and 39 GeV~\cite{Adamczyk:2017nof}.  At intermediate \pT the shape of \RAA with \pT is mass dependent with heavier particles approaching the light particle suppression level at higher momenta~\cite{Agakishiev:2011dc}.  However, even hadrons containing heavy quarks are suppressed at levels similar to light hadrons~\cite{ALICE:2012ab}.

QCD-motivated models are generally able to describe inclusive single particle \RAA qualitatively, however, for each model the details of the calculations make it difficult to compare results between models directly and extract quantitative information about the properties of the medium from such comparisons~\cite{Adare:2008cg}.
The JET collaboration was formed explicitly to make such comparisons between models and data and their extensive studies determined that for a 10 \GeV hadron the jet transport coefficient is \qhat~=~1.2$\pm$~0.3~GeV$^2$ in \Au collisions at \sNN = 200 GeV and \qhat~=~1.9$\pm$~0.7~GeV$^2$ in \Pb collisions at \sNN = 2.76 TeV~\cite{Burke:2013yra}.  

These detailed comparisons between data and energy loss models are one of the most important results in heavy ion physics and are one of the few results that directly constrain the properties of the medium.  We emphasize that these constraints came from a careful comparison of a straightforward observable to various models.  While we discuss measurements of more complicated observables later, this highlights the importance of both precision measurements of straightforward observables and careful, systematic comparisons of data to theory.  Similar approaches are likely needed to further constrain the properties of the medium.

\begin{figure*}
\begin{center}
\rotatebox{0}{\resizebox{18cm}{!}{
      \includegraphics{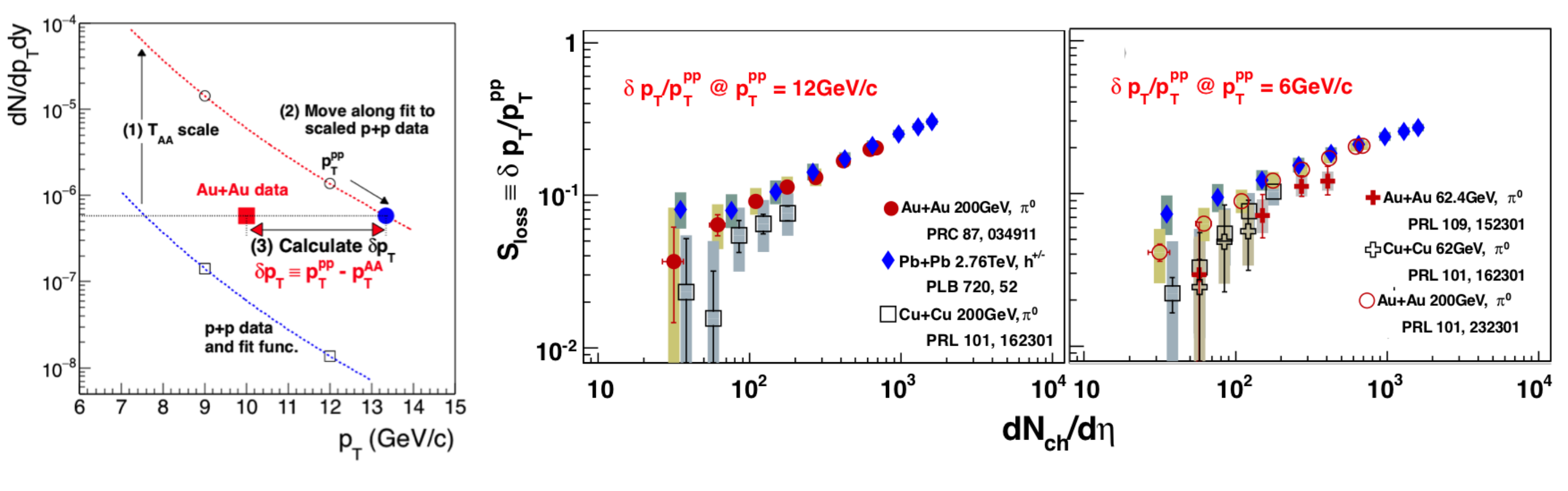}
      }}\caption{
Figure is a modified presentation of plots from PHENIX~\cite{Adare:2015cua}.  The first plot (left) is a cartoon demonstrating how $\delta p_{T}$ is determined.  The fractional energy loss, $S_{loss}$ measured as a function of the multiplicity, $dN_{ch}/d\eta$ is plotted for several heavy ion collision energies for hadrons with $p_{T}^{pp}$ of 12 GeV (middle) and 6 GeV/c (right) where $p_{T}^{pp}$ refers to the transverse momentum measured in \pp collisions.  The Pb+Pb data are from ALICE measured over $|\eta| < 0.8$ while all other data are from PHENIX which measures particle in the range $|\eta|<0.35$. These results indicate that the fractional energy loss scales with the energy density of the system.}
\label{Fig:SLoss}
\end{center}
\end{figure*}

It is remarkable that the \RAA values for hadrons at RHIC and the LHC are so similar since one would expect energy loss to increase with increased energy density which should result in a lower \RAA at the LHC with its higher collision energies. However, the hadrons in a particular \pT range are not totally quenched but rather appear at a lower \pT, so it is useful to study the shift of the hadron \pT spectrum in \AplusA collisions to \pp collisions rather than the ratio of yields. Note that the spectral shape also depends on the collisional energy. Spectra generally follow a power law trend described by $\frac{dN}{dp_{T}} \propto p_{T}^{-n}$  at high momenta. The spectra of hadrons is steeper in 200 GeV than in 2.76 TeV collisions ($n\approx8$ and $n\approx6.0$ repectively for the $p_{T}$ range 7-20~GeV/c)~\cite{Adare:2012uk,Adare:2012wg}. Therefore, for \RAA, greater energy loss at the LHC could be counteracted by the flatter spectral shape. To address this, another quantity, the fractional momentum loss, ($S_{loss}$) has also been measured to better probe a change in the fractional energy loss of partons $\Delta E/E$ as a function of collision energy. This quantity is defined as 

\begin{equation}
S_{loss}\equiv\frac{\delta p_{T}}{p_{T}}=\frac{p_{T}^{pp}-p_{T}^{AA}}{p_{T}^{pp}}\sim\Bigg\langle\frac{\Delta E}{E} \Bigg\rangle,
\end{equation}
where $p_{T}^{AA}$ is the $p_{T}$ of the \AplusA measurement. $p_{T}^{pp}$ is determined by first scaling $p_{T}$ spectrum measured in \pp collisions by the nuclear overlap function, $T_{AA}$ of the corresponding \AplusA centrality class and then determining the $p_{T}$ at which the yield of the scaled spectrum matches the yield measured in \AplusA at the $p_{T}^{AA}$ point of interest. This procedure is illustrated pictorially in \Fref{Fig:SLoss}. 

Indeed a greater fractional momentum loss was observed for the most central 2.76 TeV \Pb collisions compared to the 200 GeV \Au collisions \cite{Adare:2015cua}. The analysis found that $S_{loss}$ scales with energy density related quantities such as multiplicity ($dN_{ch}/d\eta$), as shown in \Fref{Fig:SLoss}, and $dE_{T}/dy/A_T$ where $A_T$ is the transverse area of the system. The latter quantity can be written in terms of Bjorken energy density, $\epsilon_{B_{j}}$ and the equilibrium time, $\tau_{0}$ such that $dE_{T}/dy/A_T=\epsilon_{B_{j}}\tau_{0}$ and has been shown to scale with $dN_{ch}/d\eta$~\cite{Adare:2015bua}. On the other hand, $S_{loss}$ does not scale with system size variables such as $N_{part}$. 
Assuming that $S_{loss}$ is a reasonable proxy for the mean fractional energy loss of the partons the scaling observations implies that fractional energy loss of partons scales with the energy density of the medium for these collision energies. 

\subsubsection{Jet \RAA}
\begin{figure}
\begin{center}

\rotatebox{0}{\resizebox{8cm}{!}{
             \includegraphics{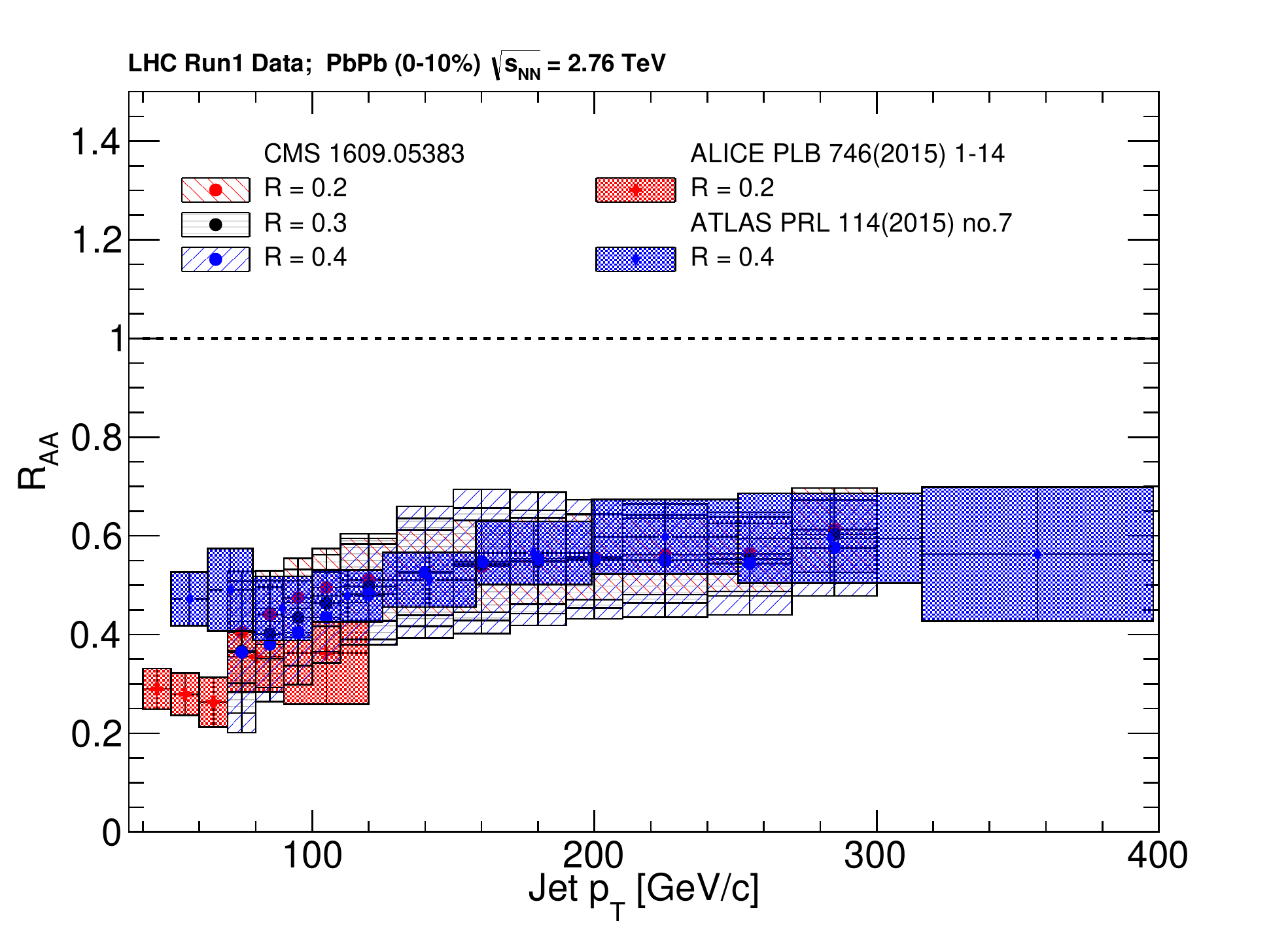}
}}\caption{
Reconstructed \akT jet \RAA from ALICE~\cite{Adam:2015ewa} with R = 0.2 for $|\eta|<$ 0.5, ATLAS~\cite{Aad:2014bxa} with R = 0.4 for $|\eta|<$ 2.1, and CMS~\cite{Khachatryan:2016jfl} with R = 0.2, 0.3 and 0.4 for $|\eta|<$ 2.0.  The ALICE and CMS data are consistent within uncertainties while the ATLAS data are higher.  This may be due to the ATLAS technique, which could impose a survivor bias and lead to a higher jet \RAA at low momenta.  Figure courtesy of Raghav Elayavalli Kunnawalkam.}
\label{Fig:JetRAA}
\end{center}
\end{figure}

Measurements of hadronic observables blur essential physics due to the complexity of the theoretical description of hadronization and the sensitivity to non-perturbative effects. In principle, measurements of reconstructed jets are expected to be less sensitive to these effects.
Next to leading order calculations demonstrate the sensitivity of \RAA measurements to the properties of the medium-induced gluon radiation~\cite{Vitev:2008rz}. These measurements can differentiate between competing models of parton energy loss mechanisms, reducing the large systematic uncertainties introduced by different theoretical formalisms~\cite{majumder}.  \Fref{Fig:JetRAA} shows the reconstructed \akT jet \RAA from ALICE~\cite{Adam:2015ewa} with R = 0.2 for $|\eta|<$ 0.5, ATLAS~\cite{Aad:2014bxa} with R = 0.4 for $|\eta|<$ 2.1, and CMS~\cite{Khachatryan:2016jfl,Khachatryan:2016jfl} with R = 0.2, 0.3, and 0.4 for $|\eta|<$ 2.0.  At lower momenta, the ALICE data are consistent with the CMS data for all radii, while the ATLAS \RAA is higher than that of ALICE. At higher momenta, all measurements of jets from all three experiments agree within the experimental uncertainties of the jet measurements.

A jet is defined by the parameters of the jet finding algorithm and selection criteria such as those that are used to identify background jets due to fluctuations in heavy ion events.  When making comparisons of jet observables between different experiments and to theoretical predictions, not only jet definitions but also the effects of selection criteria need to be considered carefully.  While the difference between the pseudorapidity coverage is unlikely to lead to the difference between the ATLAS and ALICE results given the relatively flat distribution at mid-rapidity, the resolution parameter $R$ as well as the different selection criteria could cause a difference as observed at low transverse momenta. 
The ATLAS approach to the combinatorial background, which favors jets with hard constituents, may bias the jet sample to unmodified jets, particularly at low momenta where the ATLAS and ALICE measurements overlap.  ATLAS and CMS jet measurements agree at high momenta where jets are expected to be less sensitive to the measurement details.  We therefore interpret the difference between the jet \RAA measured by the different experiments not as an inconsistency, but as different measurements due to different biases.  We implore the collaborations to construct jet observables using the same approaches to background subtraction and suppression of the combinatorial background so that the measurements could be compared directly.
Ultimately the overall consistency of \RAA at high \pT, even with widely varying jet radii and inherent biases in the jet sample, indicate that more sensitive observables are required to understand jet quenching quantitatively.

Although, the observation of jet quenching through $R_{AA}$ was a major feat, it still leaves several open questions about hard partons' interactions with the medium.  {\it How} do jets lose energy?  Through collisions with the medium, gluon bremsstrahlung, or both?  Where does that energy go?  Are there hot spots or does the energy seem to be distributed isotropically in the event?  Few experimental observables can compete with \RAA for overall precision, however, more differential observables may be more sensitive to the energy loss mechanism.

\subsubsection{Dihadron correlations}

\begin{figure}
\begin{center}

\rotatebox{0}{\resizebox{!}{8cm}{
	\includegraphics{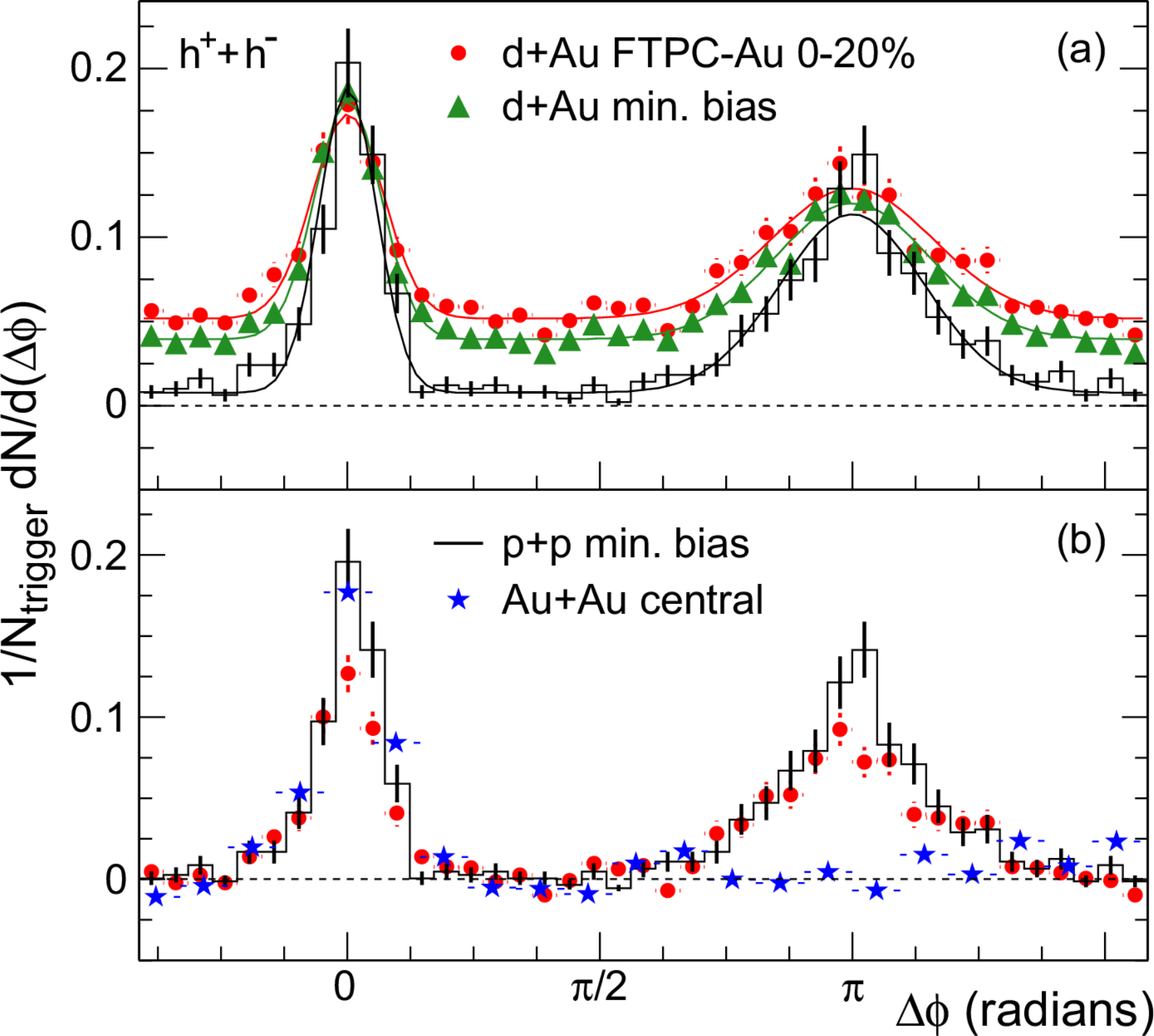}
}}\caption{
Figure from STAR~\cite{Adams:2003im}. (a) Dihadron correlations before background subtraction in \pp and \dAu and  (b)  Comparison of dihadron correlations after background subtraction in \pp, \dAu, and \Au at \sNN = 200 GeV  for associated momenta $2.0$~GeV/$c$ $<$ $p_T^{\mathrm{a}}$ $<$ $p_T^{\mathrm{t}}$ and trigger momenta $4$~$<$ $p_T^{\mathrm{t}}$ $<$ $6$~GeV/$c$.  This measurement is now understood to be quantitatively incorrect because of erroneous assumptions in the background subtraction.  We now see only partial suppression on the \as~\cite{Nattrass:2016cln}.
}\label{Fig:DihadronIconic}
\end{center}
\end{figure}

The precise mechanism responsible for modification of \dhcs cannot be determined based on these studies alone because there are many mechanisms which could lead to modification of the correlations.  This includes not only energy loss and modification of jet fragmentation but also modifications of the underlying parton spectra.  However, they are less ambiguous than spectra alone because the requirement of a high momentum trigger particle enhances the fraction of particles from jets.
\Fref{Fig:DihadronIconic} shows \dhcs in \pp, \dAu, and \Au at \sNN = 200 GeV, demonstrating suppression of the \as peak in central \Au collisions.
The first measurements of \dhcs showed complete suppression of the \as peak and moderate enhancement of the \ns peak~\cite{Adler:2002tq,Adams:2003im,Adams:2004wz}.  However, as noted above, a majority of \dhc studies did not take the odd \vn due to flow into account, including those in \Fref{Fig:DihadronIconic}.  A subsequent measurement with similar kinematic cuts including higher order \vn shows that the \as is not completely suppressed, as shown in \Fref{Fig:DihadronIconic}, but rather that there is a visible but suppressed \as peak~\cite{Nattrass:2016cln}.  Studies at higher momenta also see a visible but suppressed \as peak~\cite{Adams:2006yt}.

The suppression is quantified by
\begin{equation}
 I_{AA} = Y_{AA}/Y_{pp}.\label{Eqtn:IAA}
\end{equation}
\noindent where $Y_{AA}$ is the yield in \AplusA collisions and $Y_{pp}$ is the yield in \pp collisions.  The yields must be defined over finite \dphi and \deta ranges and are usually measured for a fixed range in associated momentum, \ptassoc.  Similar to \RAA, an \IAA greater than one means that there are more particles in the peak in \AplusA collisions than in \pp collisions and an \IAA less than one means that there are fewer.  
Gluon bremsstrahlung or collisional energy loss would result in more particles at low momenta and fewer particles at high momenta, leading to an \IAA greater than one at low momenta and an \IAA less than one at high momenta, at least as long as the lost energy does not reach equilibrium with the medium.  Both radiative and collisional energy loss would lead to broader correlations.  Partonic energy loss before fragmentation would lead to a suppression on the \as but no modification on the \ns and no broadening because the \ns jet is biased towards the surface of the medium.  Changes in the parton spectra can also impact \IAA because harder partons hadronize into more particles and higher energy jets are more collimated.

No differences between \dAu and \pp collisions are observed on either the near- or \as at midrapidity~\cite{Adler:2005ad,Adler:2006hi}, indicating that any modifications observed are due to hot nuclear matter effects.
The \ns yields at midrapidity in \AplusA, \dAu, and \pp collisions are within error at RHIC~\cite{Adams:2006yt,Abelev:2009ah,Adare:2008ae}, even at low momenta~\cite{Abelev:2009af,Agakishiev:2011st}, indicating that the \ns jet is not substantially modified, although the data are also consistent with a slight enhancement~\cite{Nattrass:2016cln}.  A slight enhancement of the \ns is observed at the LHC~\cite{Aamodt:2011vg} and a slight broadening is observed at RHIC~\cite{Agakishiev:2011st,Nattrass:2016cln,Adare:2008ae}.  The combination of broadening and a slight enhancement favors moderate partonic energy loss rather than a change in the underlying jet spectra since higher energy jets are both more collimated and contain more particles.

The \as is suppressed at high momenta at both RHIC~\cite{Adams:2006yt,Abelev:2009ah} and the LHC~\cite{Aamodt:2011vg}.  A reanalysis of reaction plane dependent \dhcs from STAR~\cite{Agakishiev:2010ur,Agakishiev:2014ada} at low momenta using a new background method which takes odd \vn into account~\cite{Sharma:2015qra} observed suppression on the \as but no broadening, even though broadening was observed on the \ns at the same momenta~\cite{Nattrass:2016cln}.  This may indicate that the \as width is less sensitive because the width is broadened by the decorrelation between the near- and \as jet axes rather than indicating that these effects are not present.  Reaction plane dependent studies can constrain the path length dependence of energy loss because, as shown in \Fref{Fig:flowschematic}, partons traveling in the reaction plane (in-plane) traverse less medium than those traveling perpendicular to the reaction plane (out-of-plane).  The \IAA is highest for low momentum particles and is at a minimum for trigger particles at intermediate angles relative to the reaction plane rather than in-plane or out-of-plane.  This likely indicates an interplay between the effects of surface bias and partonic energy loss.  

Energy loss models are generally able to describe \IAA qualitatively, however, there has been no systematic attempt to compare data to models, as was done for \RAA.  Simultaneous comparisons of \RAA and \IAA are expected to be highly sensitive to the jet transport coefficient \qhat~\cite{Jia:2011pi,Zhang:2007ja}.  Such a theoretical comparison is partially compounded by the wide range of kinematic cuts used in experimental measurements and the fact that most measurements neglected the odd \vn in the background subtraction.  

\subsubsection{Dijet imbalance}
\begin{figure*}
\begin{center}

\rotatebox{0}{\resizebox{15cm}{!}{
        \includegraphics{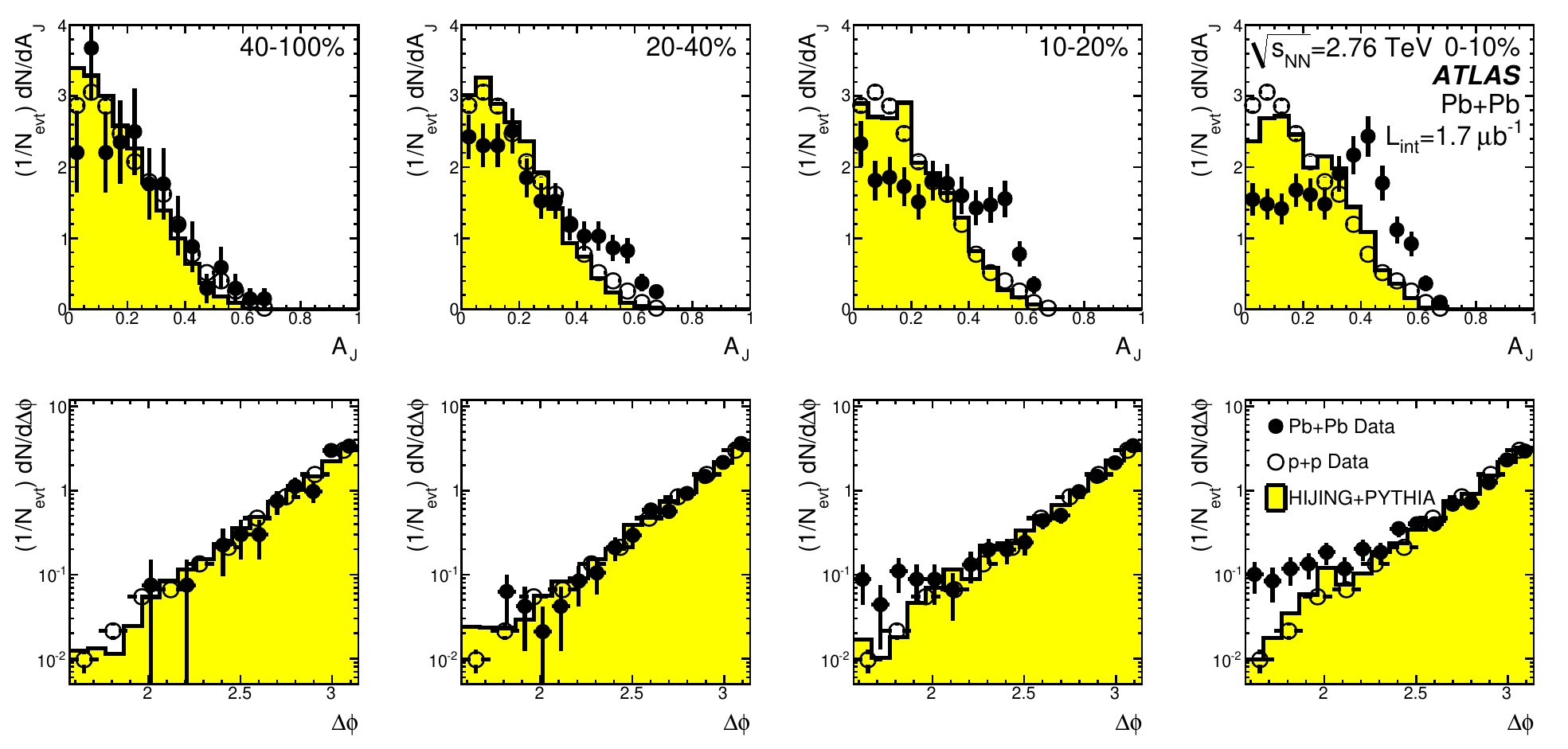}
}}\caption{
Figure from ATLAS~\cite{Aad:2010bu}.  The top row shows comparisons of \Aj~$=(E_{T1}-E_{T2})/(E_{T1}+E_{T2})$ from \pp and \Pb collisions at \sNN = 2.76 TeV with leading jets above \pT $>$ 100 GeV and subleading jets above 25 GeV.  The bottom row shows the angular distribution of the jet pairs.  This shows that the momenta of jets in jet pairs is not balanced in central \AplusA collisions, indicating energy loss.
}\label{Fig:Aj}
\end{center}
\end{figure*}

The first evidence of jet quenching in reconstructed jets at the LHC was observed by measuring the dijet asymmetry, \Aj. This observable measures the energy or momentum imbalance between the
leading and sub-leading or opposing jet in each event. Due to kinematic and detector effects, the energy of dijets will not be perfectly balanced, even in \pp collisions. Therefore to interpret this measurement in
heavy ion collisions, data from \AplusA collisions must be compared to the distributions in \pp
collisions.
\Fref{Fig:Aj} shows the dijet asymmetry measurement from the ATLAS experiment where \Aj~$=\frac{E_{T1}-E_{T2}}{E_{T1}+E_{T2}}$~\cite{Aad:2010bu}.
The left panel on the top row shows the
\Aj distribution for peripheral \Pb collisions and demonstrates that it is similar to that from \pp
collisions. However, dijets in central \Pb collisions are more likely to have
a higher $A_{J}$ value than dijets in \pp collisions, consistent with expectations from energy loss.  The bottom panel shows that these jets retain a similar angular correlation with the leading jet, even as they lose energy.  The CMS measurement of $A_{J}=\frac{p_{T1}-p_{T2}}{p_{T1}+p_{T2}}$~\cite{Chatrchyan:2011sx} shows similar trends.  The structure in the distribution of \Aj is partially due to the 100 GeV lower limit on the leading jet and the 25 GeV lower limit on the subleading jet and partially due to detector effects and background in the heavy ion collision.  These measurements are not corrected for detector effects or distortions in the observed jet energies due to fluctuations in the background.  Instead the jets from \pp collisions are embedded in a heavy ion event in order to take the effects of the background into account.  

Recently ATLAS has measured \Aj, and unfolded the distribution in order to take background and detector effects into account~\cite{ATLAS-CONF-2015-052}, with similar conclusions.  For jets above 200 GeV, the asymmetry is observed to be consistent with those observed in \pp, indicating that sufficiently high momentum jets are unmodified.  This is consistent with observation that the \RAA consistent with one for hadrons at \pT~$\approx$~100 \GeV~\cite{CMS-PAS-HIN-15-015}, indicating that very high momentum jets are not modified.

Energy and momentum must be conserved, so the balance should be restored if jets can be reconstructed in such a way that the particles carrying the lost energy are included.  For jets reconstructed with low momentum constituents, the background due to combinatorial jets is non-negligible, but requiring the jet to be matched to a jet constructed with higher momentum jet constituents, as well as a higher momentum jet will suppress the combinatorial jet background.  STAR measurements of \Aj using a high momentum constituent selection (\pT~$>$~2~\GeV) observed the same energy imbalance seen by ATLAS and CMS.  However, the energy balance was recovered by matching these jets reconstructed with high \pT constituents, to jets reconstructed with low momentum constituents (\pT~$>$~150~\MeV) and then constructing \Aj from the jets with the low momentum constituents~\cite{Adamczyk:2016fqm}.

\subsubsection{$\gamma$-hadron, $\gamma$-jet and $Z$-jet correlations}\label{sec:bosonjet}
\begin{figure}
\begin{center}

\rotatebox{0}{\resizebox{8cm}{!}{
        \includegraphics{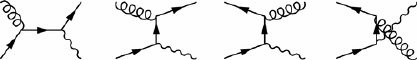}
}}\caption{
Figure from PHENIX~\cite{Adare:2010yw}.
The left two Feynman diagrams show direct photon production through Compton scattering and the right two diagrams show direct photon production through quark-antiquark annihilation.
These are the leading order processes which contribute to the production of a gamma and a jet approximately 180$^{\circ}$ apart.
}\label{fig:comptondiagram}
\end{center}
\end{figure}

At leading order, direct photons are produced via Compton scattering, q+g $\rightarrow$ q+$\gamma$, and quark-antiquark annihilation, as shown in the left two and right two Feynman diagrams in \Fref{fig:comptondiagram}, respectively. Due to the dearth of anti-quarks and abundance of gluons in the proton, Compton scattering is the dominant production mechanism for direct photons in \pp and \AplusA collisions.   Therefore jets recoiling from a direct photon at midrapidity are predominantly quark jets.  In the center of mass frame at leading order, the photon and recoil quark are produced heading precisely 180$^{\circ}$ away from each other in the transverse plane with the same momentum.  At higher order, fragmentation photons and gluon emission impact the correlation such that the momentum is not entirely balanced and the back-to-back positions are smeared, even in \pp collisions.  Since photons do not lose energy in the QGP, the photon will escape the medium unscathed and the energy of the opposing quark can be determined from the energy of the photon.  This channel is called the ``Golden Channel'' for jet tomography of the QGP because it is possible to calculate experimental observables with less sensitivity to hadronization and other non-perturbative effects than \dhcs and measurements of reconstructed jets.  Additionally, direct photon analyses remove some of the ambiguity with respect to differences between quarks and gluons since the outgoing parton opposing the direct photon is predominantly a quark.

\begin{figure}
\begin{center}
\rotatebox{0}{\resizebox{8cm}{!}{
        \includegraphics{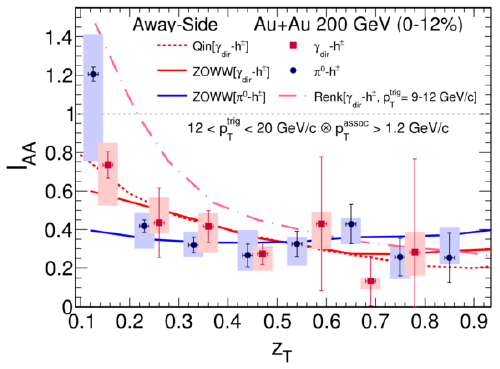}
}}\caption{
Figure from STAR~\cite{STAR:2016jdz}.
The \as $I_{AA}$ for direct photon-hadron correlations (red squares) and $\pi^{0}$-hadron correlations (blue circles) plotted as a function of $z_{T}=p_{T,h}/p_{T,trig}$ as measured by STAR in central 200 GeV Au+Au collisions.  This shows the suppression of hadrons 180$^{\circ}$ away from a direct photon.  The data are consistent with theory calculations which show the greatest suppression at high $z_{T}$ and less suppression at low $z_{T}$. The curves are theory calculations from Qin~\cite{Qin:2009bk}, Renk~\cite{Renk:2009ur} and ZOWW~\cite{Zhang:2009rn,Chen:2010te}. }\label{Fig:starghIAA}
\end{center}
\end{figure}

Correlations of direct photons with hadrons can be used to calculate \IAA, as for \dhcs.  Studies of $\gamma$-h at RHIC led to similar conclusions to those reached by \dhcs, as shown in Figure \ref{Fig:starghIAA}, demonstrating suppression of the \as jet~\cite{Adare:2009vd,Adare:2010yw,Abelev:2009gu,STAR:2016jdz}.  In addition, $\gamma$-h correlations can measure the fragmentation function of the \as jet assuming the jet energy is the photon energy. This is discussed in \Sref{Sec:FragFuncsBosonJet}. It should be noted that nonzero photon \vnum{2} and \vnum{3} have been observed~\cite{Adare:2015lcd,Adare:2011zr}, leading to a correlated background.  The physical origin of this \vtwo is unclear, since photons do not interact with the medium, so it is also unclear if \vthree and higher order \vn impact the background. Measurements at high momenta are robust because the background is small and the photon \vtwo appears to decrease with $p_{T}$. In \cite{Adare:2012qi}, the systematic uncertainty due to \vthree was estimated and included in the total systematic uncertainty. Since the direct photon-hadron correlations are extracted by subtracting photon-hadron correlations from decays (primarily from $\pi^0 \rightarrow \gamma \gamma$) from inclusive photon-hadron correlations, the impact of the \vn in the final direct photon-hadron correlations is reduced as compared to dihadron and jet-hadron correlations.  

\begin{figure*}
\begin{center}
\rotatebox{0}{\resizebox{!}{5cm}{
	\includegraphics{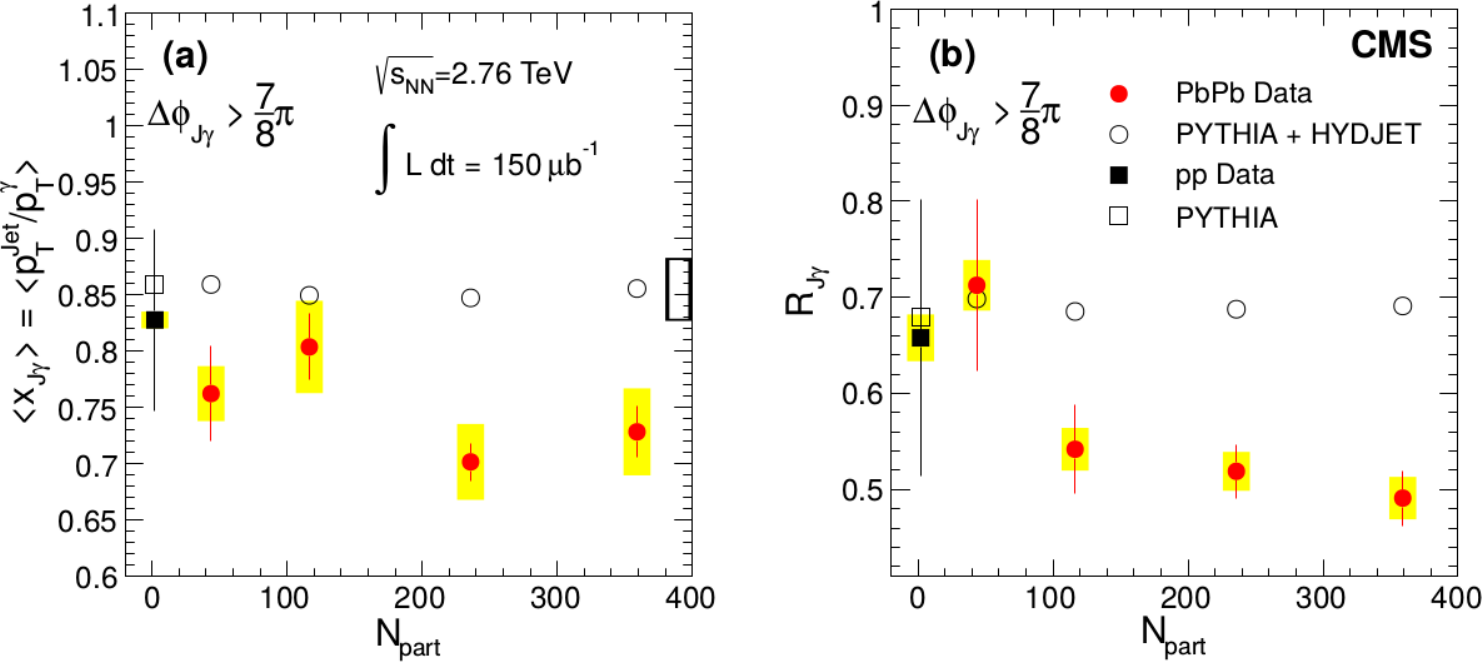}
}}
\caption{
Figure from CMS~\cite{Chatrchyan:2012gt} for isolated photons with $p_{T} > 60$ \GeV and associated jets with $p_{T} > 30$ \GeV. 
(a) Average ratio of jet transverse momentum to photon transverse momentum, $\langle x_{J\gamma} \rangle$ , as a function of the number of participating nucleons $N_{part}$.  
(b)  Average fraction of isolated photons with an associated jet above 30 \GeV, $R_{J\gamma}$, as a function of $N_{part}$.  This demonstrates that the quark jet 180$^{\circ}$ away from a direct photon loses energy, with the energy loss increasing with increasing centrality.
}\label{Fig:CMSGammaJet}
\end{center}
\end{figure*}

Direct photons can also be correlated with a reconstructed jet.  In principle, this is a direct measurement of partonic energy loss.  \Fref{Fig:CMSGammaJet}(a) shows measurements of the energy imbalance between a photon with energy E $>$ 60 GeV and a jet at least $\frac{7}{8}\pi$ away in azimuth with at least  $E_{jet}>$ 30 GeV.  Even in \pp collisions, the jet energy does not exactly balance the photon energy because of next-to-leading order effects and because some of the quark's energy may extend outside of the jet cone.  The lower limit on the energy of the reconstructed jet is necessary in order to suppress background from combinatorial jets, but it also leads to a lower limit on the fraction of the photon energy observed.  \Fref{Fig:CMSGammaJet}(a) demonstrates that the quark loses energy in \Pb collisions.  \Fref{Fig:CMSGammaJet}(b) shows the average fraction of isolated photons matched to a jet, $R_{J\gamma}$.  In \pp collisions nearly 70\% of all photons are matched to a jet, but in central \Pb collisions only about half of all photons are matched to a jet.  These measurements provide unambiguous evidence for partonic energy loss.  However, the kinematic cuts required to suppress the background leave some ambiguity regarding the amount of energy that was lost.  Some of the energy could simply be swept outside of the jet cone. The preliminary results of an analysis with higher statistics for the \pp data and the addition of \pPb collisions also shows no significant modification, confirming that the \Pb imbalance does not originate from cold nuclear matter effects~\cite{CMS:photon2}.

By construction, measurements of the process q+g $\rightarrow$ q+$\gamma$ can only measure interactions of quarks with the medium.  Since there are more gluons in the initial state and quarks and gluons may interact with the medium in different ways, studies of direct photons alone cannot give a full picture of partonic energy loss.

With the large statistics data collected during the 2015 \Pb running of the LHC at 5 TeV, another ``Golden Probe'' for jet tomography of the QGP, the coincidences of a $Z^{0}$ and a jet, became experimentally accessible~\cite{Wang:1996pe,Neufeld:2010fj}. 
While this channel has served as an essential calibrator of jet energy in TeV \pp collisions, in heavy ion collisions it can be used to calibrate in-medium parton energy loss as the $Z^{0}$ carries no color charge and is expected to escape the medium unattenuated like the photon. However, photon measurements at higher momentum are limited due to the large background from decay photons in experimental measurements. Recent measurements of Z boson tagged jets in \Pb collisions at \sNN = 5.02 TeV~\cite{CMS:ZJet} show that angular correlations between Z bosons and jets are mostly preserved in central \Pb collisions.  However, the transverse momentum of the jet associated with that Z boson appears to be shifted to lower values with respect to the observations in \pp collisions, as expected from jet quenching. 

\subsubsection{Hadron-jet correlations}
\begin{figure}
\begin{center}
\rotatebox{0}{\resizebox{8cm}{!}{
        \includegraphics{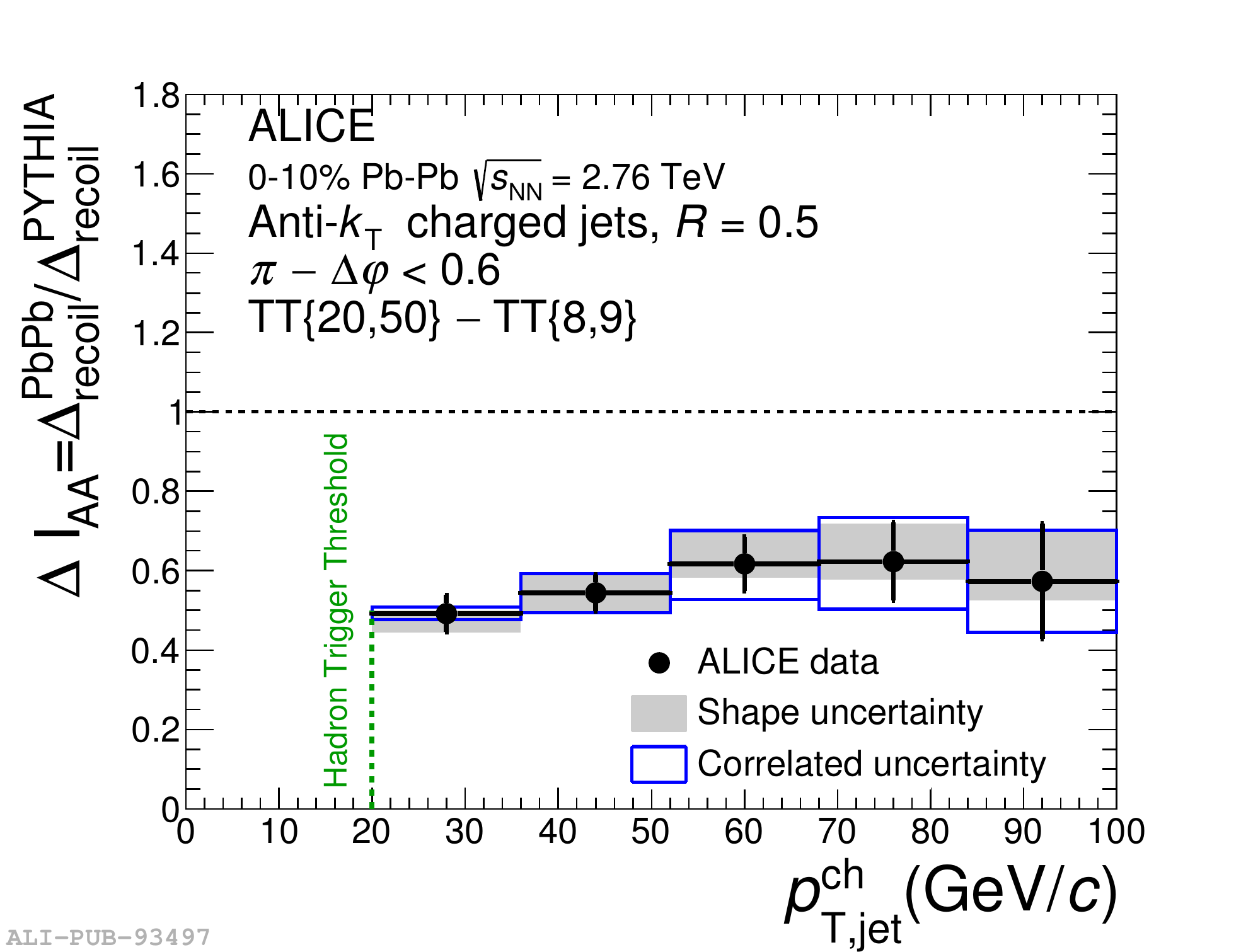}
}}\caption{
Figure from ALICE~\cite{Adam:2015doa}.  $\Delta I_{AA} = \Delta_{recoil}^{PbPb}/\Delta_{recoil}^{PYTHIA}$ where $\Delta_{recoil}$ is the difference between the number of jets within $\pi - \Delta\phi < 0.6$  of a hadron with $20< p_{T}<50$ \GeV and a hadron with $8< p_{T}<9$ \GeV.  The green line indicates the momentum of the higher momentum hadron, an approximate lower threshold on the jet momentum.  This demonstrates the suppression of a jet 180$^{\circ}$ away from a hard hadron.
}\label{Fig:HadronJet}
\end{center}
\end{figure}

Correlations between a hard hadron and a reconstructed jet were measured to overcome the downside of an explicit bias imposed by the background suppression techniques described in \Sref{Sec:JetMethod}.  Similar to \dhcs, a reconstructed hadron is selected and the yield of jets reconstructed within $|\pi - \Delta\phi| < 0.6$ relative to that hadron is measured in~\cite{Adam:2015doa}.  For sufficiently hard hadrons, a large fraction of the jets correlated with those hadrons would be jets that originated from a hard process, however, for low momentum hadrons, the yield will be dominated by combinatorial jets.  The yield of combinatorial jets should be independent of the hadron momentum, so the difference between the yields, $\Delta_{recoil}$, is calculated to subtract the background from the ensemble of jet candidates.  This difference in yields is then compared to the same measurement in \pp collisions. 

Since the requirement of a hard hadron is opposite the jet being studied, no fragmentation bias is imposed on the reconstructed jet. Therefore, this measurement may be more sensitive to modified jets than observables that require selection criteria on the jet candidates themselves.  \Fref{Fig:HadronJet} shows the ratio of $\Delta_{recoil}$ in \Pb collisions to that in \pp collisions, $\Delta I_{AA} = \Delta_{recoil}^{PbPb}/\Delta_{recoil}^{PYTHIA}$.  PYTHIA is used as a reference rather than data due to limited statistics available in the data at the same collision energy.  PYTHIA agrees with the data from \pp collisions at \sqrts = 7 TeV.  These data demonstrate that there is substantial jet suppression, consistent with the results discussed above.  

Measurements of hadron-jet correlations by STAR~\cite{Adamczyk:2017yhe} used a novel mixed event technique for background subtraction in order to extend the measurement to low momenta.  The conditional yield correlated with a high momentum hadron was clearly suppressed in central \Au collisions relative to that observed in peripheral collisions, though substantially less so at the lowest momenta.
A benefit of this method is that, in principle, the conditional yield of jets correlated with a hard hadron can be calculated with perturbative QCD. 

\subsubsection{Path length dependence of inclusive \RAA and jet $v_{n}$}

\begin{figure*}
\begin{center}
\rotatebox{0}{\resizebox{15cm}{!}{
        \includegraphics{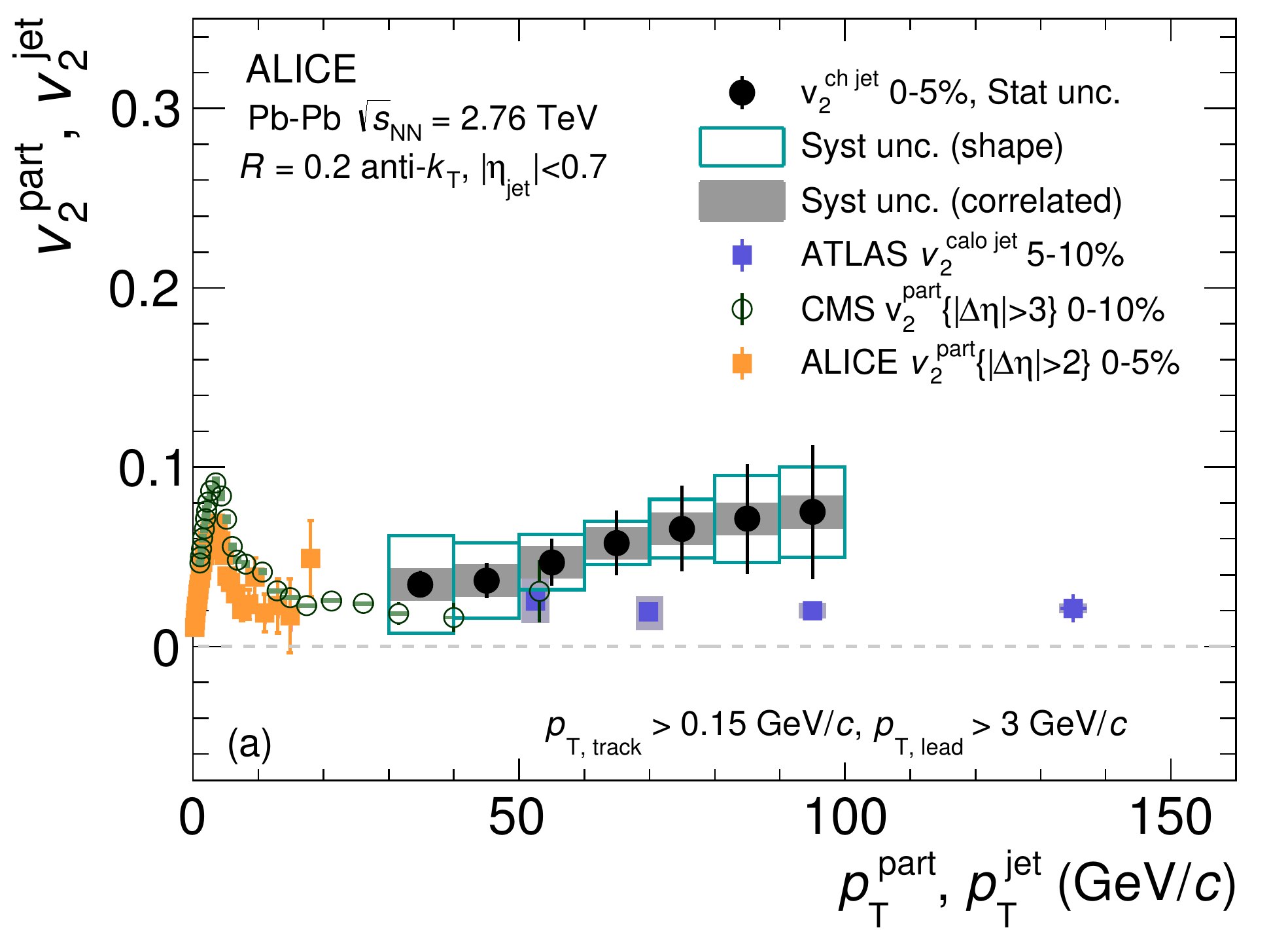}
        \includegraphics{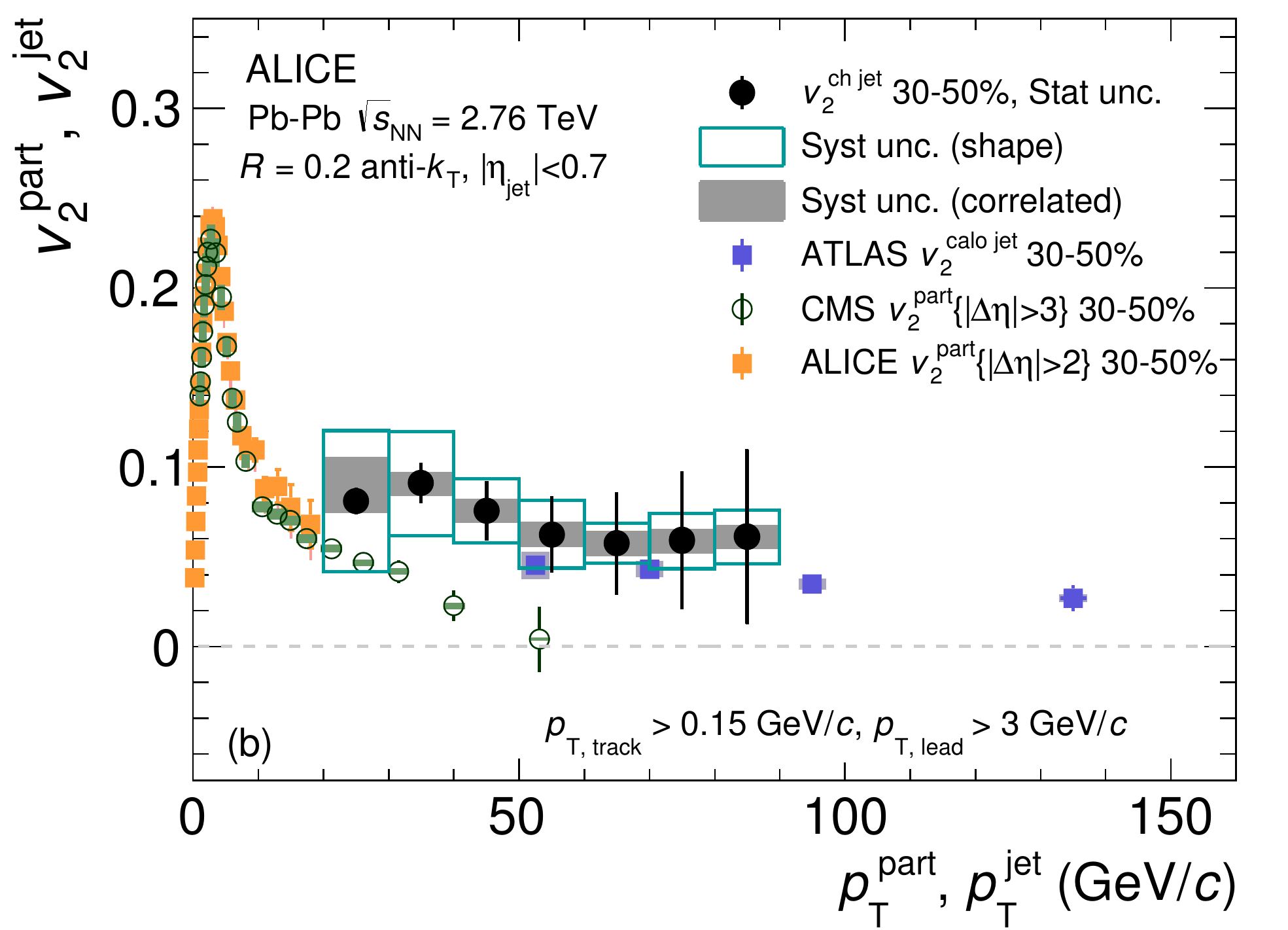}
}}\caption{
Figure from ALICE~\cite{Adam:2015mda}.
Jet \vnum{2} from charged jets by ALICE~\cite{Adam:2015mda} and calorimeter jets by ATLAS~\cite{Aad:2013sla} compared to the charged hadron \vnum{2} for 5--10\% (left) and 30--50\% collisions~\cite{Chatrchyan:2012xq,Abelev:2012di}.  This demonstrates that partonic energy loss is path length dependent.
}\label{Fig:Jetvn}
\end{center}
\end{figure*}

The azimuthal asymmetry shown in \Fref{Fig:flowschematic} provides a natural variation in the path length traversed by hard partons and the orientation of the reaction plane can be reconstructed from the distribution of final state hadrons.  The correlations with this reaction plane can therefore be used to investigate the path length of partonic energy loss.  The reaction plane dependence of inclusive particle \RAA demonstrates that energy loss is path length dependent~\cite{Adler:2006bw}, as expected from models.  The path length changes with collision centrality, system size, and angle relative to the reaction plane, however, the temperature and lifetime of the QGP also change when the centrality and system size are varied.  When particle production is studied relative to the reaction plane angle, the properties of the medium remain the same while only the path length is changed.  Because the eccentricity of the medium and therefore the path length can only be determined in a model, any attempt to determine the absolute path length is model dependent.
Attempts to constrain the path length dependence of \RAA were explored in~\cite{Adler:2006bw}.  While these studies were inconclusive, they showed that \RAA is constant at a fixed mean path length and that there is no suppression for a path length below L = 2 fm, indicating that there is either a minimum time a hard parton must interact with the medium or there must be substantial effects from surface bias.  More conclusive statements would require more detailed comparisons to models.

At high \pT, the single particle \vn in \eref{Eqtn:FlowFourierDecomposition} are dominated by jet production and a non-zero \vtwo indicates path length dependent jet quenching.  Above 10 \GeV, a non-zero \vtwo is observed at RHIC~\cite{Adare:2013wop} and the LHC~\cite{Abelev:2012di,Chatrchyan:2012xq} and can be explained by energy loss models~\cite{Abelev:2012di}.  
Above 10 \GeV, \vthree in central collisions is consistent with zero~\cite{Abelev:2012di}.  The \vn of jets themselves can be measured directly, however, only jet \vtwo has been measured~\cite{Adam:2015mda,Aad:2013sla}.  \Fref{Fig:Jetvn} compares jet and charged particle \vtwo from ATLAS and ALICE.  ALICE measurements are of charged jets, which are only constructed with charged particles and not corrected for the neutral component, with $R$ = 0.2 and $|\eta| <$ 0.7 and ATLAS measurements are reconstructed jets with R = 0.2 and $|\eta| <$ 2.1.  The \vtwo observed by ALICE is higher than that observed by ATLAS, although consistent within the large uncertainties.  The ALICE measurement is unfolded to correct for detector effects, but it is not corrected for the neutral energy contribution.
Both measurements use methods to suppress the background which could lead to greater surface bias or bias towards unmodified jets.  The ALICE measurement requires a track above 3 \GeV in the jet to reduce the combinatorial background.  The ATLAS measurement requires the calorimeter jets used in the measurement to be matched to a 10 GeV track jet or to contain a 9 GeV calorimeter cluster. Because of the higher momentum requirement the ATLAS measurement has a greater bias than the ALICE sample of jets.  

These measurements provide some constraints on the path length dependence, however,  this is not the only relevant effect.  Theoretical calculations indicate that 
both event-by-event initial condition fluctuations and jet-by-jet energy loss fluctuations play a role in \vn at high \pT~\cite{Zapp:2013zya,Noronha-Hostler:2016eow,Betz:2016ayq}.
This is perhaps not surprising, analogous to the importance of fluctuations in the initial state for measurements of the \vn due to flow.
However, it does indicate that much more insight into which observables are most sensitive to path length dependence and the role of fluctuations in energy loss is needed from theory.

\subsubsection{Heavy quark energy loss}\label{sec:hfjet}

The jet quenching due to radiative energy loss is expected to depend upon the species of the fragmenting parton~\cite{Horowitz:2007su}.  The simplest example is gluon jets, which are expected to lose more energy in the medium than quark jets due to their larger color factor. Similarly,  the mass of the initial parton also plays a role and the interpretation of this effect depends on the theoretical treatment of parton-medium interactions.  Strong coupling calculations based on AdS/CFT correspondence predict large mass effects at all transverse momenta and in weak-coupling calculations based on pQCD mass effects may arise from the ``dead-cone'' effect~\cite{Dokshitzer:2001zm}, the suppression of gluon emission at small angles relative to a heavy quark, but may be limited to a small range of heavy-quark transverse momenta comparable to the heavy-quark mass. However, the relevance of the dead-cone effect in heavy ion collisions is debated~\cite{Aurenche:2009dj}.

Searches for a decreased suppression of heavy flavor using single particles are still inconclusive due to large uncertainties, although they indicate that heavy quarks may indeed lose less energy in the medium.  As shown in \Fref{Fig:PHENIXtshirt}, the \RAA of single electrons from decays of heavy flavor hadrons is within uncertainties of that of hadrons containing only light quarks.  Measurements of single leptons are somewhat ambiguous because of the difference between the momentum of the heavy meson and the decay lepton.  Since the mass effect is predicted to be momentum dependent with negligible effects for \pT~$\gg m$, the decay may wash out any mass effect.  The \RAA of D mesons is within uncertainties of the light quark \RAA~\cite{Adam:2015nna,Adam:2015sza,Adamczyk:2014uip}.  Particularly at the LHC, these results may be somewhat ambiguous because D mesons may also be produced in the fragmentation of light quark or gluon jets.  B mesons are much less likely to be produced by fragmentation.  Preliminary measurements of B meson \RAA show less suppression than for light mesons, although the uncertainties are large and prohibit strong conclusions~\cite{CMS-PAS-HIN-16-011}.

\begin{figure}
\begin{center}

\rotatebox{0}{\resizebox{8cm}{!}{
        \includegraphics{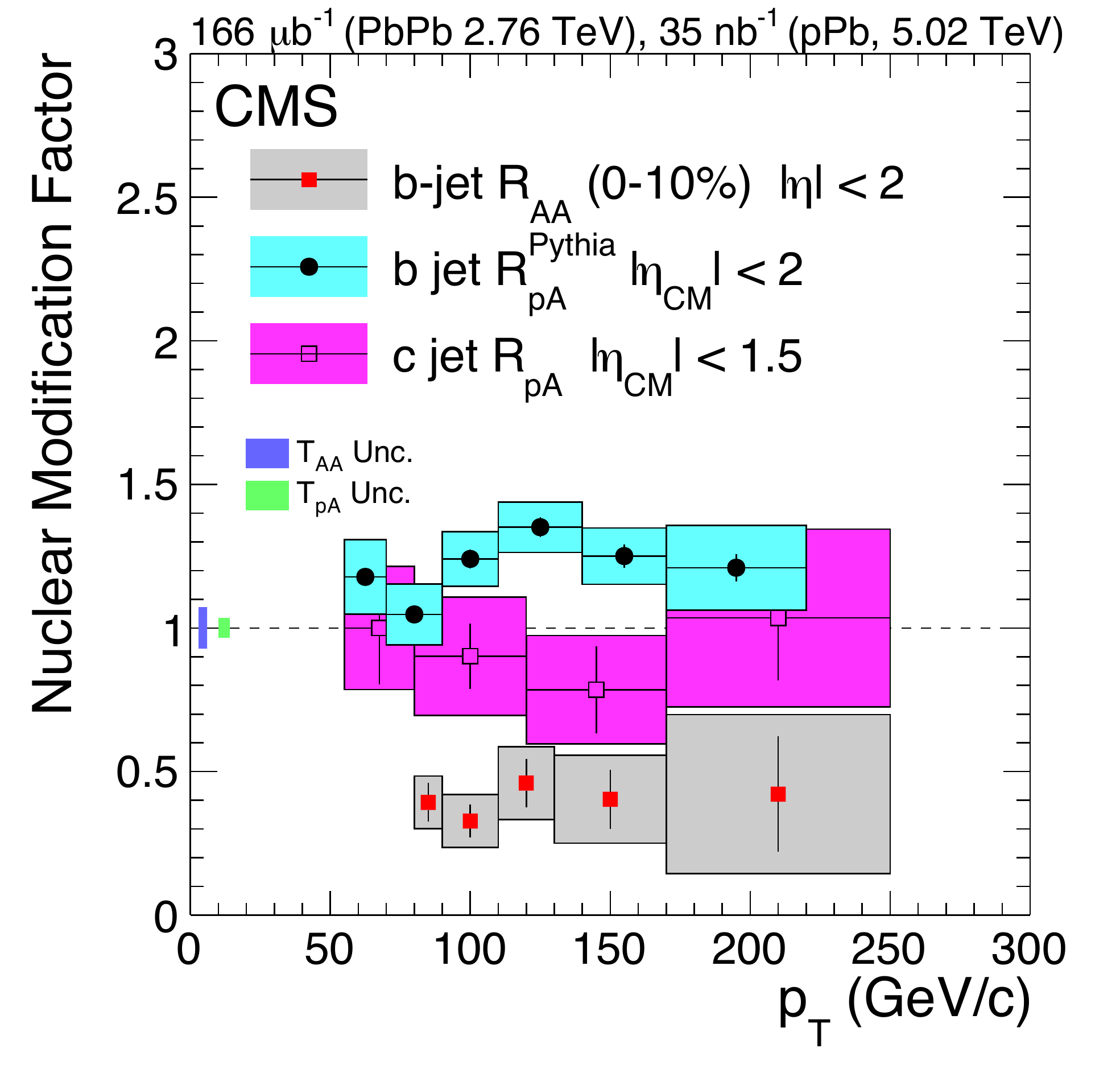}
}}\caption{The \RAA and \RpPb of heavy flavor associated jets measured by the CMS Collaboration~\cite{Chatrchyan:2013exa,Khachatryan:2015sva,Sirunyan:2016fcs}.  This shows that b quarks lose energy in the medium.  Figure courtesy of Kurt Jung.}
\label{Fig:CMSHFJets}
\end{center}
\end{figure}

Experimentally, heavy flavor jets are primarily identified using the relative long lifetimes of hadrons containing heavy quarks, resulting in decay products significantly displaced from the primary vertex.  
A variant of the secondary vertex mass, requiring three or more charged tracks, is also used to extract the relative contribution of charm and bottom quarks to various heavy flavor jet observables.
However these methods cannot discriminate between heavy quarks from the original hard scattering, which then interact with the medium and lose energy, and those from a parton fragmenting into bottom or charm quarks~\cite{Huang:2013vaa}.  A requirement of an additional B-meson in the event could ensure a purer sample of bottom tagged jets~\cite{Huang:2015mva}, however, this is not currently experimentally accessible due to the limited statistics. \Fref{Fig:CMSHFJets} shows a compilation of all current measurements of heavy flavor jets at LHC~\cite{Chatrchyan:2013exa,Khachatryan:2015sva,Sirunyan:2016fcs}.  The \RAA of bottom quark tagged jets is measured utilizing the \Pb and \pp data collected at \sNN = 2.76 TeV. Bottom tagged jet measurements in \pPb collisions are also performed to study cold nuclear matter effects in comparison to expectations from PYTHIA at the 5 TeV center of mass energy~\cite{Khachatryan:2015sva}. Jets which are associated with the charm quarks in \pPb collisions are also studied with a variant of the bottom tagging algorithm~\cite{Sirunyan:2016fcs}.    A strong suppression of \RAA of jets associated with bottom quarks is observed in \Pb collisions while the \RpPb is consistent with unity. These CMS measurements demonstrate that jet quenching does not have a strong dependence on parton mass and flavor, at least in the jet \pT range studied~\cite{Chatrchyan:2013exa,Khachatryan:2016jfl}.  The charm jet \RpPb also shows consistent results with negligible cold nuclear matter effects when compared with the measurements from \pp collisions.

\subsubsection{Summary of experimental evidence for partonic energy loss in the medium}
Partonic energy loss in the medium is demonstrated by numerous measurements of jet observables.  To date, the most precise quantitative constraints on the properties of the medium come from comparisons of \RAA to models by the JET collaboration~\cite{Burke:2013yra}.  The interpretation of \RAA as partonic energy loss is confirmed by measurements of dihadron, gamma-hadron, jet-hadron, hadron-jet, and jet-jet correlations.  The assumption about the background contribution and the biases of these measurements vary widely, so the fact that they all lead to a coherent physical interpretation strengthens the conclusion that they are due to partonic energy loss in the medium.  This energy loss scales with the energy density of the system rather than the system size.  

Reaction plane dependent inclusive particle \RAA, inclusive particle \vnum{2}, and jet \vnum{2} indicate that this energy loss is path length dependent, perhaps requiring a parton to traverse a minimum of around 2 fm of QGP to lose energy.  Comparison of jet \vn to models indicates that jet-by-jet fluctuations in partonic energy loss impacts reaction plane dependent measurements significantly, however, this is not yet fully understood theoretically.  

Measurements of heavy quark energy loss are consistent with expectations from models, however, they are also consistent with the energy loss observed for gluons and light quarks.
Studies of heavy quark energy loss will improve substantially with the slated increases in luminosity and detector upgrades.  The STAR heavy flavor tracker has already enabled higher precision measurements of heavy flavor at RHIC and one of the core goals of the proposed detector upgrade, sPHENIX, is precision measurements of heavy flavor jets.  Run 3 at the LHC will enable higher precision measurements of heavy flavor, including studies of heavy flavor jets in the lower momentum region which may be more sensitive to mass effects.

The key question for the field is how to constrain the properties of the medium further.  
The Monte Carlo models the Jetscape collaboration is developing will include both hydrodynamics and partonic energy loss and the Jetscape collaboration plans Bayesian analyses similar to~\cite{Novak:2013bqa,Bernhard:2016tnd} incorporating jet observables.  These models will also enable the exact same analysis techniques and background subtraction methods to be applied to data and theoretical calculations.  We propose including single particle \RAA (including particle type dependence), jet \RAA (with experimental analysis techniques applied), high momentum single particle \vtwo, jet \vtwo, hadron-jet correlations, and \IAA from both $\gamma$-hadron and dihadron correlations. The analysis method for all of these observables should be replicable in Monte Carlos.  We omit \Aj because a majority of these measurements are not corrected for detector effects.  Bayesian analyses comparing theoretical calculations to data may be the best avenue for constraining the properties of the medium using measurements of jets.  This is likely to improve our understanding of which observables are most useful for constraining models.

 \subsection{Influence of the medium on the jet}\label{Sec:ResultsFragmentation}
\Sref{Sec:EnergyLoss} examined the evidence that partons lose energy in the medium, but did not examine how partons interact with the medium.  
Understanding modifications of the jet by the medium requires a bit of a paradigm shift.  As highlighted in \Sref{Sec:ExpMethods}, a measurement of a jet is not a measurement of a parton but a measurement of final state hadrons generated by the fragmentation of the parton.  Final state hadrons are grouped into the jet (or not) based on their spatial correlations with each other (and therefore the parton).
Whether or not the lost energy retains its spatial correlation with the parent parton depends on whether or not the lost energy has had time to equilibrate in the medium.  
If a bremsstrahlung gluon does not reach equilibrium with the medium, when it fragments it will be correlated with the parent parton.  
Interactions with the medium shift energy from higher momentum final state particles to lower momentum particles and broadens the jet.  Similar apparent modifications could occur if partons from the medium become correlated with the hard parton through medium interactions~\cite{Casalderrey-Solana:2016jvj}.  Whether or not this lost energy is reconstructed as part of a jet depends on the jet finding algorithm and its parameters.

Whereas the observation that energy is lost is relatively straightforward, there are many different ways in which the jet may be modified, and we cannot be sure which mechanisms actually occur in which circumstances until we have measured observables designed to look for these effects.  There are several different observables indicating that jets are indeed modified by the medium, each with different strengths and weaknesses.  We distinguish between mature observables -- those which have been measured and published, usually by several experiments -- and new observables -- those which have either only been published recently or are still preliminary.  Mature observables largely focus on the average properties of jets as a function of variables which we can either measure directly or are straightforward to calculate, such as momentum and the position of particles in a jet.  This includes dihadron correlations (h-h); correlations of a direct photon or Z with either a hadron or a reconstructed jet ($\gamma$-h and $\gamma$-jet); the jet shape ($\rho(r)$); the dijet asymmetry (\Aj); the momentum distribution of particles in a reconstructed jet, called the fragmentation function ($D_{jet}(z)$ where $z=p_T/E_{jet}$); identification of constituents (PID), and heavy flavor jets (HF jets).  Where our experimental measurements of these observables have limited precision, this is either due to the limited production cross section (heavy flavor jets and correlations with direct photons) or due to limitations in our understanding of the background (identified particles).

Our improving understanding of the parton-medium interactions has largely motivated the search for new, more differential observables.  Partonic energy loss is a statistical process so ensemble measurements such as the average distribution of particles in a jet, or the average fractional energy loss, are important but can only give a partial picture of partonic energy loss.  Just as fluctuations in the initial positions of nucleons must be understood to properly interpret the final state anisotropies of the medium, fluctuations play a key role in partonic interactions with the medium.  The average shape and energy distribution of a jet is smooth, but each individual jet is a lumpy object.  These new observables include the jet mass $M_{jet}$, subjettiness ($N_{subjettiness}$), LeSub, the splitting function $z_g$, the dispersion ($\rm p_T^D$), and the girth (g).  We leave the definitions of these variables to the following sections and we focus our discussion on observables which have been measured in heavy ion collisions, omitting those which have only been proposed to date.  In general these observables are sensitive to the properties and structure of individual jets, and they are adapted from advances in jet measurements from particle physics.  Investigations of new observables are important because they will allow access to well defined pQCD observables, which increases the sensitivity of our measurements to the properties of the QGP.  The goal of each new observable is to construct something that is sensitive to properties of the medium that our mature observables are not sufficiently sensitive to, or to be able to disentangle physics processes that are not directly related to the medium properties, such as the difference in fragmentation between quark and gluon jets.  Most measurements of these new observables are still preliminary and we therefore avoid drawing strong conclusions from them.  Our understanding of these observables is still developing, particularly our understanding of how they are impacted by analysis cuts and the approach to the approach used to remove background effects.  An observable which is highly effective for, say, distinguishing between quark and gluon jets in \pp collisions, may not be as effective in heavy ion collisions.

\begin{table*}
\begin{center}
\caption{Summary of measurements sensitive to fragmentation in heavy ion collisions. Preliminary measurements are denoted with a (P).  New observables are separated from mature observables by a line. The first two columns after the observable describe biases inherent to the observable, while the next four columns refer to observations made from the measured results.  We refer the readers to each section for details of measurements of each observable.}
\label{Tab:FragmentationObservables}
\begin{tabular}{|C{2cm} | c| c |C{2cm} |C{2cm} | C{2cm} | c|c  |}
\hline
Observable        &kinematics   & q/g bias & evidence of modification & evidence of broadening& evidence of softening & measured by & Discussion \\ \hline
$D_{jet}(z)$     &constrained  &q bias     & yes  & insensitive &yes&CMS, ATLAS  & \ref{Sec:FragFuncsJets} \\
$\gamma$-h                &very well    &q only     & yes & yes&yes&STAR, PHENIX & \ref{Sec:FragFuncsBosonJet}   \\
$\gamma$-jet              &very well    &q only     & yes &           & &CMS& \ref{Sec:FragFuncsBosonJet}    \\
h-h                        &poor         &unknown& yes & yes&yes&STAR, PHENIX, ALICE, CMS & \ref{Sec:dihadronfrag}  \\
jet-h             &constrained  &q bias     & yes & yes&yes&ALICE(P), CMS, STAR & \ref{Sec:jethadfrag} \\
\Aj               &constrained  &q bias     & yes & insensitive&yes&STAR, ATLAS, CMS & \ref{Sec:Ajfrag}\\
$\rho(r)$         &constrained  &q bias     & yes &yes&yes&CMS & \ref{Sec:jetshape}\\ 
identified h-h               &poor &select     & no & &&STAR, PHENIX & \ref{Sec:PID}\\
HF jets          &constrained  &q          & yes & & &CMS &N/A\\  \hline
LeSub             &constrained  &unknown& no & &  &ALICE(P) &\ref{Sec:LeSub}\\
$\rm p_T^D$       &constrained  &select     & yes & & &ALICE(P) &\ref{Sec:Dispersion}\\
girth             &constrained  &select     & yes & &  &ALICE(P)&\ref{Sec:Girth} \\
$z_g$             &constrained  &unknown     & yes (CMS), no (STAR)  &          &&CMS, STAR(P)&\ref{Sec:Grooming} \\
 $\tau_{N}$&constrained  &unknown        & no & &&ALICE(P)&\ref{Sec:Nsubjettiness} \\
$M_{jet}$         &constrained  &unknown        & no &        &&ALICE &\ref{Sec:JetMass}\\
\hline
\end{tabular}
\end{center}
\end{table*}

We summarize the current status of observables sensitive to the medium modifications of jets in \Tref{Tab:FragmentationObservables}.  This list of observables also shows the evolution of the field.  Early on, due to statistical limitations, studies focused on dihadron correlations.  These measurements are straightforward experimentally, however, they are difficult to calculate theoretically because all hadron pairs contribute and the kinematics of the initial hard scattering is poorly constrained.  In contrast, as discussed in~\Sref{sec:bosonjet}, when direct photons are produced in the process q+g $\rightarrow$ q+$\gamma$, the initial kinematics of the hard scattered partons are known more precisely.  In some kinematic regions, these measurements are limited by statistics, and in others they are limited by the systematic uncertainty predominately from the subtraction of background photons from $\pi^0$ decay.  Measurements of reconstructed jets are feasible over a wider kinematic region, but the kinematics of the initial hard scattering are not constrained as well.  Nearly all measurements are biased towards quarks for the reasons discussed in \Sref{Sec:ExpMethods}, however, it may be possible to tune the bias either using identified particles or by using new observables that select for particular fragmentation patterns.  

\Tref{Tab:FragmentationObservables} summarizes whether or not modifications, particularly broadening and softening, have been observed using each observable and which experiments have measured them.  
This table demonstrates that each measurement has strengths and weaknesses and that all observations contribute to our current understanding.  Modifications to the jet structure have been observed for most observables, but not all.  Since each observable is sensitive to different modifications, all provide useful input for differentiating between jet quenching models and understanding the effects of different types of initial and final state processes.  We begin our discussion of measurements indicating modification of jets by the medium with mature observables.  For each observable we revisit these issues in a discussion stating what we have learned from that observable.

\subsubsection{Fragmentation functions with jets}\label{Sec:FragFuncsJets}
Fragmentation functions are a measure of the distribution of final state particles resulting from a hard scattering and represent the sum of parton fragmentation functions, $D_{i}^{h}$, where $i$ represents each parton type ($u, d, g, etc.$) contributing to the final distribution of hadrons, $h$. Typically, fragmentation functions are measured as a function of $z$ or $\xi$ where $z=p^{h}/p$ and $\xi=-\ln(z)$, where $p$ is the momentum of parton produced by the hard scattering. Jet reconstruction can be used to determine the jet momentum, $p^{jet}$ to approximate the parton momentum $p$, while the momentum of the hadrons, $p^{h}$, are measured for each hadron that is clustered into the jet by the jet reconstruction algorithm. 
In collider experiments, the transverse momentum,\pT, is typically substituted for the total momentum p in the fragmentation function.
It should be noted that this is not precisely the same observable as what is commonly referred to as the fragmentation function by theorists.

\begin{figure*}
\begin{center}

\rotatebox{0}{\resizebox{13cm}{!}{
        \includegraphics{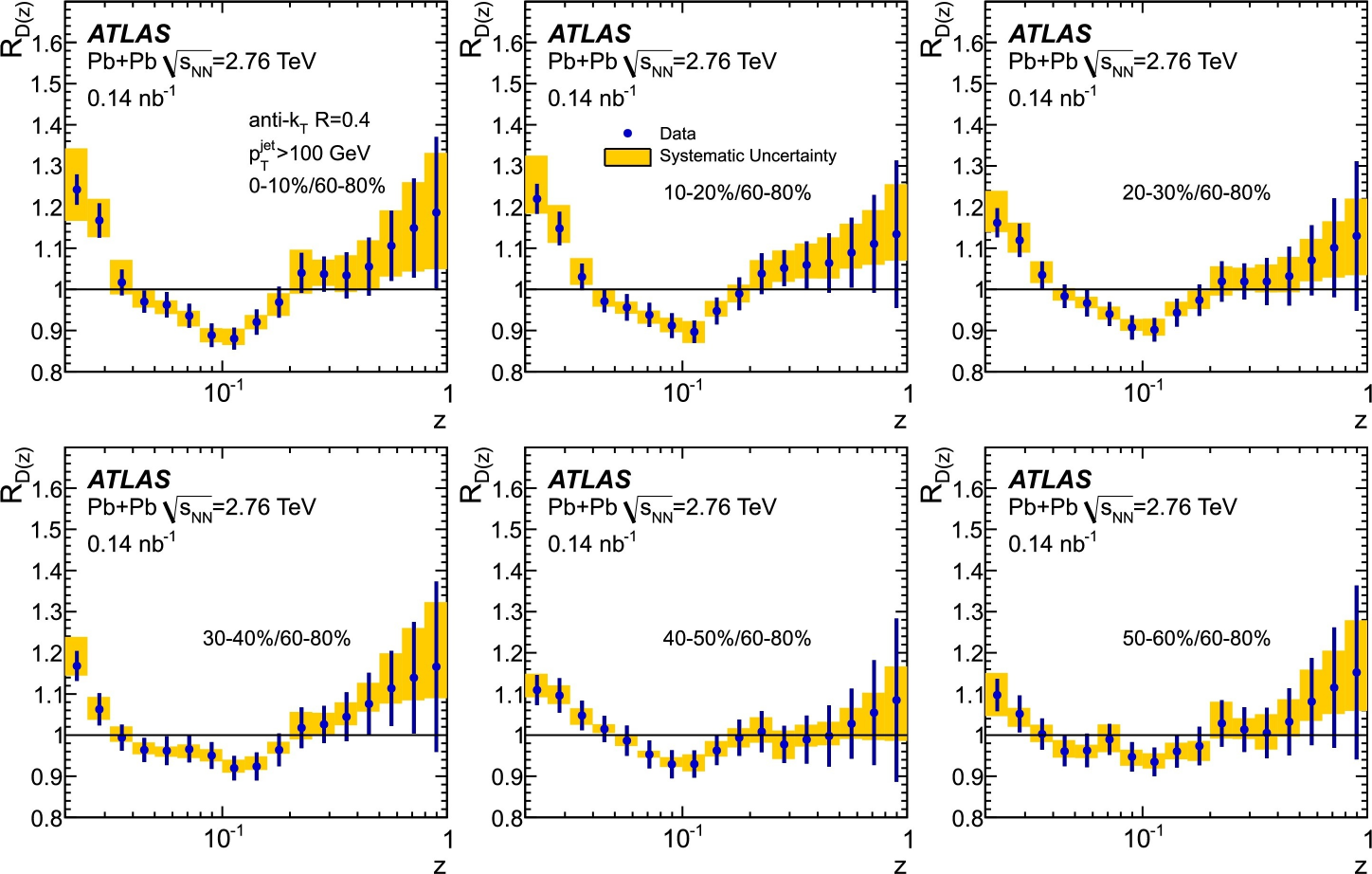}
}}\caption{
Figure from ATLAS~\cite{Aad:2014wha}. Ratio of fragmentation functions from reconstructed jets measured by ATLAS for jets in \Pb collisions at various centralities to those in 60-80\% central collisions at \sNN = 2.76 TeV.  This shows that fragmentation functions are modified in \AplusA collisions, with an enhancement at low momenta (low $z$) and a depletion at intermediate momenta (intermediate $z$), with the modification increasing from more peripheral to more central collisions.
}\label{Fig:LHCFF}
\end{center}
\end{figure*}

The fragmentation functions for jets in \Pb collisions at \sNN = 2.76 TeV have been measured by the ATLAS~\cite{Aad:2014wha} and CMS~\cite{Chatrchyan:2014ava,Chatrchyan:2012gw} Collaborations. The ratios of the fragmentation functions for several different centrality bins to the most peripheral centrality bin are shown in Figure \ref{Fig:LHCFF}. The most central collisions show a significant change in the average fragmentation function relative to peripheral collisions. At low z there is a noticeable enhancement followed by a depletion at intermediate z. This suggests that the energy loss observed for mid to high momentum hadrons is redistributed to low momentum particle production.  We note that this corresponds to only a few additional particles and is a small fraction of the energy that \RAA, \Aj and the other energy loss observables discussed in \Sref{Sec:EnergyLoss} indicate is lost. Arguably, this is the most direct observation of the softening of the fragmentation function expected from partonic energy loss in the medium. However, the definition of a fragmentation function in \Eref{Eq:Factorization} uses the momentum of the initial parton and, as discussed in \Sref{Sec:ExpMethods}, a jet's momentum is not the same as the parent parton's momentum. 
Fragmentation functions measured with jets with large radii are approximately the same as the fragmentation functions in \Eref{Eq:Factorization}, but this is not true for the jets with smaller radii measured in heavy ion collisions.

\begin{figure}
\begin{center}

\rotatebox{0}{\resizebox{8cm}{!}{
        \includegraphics{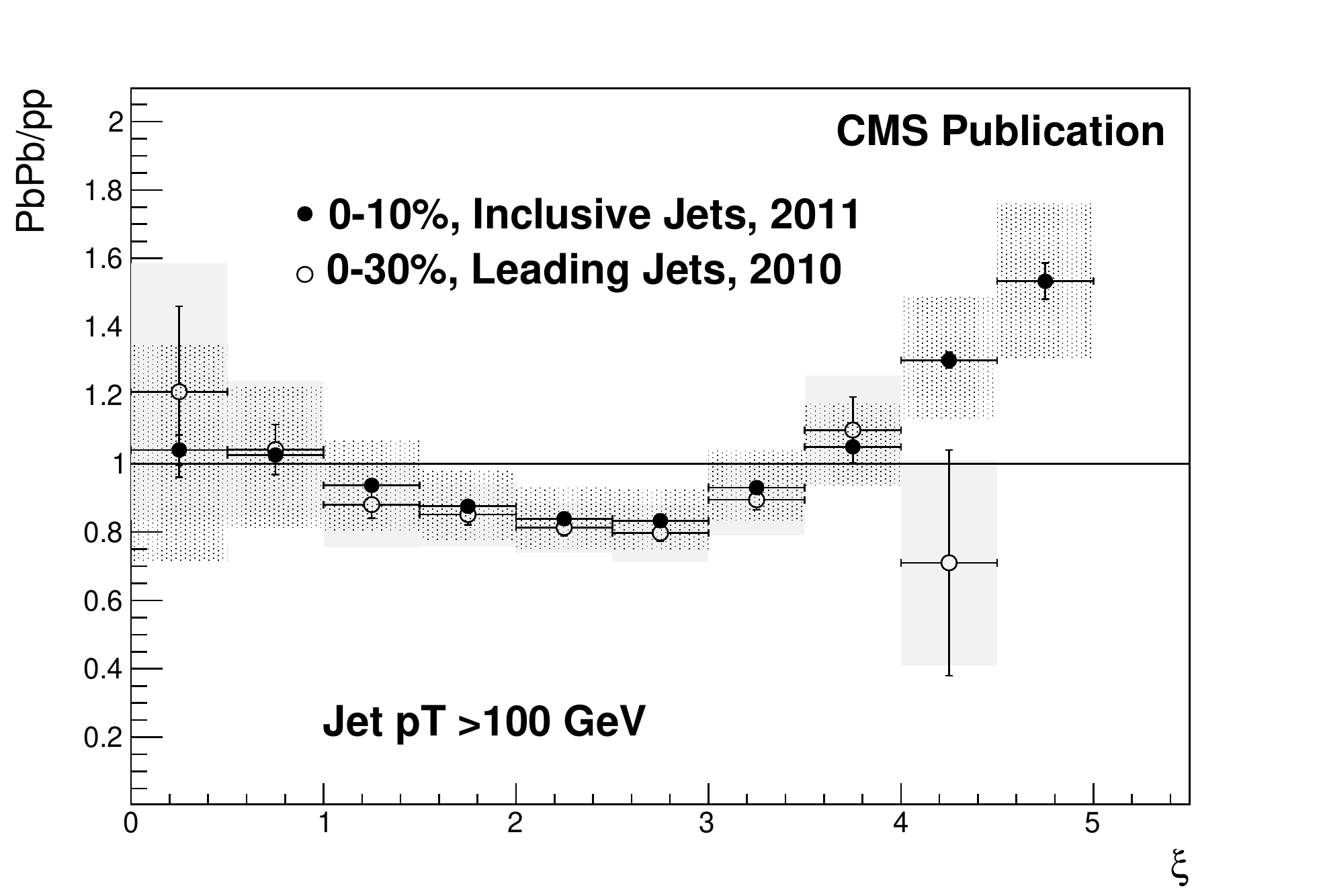}
}}\caption{Comparison of CMS measurements of fragmentation functions in \Pb over pp from reconstructed jets for jets in \Pb collisions at \sNN = 2.76 TeV from 2010 and 2011 data~\cite{Chatrchyan:2014ava,Chatrchyan:2012gw}.  Even though the two measurements are consistent, the 2010 data in isolation indicate that fragmentation is not modified while the 2011 data, which extend to lower momenta and use a less biased jet sample, clearly show modification at low momenta (high $\xi$).  This highlights the difficulty in drawing conclusions from a single measurement, particularly when neglecting possible biases.  
}\label{Fig:LHCFFCMS}
\end{center}
\end{figure}

It is important to note that initial fragmentation measurements from the LHC used only dijets samples with large momenta (\pT $>4$ GeV/c) constituents, which indicated that there was no modification of fragmentation functions~\cite{Chatrchyan:2012gw}.   With increased statistics and improved background estimation techniques these fragmentation measurements were remeasured later with inclusive jets with constituent tracks with \pT $ >1$ GeV/c utilizing the 2011 data. \Fref{Fig:LHCFFCMS} compares the measurements from CMS from two different measurements using 2010 and 2011 data.  The initial 2010 analysis did not include lower momentum jet constituents due to the difficulty with background subtraction in that kinematic region and focused on leading and subleading jets. While the two measurements are consistent, the conclusion drawn from the 2010 data alone was that there was no apparent modification of the jet fragmentation functions.  This highlights how critical biases are to the proper interpretation of measurements. The high momentum of these jets combined with the background subtraction and suppression techniques also means that the data in both \Fref{Fig:LHCFF} and \Fref{Fig:LHCFFCMS} are likely biased towards quark jets.

\subsubsection{Boson tagged fragmentation functions}\label{Sec:FragFuncsBosonJet}

\begin{figure}
\begin{center}
\rotatebox{0}{\resizebox{8cm}{!}{
        \includegraphics{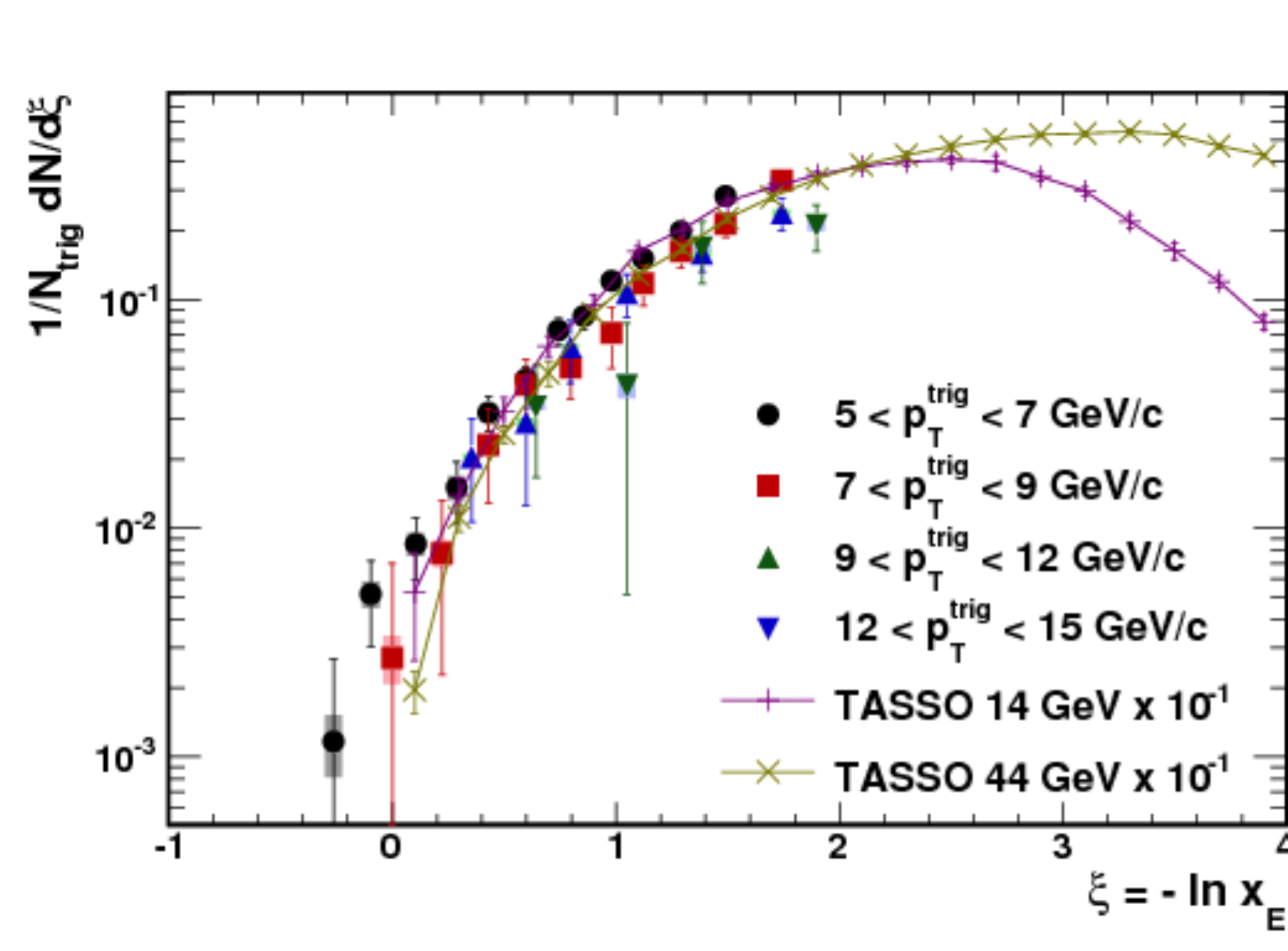}
}}\caption{
Figure from PHENIX~\cite{Adare:2010yw}. $\xi=-\ln(x_E)$ distributions where $x_E = -|p_{T}^{a}/p_{T}^{t}|\cos(\Delta\phi)\approx z$ for isolated direct photon-hadron correlations for several photon \pT ranges from \pp collisions at \sqrts = 200 GeV compared to TASSO measurements in \ee collisions at \sqrts =14 and 44 GeV.  This demonstrates that direct photon measurements can be used reliably to extract quark fragmentation functions in \pp collisions and that fragmentation functions are the same in \ee and \pp collisions.
}\label{Fig:PHENIXGammaFF}
\end{center}
\end{figure}

\begin{figure}
\begin{center}
\rotatebox{0}{\resizebox{8cm}{!}{
        \includegraphics{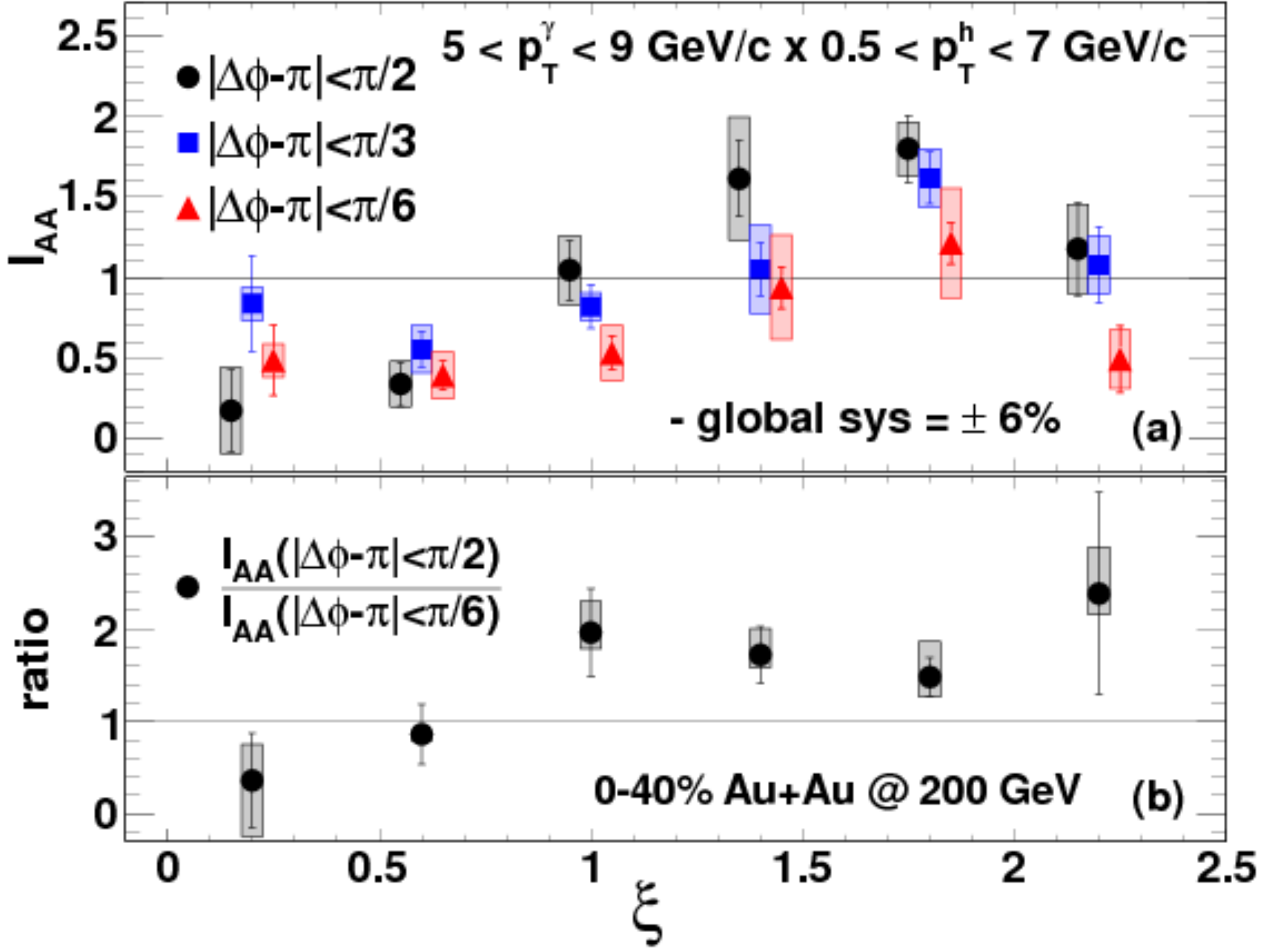}
}}\caption{
Figure from PHENIX~\cite{Adare:2012qi}. The top panel shows \IAA for the \as as a function of  $\xi=\log(\frac{1}{z}) = \log(\frac{p^{jet}}{p^{had}})$. The points are shifted for clarity. The bottom panel shows the ratio of the \IAA for $|\Delta\phi-\pi|<\pi/2$ to $|\Delta\phi-\pi|<\pi/6$.  This demonstrates the enhancement at low momentum combined with a suppression at high momentum, a shift consistent with expectations from energy loss models.  The change is largest for wide angles from the direct photon.
}\label{Fig:ppg113IAA}
\end{center}
\end{figure}

As described previously, bosons can be used to tag the initial kinematics of the hard scattering. For fragmentation functions, this gives access to the initial parton momentum in the calculation of the fragmentation variable $z$.
At the top \Au  collision energy at RHIC, \sNN = 200 GeV, there have been no direct measurements of fragmentation functions from reconstructed jets so far, however, $\gamma$-hadron correlations have been measured both in \pp and \Au collisions. The fragmentation function was measured in \pp collisions at RHIC as a function of $x_E = -|\frac{p_{T}^{a}}{p_{T}^{t}}|\cos(\Delta\phi)\approx z$~\cite{Adare:2010yw} and is shown in \Fref{Fig:PHENIXGammaFF}. The \pp
results agree well with the TASSO measurements of the quark fragmentation function in electron-positron collisions, which is consistent with the production of a quark jet opposite the direct photon as expected in 
Compton scattering. Using the \pp results as a reference, direct photon-hadron correlations were measured in \Au collisions at RHIC~\cite{Adare:2012qi}. The \IAA are shown in
Figure \ref{Fig:ppg113IAA}. A suppression is observed for $\xi<1$ ($z>0.4$) while an enhancement is observed for $\xi>1$ ($z<0.4$). This suggests that energy loss at high z is redistributed to low $z$. Comparing these results to the results from STAR~\cite{Abelev:2009gu,STAR:2016jdz} suggests that this is not a $z_{T}$ dependent effect but rather a \pT dependent effect. STAR measured direct photon-hadron correlations for a similar $z_{T}$ range but does not observe the clear enhancement exhibited in the PHENIX measurement. However, STAR is able to measure low values of $z_{T}$ by increasing the trigger photon \pT, while PHENIX goes to low $z_{T}$ by decreasing the associated hadron \pT. Preliminary PHENIX results as a function of photon \pT are consistent with the conclusion that modifications of fragmentation depend on associated particle \pT rather than $z_{T}$. Furthermore, STAR does observe an enhancement for jet-hadron correlations with hadrons of \pT~$< 2$ GeV/c which is consistent with the PHENIX direct photon-hadron observation. 

The direct photon-hadron correlations also suggest that the low \pT enhancement occurs at wide angles with respect to the axis formed by the hard scattered partons. Figure
\ref{Fig:ppg113IAA} shows the yield measured by PHENIX for different $\Delta\phi$ windows on the \as. The enhancement is most significant for the
widest window, $|\Delta\phi-\pi|<\pi/2$.

\subsubsection{Dihadron correlations}\label{Sec:dihadronfrag}

Measurements of \dhcs are sensitive to modifications in fragmentation, although the interpretation is complicated because the initial kinematics of the hard scattering are poorly constrained.  Differences observed in the correlations can either be due to medium interactions or due to changes in the parton spectrum.  At high $p_{T}$, there are no indications of modification of the near- or \as at midrapidity in \dAu collisions~\cite{Adler:2005ad,Adler:2006hi} so any effects observed in \AplusA are hot nuclear matter effects and either \dAu or \pp can be used as a reference for \AplusA collisions.  

\begin{figure}
\begin{center}
\rotatebox{0}{\resizebox{8cm}{!}{
	\includegraphics{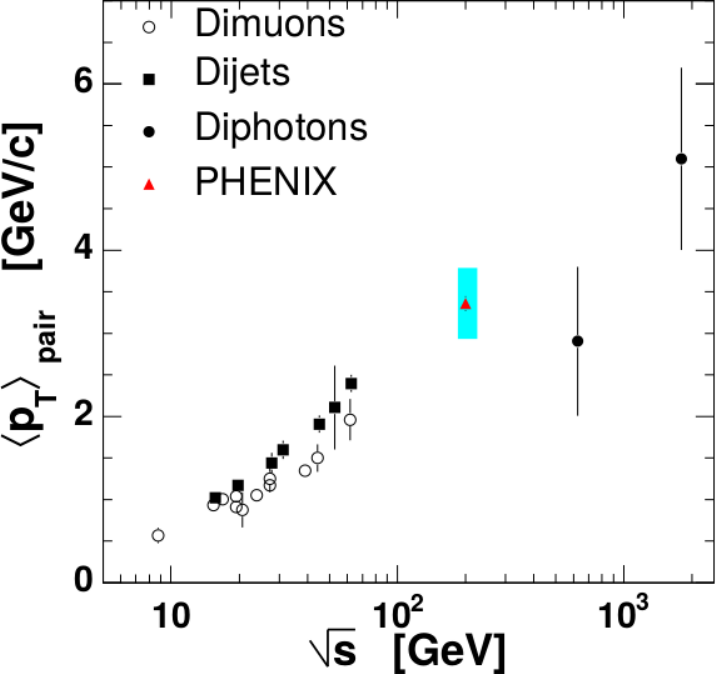}
}}\caption{
Figure from PHENIX~\cite{Adler:2006sc}.  Compilation of $\langle p_{T} \rangle _{pair} = \sqrt{2}$\kT measurements where \kT is the acoplanarity momentum vector. Dihadron correlation measurements in \pp collisions from PHENIX are consistent with the trend from dimuon, dijet and diphoton measurements at other collision energies. Dimuon and dijet measurements are from fixed target experiments and the diphoton measurements are from the Tevetron. 
 }\label{Fig:ktvss}
\end{center}
\end{figure}

The \ns peak can be used to study the angular distribution of momentum and particles around the triggered jet. The \as peak is wider than the \ns due to the resolution of the triggered jet peak axis and the effect of the acoplanarity momentum vector, \kT. Dihadron correlations have been measured in \pp collisions to determine the intrinsic \kT. Measurements of $\langle p_{T} \rangle _{pair} = \sqrt{2}$\kT as a function of \sqrts are shown in \Fref{Fig:ktvss}. 

The effect of the nucleus on \kT has been studied in \dAu collisions at 200 GeV \cite{Adler:2005ad} and in \pPb collisions at 5.02 TeV \cite{Adam:2015xea} via dihadron correlations and reconstructed jets respectively. The dihadron measurements in \dAu are consistent with the PHENIX \pp measurements shown in \Fref{Fig:ktvss}, while the \pPb dijet results agree with PYTHIA expectations. Since no broadening has been observed in \pPb or \dAu collisions, any broadening of the \as jet peak in \AplusA collisions would be the result of modifications from the QGP. Assuming this is purely from radiative energy loss, the transport coefficient \qhat can be extracted directly from a measurement of \kT according to \qhat $\propto \langle k_{T}^{2} \rangle$ \cite{Tannenbaum:2017afg}.

\begin{figure*}
\begin{center}
\rotatebox{0}{\resizebox{15cm}{!}{
	\includegraphics{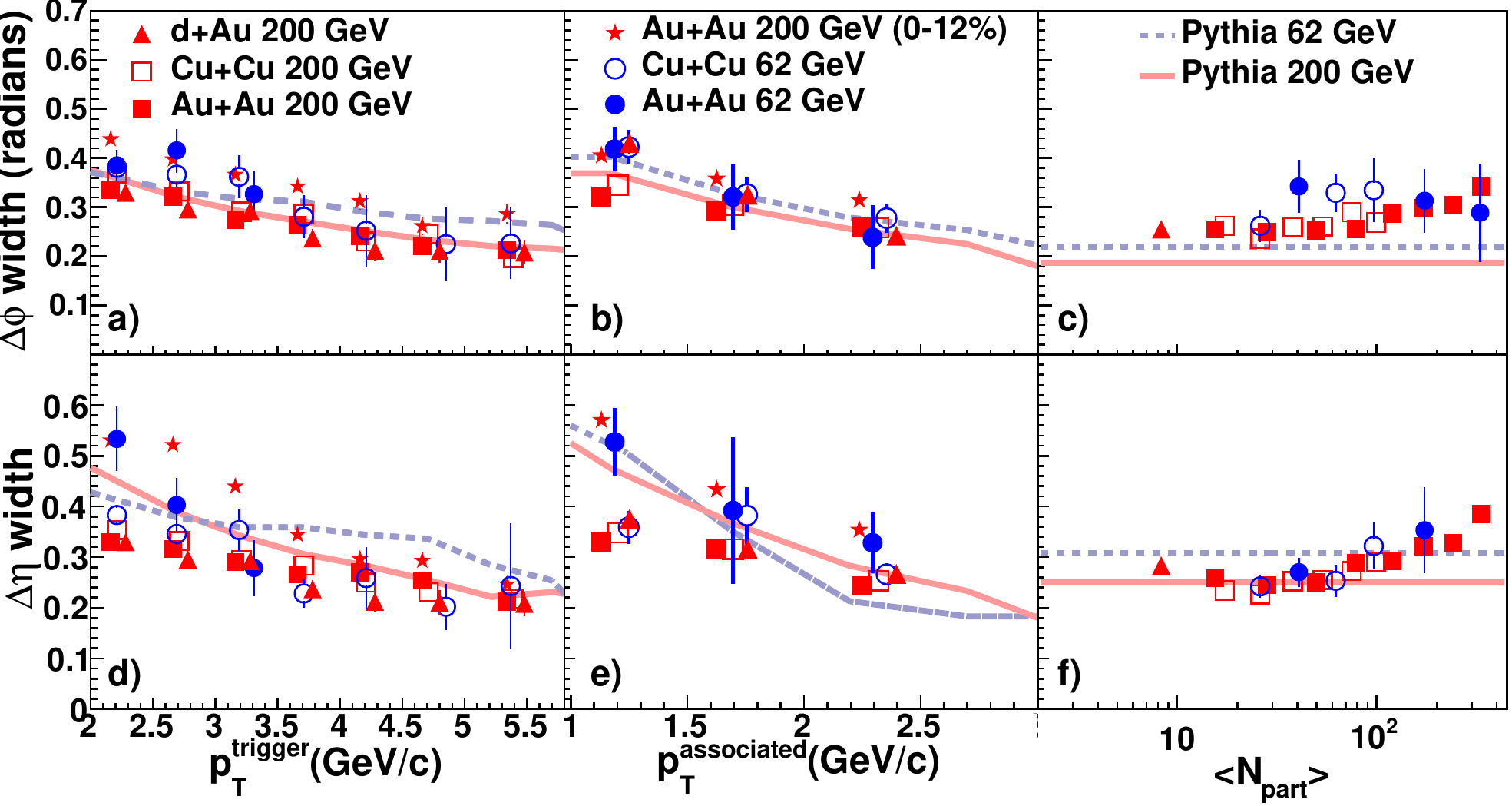}
}}\caption{
Figure from STAR~\cite{Agakishiev:2011st}. Dependence of the Gaussian widths in $\Delta\phi$ and $\Delta\eta$ on 
\pttrig for $1.5$~GeV/$c$ $<$ $p_T^{\mathrm{a}}$ $<$ $p_T^{\mathrm{t}}$, \ptassoc for $3$~$<$ $p_T^{\mathrm{t}}$ $<$ $6$~GeV/$c$, and 
$ \langle N_{\mathrm{part}}\rangle$ for $3$~$<$ $p_T^{\mathrm{t}}$ $<$ $6$~GeV/$c$ and $1.5$~GeV/$c$ $<$ $p_T^{\mathrm{a}}$ $<$ $p_T^{\mathrm{t}}$ for 0-95\% \dAu, 0-60\% \Cu at \sNN = 62.4 GeV and \sNN = 200 GeV, 0-80\% \Au at \sNN = 62.4 GeV, and 0-12\% and 40-80\% \Au at \sNN = 200 GeV.  This demonstrates that the correlation is broadened in  central \Au collisions.
 }\label{Fig:DHCSNSWidths}
\end{center}
\end{figure*}

\Fref{Fig:DHCSNSWidths} shows the widths in $\Delta\phi$ and $\Delta\eta$ on the \ns as a function of \pttrig, \ptassoc, and the average number of participant nucleons, $ \langle N_{\mathrm{part}}\rangle$ for \dAu, \Cu, and \Au collisions at \sNN = 62.4 and 200 GeV~\cite{Agakishiev:2011st}.  The \ns is broader in both $\Delta\phi$ and $\Delta\eta$ in central collisions.  This broadening does not have a strong dependence on the angle of the trigger particle relative to the reaction plane~\cite{Nattrass:2016cln}.  One interpretation of this is that the jet-by-jet fluctuations in partonic energy loss are more significant than path length dependence for this observable~\cite{Zapp:2013zya}.
Higher energy jets have higher particle yields and are more collimated, so if changes were due to an increase in the average parton energy the yield would increase but the width would decrease.  In contrast, interactions with the medium would lead to broadening and the softening of the fragmentation function which would lead to more particles.  The \ns yields are not observed to be modified~\cite{Agakishiev:2011st}, although \IAA at RHIC~\cite{Nattrass:2016cln} is also consistent with the slight enhancement seen at the LHC~\cite{Aamodt:2011vg}.  This indicates that the increase in width is most likely due to medium interactions rather than changes in the parton spectra.

Recent studies of the \as do not indicate a measurable broadening~\cite{Nattrass:2016cln}, at least for the low momenta in this study ($4$~$<$ $p_T^{\mathrm{t}}$ $<$ $6$~GeV/$c$, $1.5$~GeV/$c$ $>$ $p_T^{\mathrm{a}}$).  This is in contrast to earlier studies which neglected odd \vn in the background subtraction, indicating dramatic shape changes.  These earlier studies are discussed in greater detail in \Sref{Sec:MachCone} because the modifications observed were generally interpreted as an impact of the medium on the jet.  We note that broadening is observed on the \as for jet-hadron correlations, as discussed below.  The current apparent lack of broadening in dihadron correlations may indicate that this is not the most sensitive observable because of the decorrelation between the trigger on the \ns and the angle of the \as jet.  It may also be a kinematic effect because modifications are extremely sensitive to momentum.  The \as \IAA decreases with increasing \ptassoc, indicating a softening of the fragmentation function of surviving jets~\cite{Nattrass:2016cln}.

A large collection of experimental measurements in \ee collisions show that jets initiated by gluons exhibit differences with respect to jets from light-flavor quarks~\cite{Acton:1993jm,OPAL:1995ab,Abreu:1995hp,Barate:1996fi,Buskulic:1995sw}. First, the charged particle multiplicity is higher in gluon jets than in light-quark jets. Second, the fragmentation functions of gluon jets are considerably softer than that of quark jets. Finally, gluon jets appeared to be less collimated than quark jets. These differences have already been exploited to differentiate between gluon and quark jets in \pp collisions~\cite{CMS:2013kfa}.  The simplest and most studied variable used experimentally is the multiplicity, the total number of constituents of reconstructed jet.  Since gluon hadronization produces jets which are `wider' than jets induced by quark hadronization, jet shapes could be studied with jet width variables to distinguish quark and gluon jets.

\begin{figure*}
\begin{center}
\rotatebox{0}{\resizebox{\textwidth}{!}{
	\includegraphics{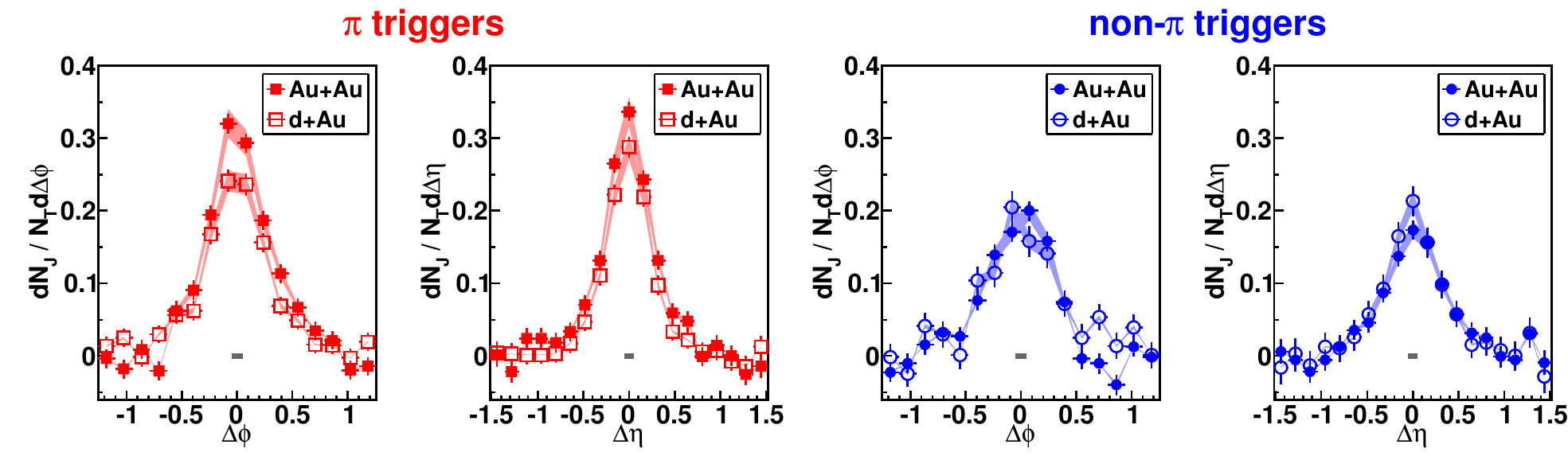}
}}
\caption{
Figure from STAR~\cite{Abdelwahab:2014cvd}.
The \dphi and \deta projections of the correlation for $|\Delta\eta|<$~0.78 and $|\Delta\phi|<\pi/4$, respectively, for pion triggers (left two panels) and non-pion triggers (right two panels). Filled symbols show data from the 0--10\% most central \Au collisions at \sNN = 200 GeV.  Open symbols show data from minimum bias \dAu data at \sNN = 200 GeV.  This figure shows that the yield is higher for pion trigger particles than non-pion trigger particles, which are mostly kaons and protons, and that there is a higher yield for pion trigger particles in central \Au collisions than in \dAu collisions.  This may be an indication of differences in partonic energy loss for quarks and gluons in the medium.
}
\label{Fig:PIDCorr}
\end{center}
\end{figure*}

Since there are significant differences in baryon and meson production in \AplusA collisions compared to \pp collisions, such differences may exist for jets.  Furthermore, energy loss is different for quark and gluon jets, so species-dependent energy loss may mean that there are differences between jets with different types of leading hadrons.  These differences may be observed through comparisons of jets with leading baryons and mesons or light and strange hadrons.  The OPAL collaboration measured the ratio of K$_{0}^{S}$ production in \ee collisions in gluon jets to that in quark jets to be 1.10~$\pm$~0.02~$\pm$~0.02 and the ratio of $\Lambda$ production in gluon jets to that in quark jets to be 1.41~$\pm$~0.04~$\pm$~0.04~\cite{Ackerstaff:1998ev}, meaning that jets containing a $\Lambda$ or a proton are somewhat more likely to arise from gluon jets than jets which do not contain a baryon.  This difference is small, however, a large difference in the interactions between quark and gluon jets in heavy ion collisions may be observable.

Measurements of \dhcs with identified leading triggers may be sensitive to these effects.  Studies of identified strange trigger particles found a somewhat higher yield in jets with a leading K$_{0}^{S}$ than those with a leading unidentified charged hadron or $\Lambda$ at the same momentum~\cite{Abelev:2016dqe}.  This was also observed in \dAu collisions, indicating that the more massive leading $\Lambda$ simply takes a larger fraction of the jet energy.  The slight centrality dependence indicates there may be medium effects, however, these could arise from differences in quark and gluon jets or from strange and non-strange jets.  Ultimately these data are inconclusive due to their low precision.  \Dhcs with identified pion and non-pion triggers~\cite{Abdelwahab:2014cvd} shown in \Fref{Fig:PIDCorr} observed a higher yield in jets with a leading pion than those with a leading kaon or proton.  This difference was larger in \Au collisions than in \dAu collisions, which~\cite{Abdelwahab:2014cvd} proposes may be impacted to fewer baryon trigger particles coming from jets due to recombination.  Both of these results could be impacted by several effects -- differences in quark and gluon jets in the vacuum, differences in energy loss in the medium for quark and gluon jets, and modified fragmentation in the medium.  Since both studies observe differences, at least some of these effects are present in the data, however, the data cannot distinguish which effects are present.

\subsubsection{Jet-hadron correlations}\label{Sec:jethadfrag}

\begin{figure*}
\begin{center}
\rotatebox{0}{\resizebox{15cm}{!}{
	\includegraphics{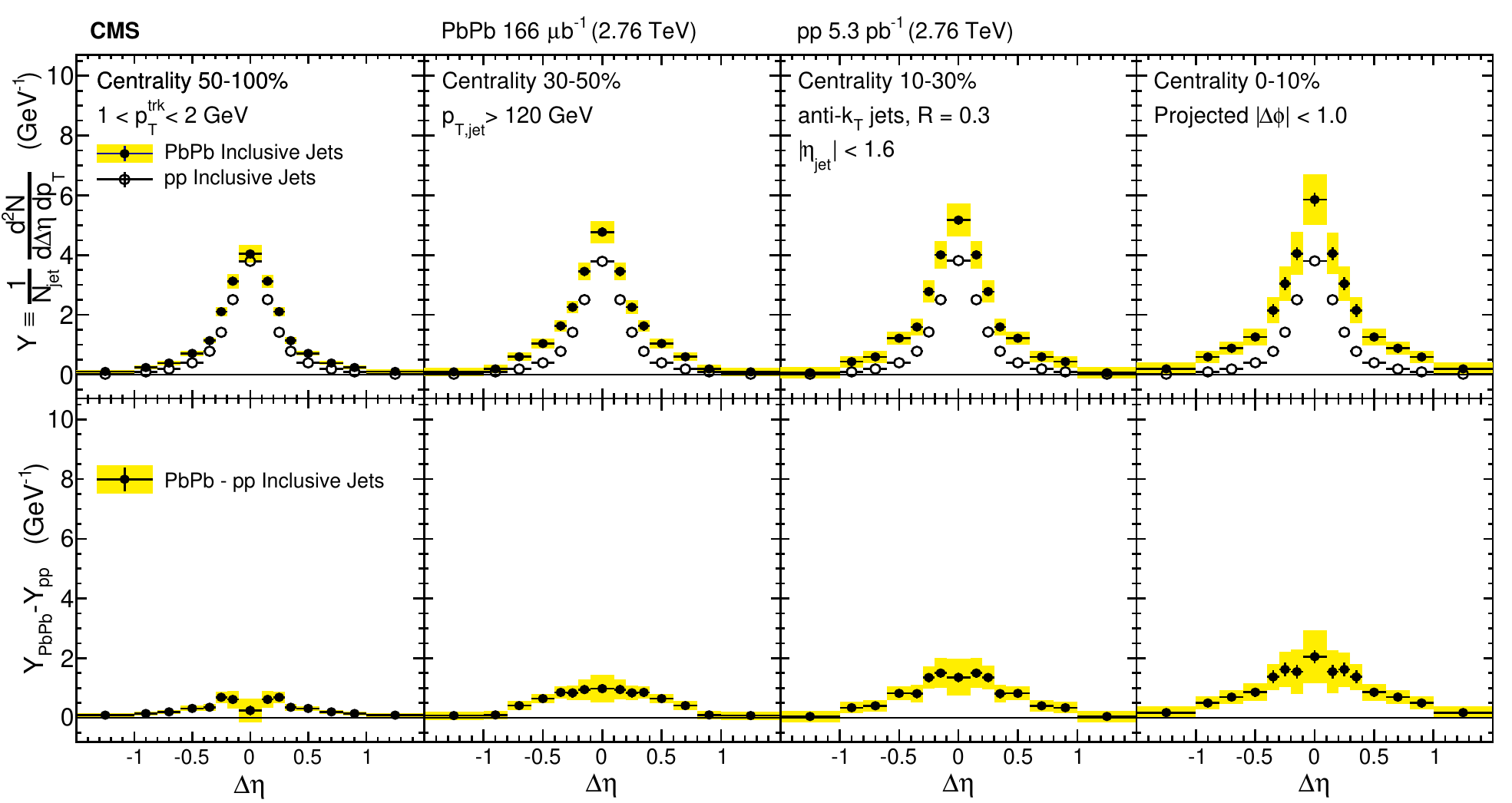}
}}\caption{
Figure from CMS~\cite{Khachatryan:2016erx}.  Symmetrized \deta distributions correlated with \Pb and \pp inclusive jets with \pT$>$ 120 GeV are shown in the top panels for tracks with 1 $<$\pT$<$ 2 GeV. The difference between per-jet yields in \Pb and \pp collisions is shown in the bottom panels.  These measurements indicate that the jet is broadened and softened, as expected from energy loss models.
 }\label{Fig:CMSTrackJet}
\end{center}
\end{figure*}

Measurements of jet-hadron correlations are sensitive to the broadening and softening of the fragmentation function, but have the advantage over dihadron correlations that the jet will be more closely correlated with the kinematics of its parent parton than a high \pT hadron.  \Fref{Fig:CMSTrackJet} shows \jhcs measured by CMS~\cite{Khachatryan:2016erx} as a function of \deta from the trigger jet.  Not shown here are the results as a function of \dphi from the trigger jet, however the conclusions were quantitatively the same.  The jets in this sample had a resolution parameter of $R$ = 0.3 and a leading jet \pT $>$ 120 \GeV in order reduce the effect of the background on the trigger jet sample.  The background removal for the jets reconstructed in \Pb was done via the HF/Voronoi method, which is described in \cite{CMS-DP-2013-018}, a slightly different method than described in \Sref{Sec:ExpMethods}.  The effect of the combinatorial background on the distribution of associated tracks was removed by a sideband method, in which the background is approximated by the measured two dimensional correlations in the range 1.5 $< |\Delta \eta| <$ 3.0.  Jets in \Pb are observed to be broader, with the greatest increase in the width for low momentum associated particles.  This is consistent with expectations from partonic energy loss.  These studies found that the subleading jet was broadened even more than the leading jet, indicating a bias towards selecting less modified jets as the leading jet.

\begin{figure}
\begin{center}
\rotatebox{0}{\resizebox{8cm}{!}{
	\includegraphics{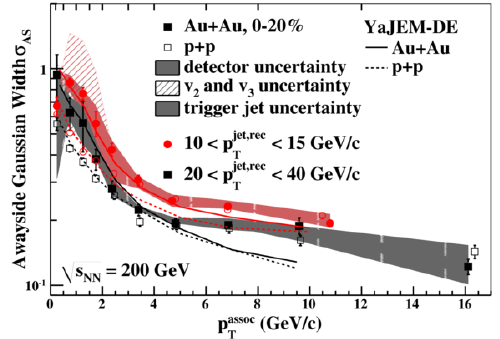}
}}
\rotatebox{0}{\resizebox{8cm}{!}{
	\includegraphics{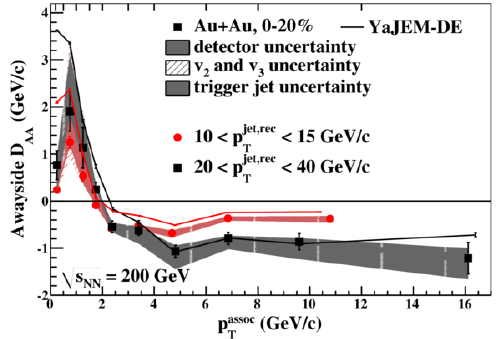}
}}
\caption{
Figure from STAR~\cite{Adamczyk:2013jei}.  
Gaussian widths of the \as peaks ($\sigma_{AS}$) for \pp collisions (open squares) and central \Au collisions (solid squares) (upper) and \as momentum difference $D_{AA}$ as defined in \Eref{Eq:DAA} (lower) are both plotted as a function of \ptassoc.   The widths (note the log scale on the y-axis) show no evidence of broadening in \Au relative to \pp due to the large uncertainties in the \Au measurement. However, $D_{AA}$ shows the suppression of high momentum particles associated with the jet is balanced by the enhancement of lower momentum associated particles.  The point at which enhancement transistions to suppression appears to occur at the same associated particle's momentum and does not depend on the jet momentum. Data are for \sNN = 200 GeV collisions and YaJEM-DE model calculations are from~\cite{Renk:2012hz}.
 }\label{Fig:STARJetHadron}
\end{center}
\end{figure}

Jet hadron correlations have also been studied at RHIC energies, where the width and yield of the \as peak, rather than the associated particle correlations themselves, can be seen in \Fref{Fig:STARJetHadron}.  This figure shows the \as widths and 
\begin{equation}
 D_{AA} = Y_{Au+Au}\langle p_{T}^{assoc} \rangle_{Au+Au} - Y_{p+p}\langle p_{T}^{assoc} \rangle_{p+p}\label{Eq:DAA}
\end{equation}
where $Y_{Au+Au}$ and $Y_{p+p}$ are the number of particles in the \as from~\cite{Adamczyk:2013jei} for two different ranges of jet \pT.  The width in \pp is consistent with that in \Au within uncertainties, although the uncertainties are large due to the large uncertainties in the \vn.  The $D_{AA}$ shows that momentum is redistributed within the jet, with suppression ($D_{AA}<0$) for \pT~$<$~2 \GeV associated particles and enhancement ($D_{AA}>0$) for ~$>$~2 \GeV.  This indicates that the suppression at high momenta was balanced by the enhancement at low momenta, which means that this change in the jet structure likely comes from modification of the jet rather than modifications of the jet spectrum.  This enhancement at low \pT is at the same associated momentum for both jet energies, which may indicate that the enhancement is not dependent on the energy of the jet but the momentum of the constituents.  

\subsubsection{Dijets}\label{Sec:Ajfrag}

The LHC \Aj measurements shown in \Fref{Fig:Aj} show a significant energy imbalance for dijets due to medium effects in central collisions~\cite{Aad:2010bu,Chatrchyan:2011sx} while RHIC \Aj measurements suggest that energy imbalance observed for jet cones of R=0.2 can be recovered within a jet cone of R=0.4 for measurable dijet events~\cite{Adamczyk:2016fqm}.  The STAR measurements demonstrate that the energy imbalance is recovered when including low \pT constituents~\cite{Adamczyk:2016fqm}, also indicating a softening of the fragmentation function. Comparing these two results is complicated since they have very different surface biases, both due to the experimental techniques and the different collision energies.  In order to interpret such comparisons and draw definitive conclusions a robust Monte Carlo generator is required because the differences in these observables are not analytically calculable.  To develop a better picture of the transverse structure of the jets, it is best to measure observables specifically designed to probe the transverse direction.

The effect on dijets along the direction transverse to the jet axis was studied by measuring the angular difference between the reconstructed jet axis of the leading and sub-leading jets~\cite{Aad:2010bu,Chatrchyan:2011sx}. These results are shown in Figure \ref{Fig:Aj} and little change to the angular deflection of the sub-leading jet in central \Pb collisions compared to \pp collisions is observed. It is important to point out that the tails in the \pp distribution may be due to 3-jet events while those pairs in \Pb events are the results of dijets undergoing energy loss.

\subsubsection{Jet Shapes}\label{Sec:jetshape}

Another observable that is related to the structure of the jet is the called the jet shape.  This observable is constructed with the idea that the high energy jets we are interested in are roughly conical.  First a jet finding algorithm is run to determine the axis of the jet, and then the sum of the transverse momentum of the tracks in concentric rings about the jet axis are summed together (and divided by the total transverse jet momentum).   The differential jet shape observable $\rho(r)$ is thus the radial distribution of the transverse momentum: 
\begin{equation}
 \rho(r) = \frac{1}{\delta r} \frac{1}{N_{\mathrm jet}} \sum_{\mathrm jets} \frac{ \sum_{{\mathrm tracks} \epsilon [r_a, r_b)} p_{T}^{\mathrm track}}{p_{T}^{\mathrm jet}}
\end{equation}
\noindent where the jet cone is divided rings of width $\delta r$ which have an inner radius $r_a$ and an outer radius $r_b$.

\begin{figure*}
\begin{center}
\rotatebox{0}{\resizebox{\textwidth}{!}{
	\includegraphics{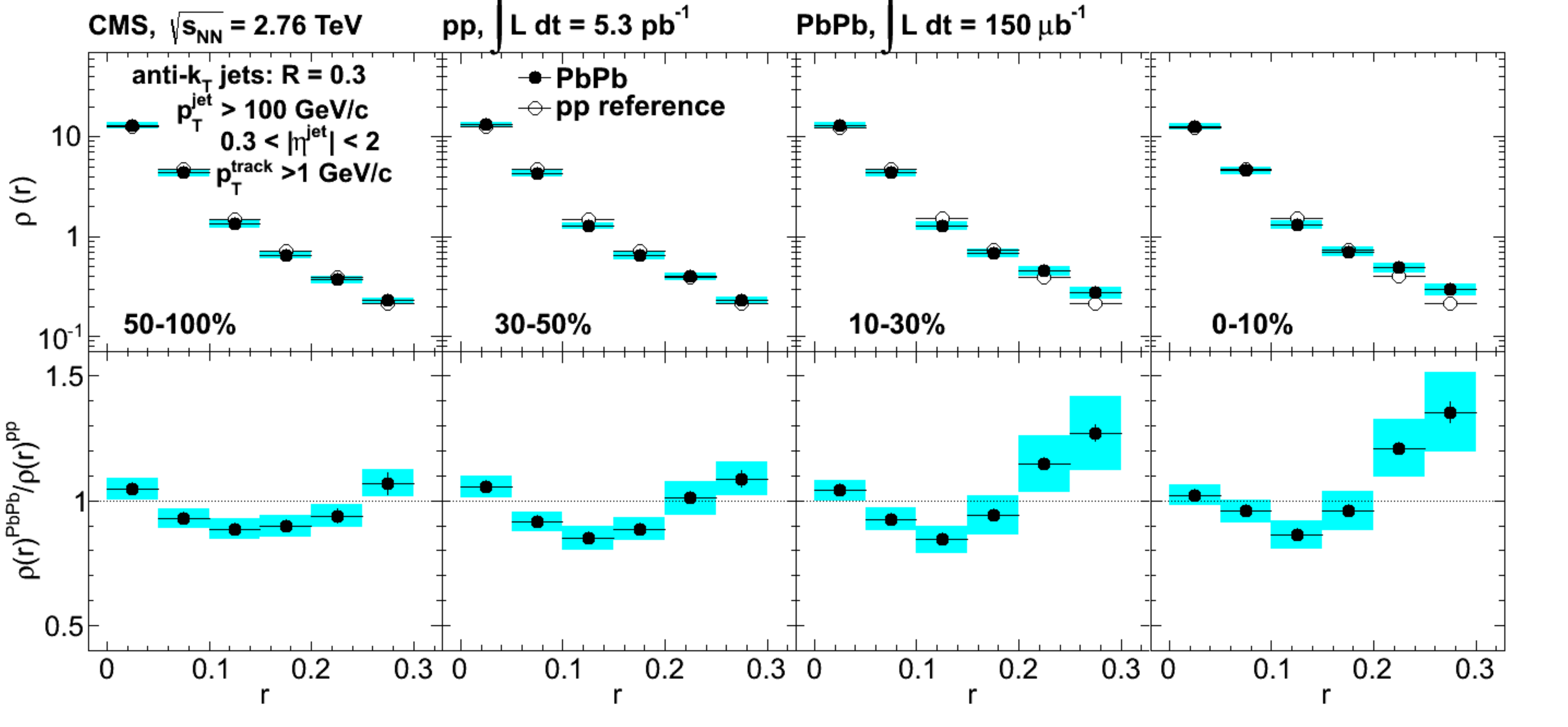}
}}\caption{
Figure from CMS~\cite{Chatrchyan:1605718}.   Differential jet shapes in \Pb and \pp collisions for four \Pb centralities. Each spectrum is normalized so that its integral is unity.  This shows that there are more particles in jets in central collisions and these modifications are primarily at large angles relative to the jet axis, as expected from partonic energy loss.
 }\label{Fig:CMSJetShapes}
\end{center}
\end{figure*}

The differential and integrated jet shape measurements measured by CMS are shown in \Fref{Fig:CMSJetShapes}.  For this CMS study, inclusive jets with $p_{T} > 100$ GeV/c, resolution parameter $R$ =  0.3 and constituent tracks with $p_{T}>1$ GeV/c were used.  The effect of the background on the signal jets was removed through the iterative subtraction technique described in \Sref{Sec:ExpMethods}.  The associated tracks were not explicitly required to be the constituent tracks, however given that the momentum selection criteria is the same and the conical nature of jets at this energy, they will essentially be the same.  The effect of the background on the distribution of the associated particles was removed via an $\eta$ reflection method, where the analysis was repeated for an $R$ = 0.3 cone with the opposite sign $\eta$ but same $\phi$.  This preserves the flow effects in a model independent way in the determination of the background.  
The differential jet shapes in the most central \Pb collisions are broadened in comparison to measurements done in \pp collisions at the same center of mass energy ~\cite{Chatrchyan:1605718}.  As shown in other measurements, the effect is centrality dependent.
These measurements demonstrate that there is an enhancement in the modification with increasing angle from the jet axis, indicating a broadening of the jet profile and a depletion near r~$\approx$~0.2.

\subsubsection{Particle composition}\label{Sec:PID}

\begin{figure}
\begin{center}
\rotatebox{0}{\resizebox{8.0cm}{!}{
	\includegraphics{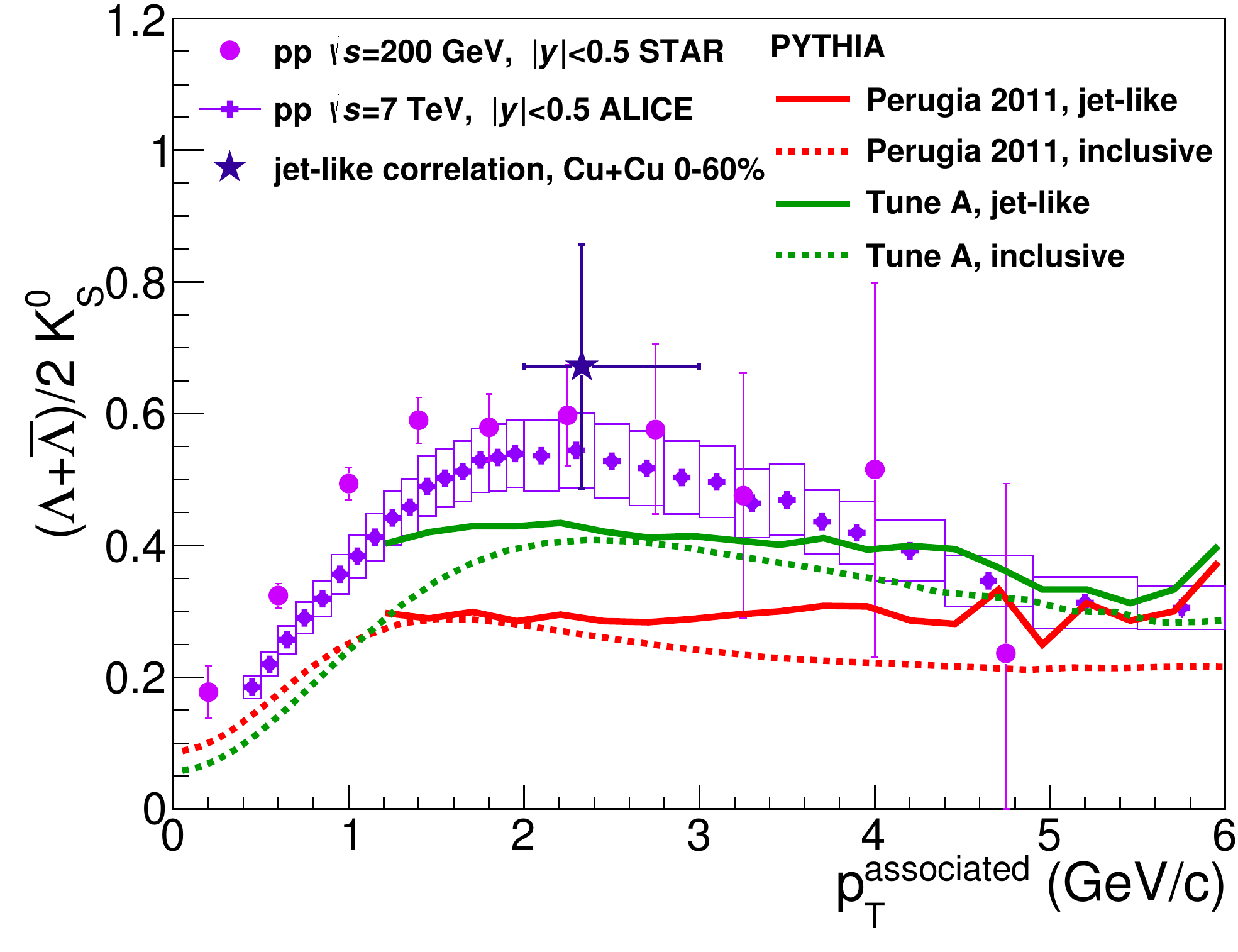}
}}
\caption{
Figure from STAR~\cite{Abelev:2016dqe}.
$\Lambda$/$K^0_S$ ratio measured in jet-like correlations in 0-60\% \Cu collisions at \sNN = 200 GeV for $3$~$<$ $p_T^{\mathrm{trigger}}$ $<$ $6$~GeV/$c$ and 2~$< p_T^{\mathrm{associated}} < $ 3~GeV/$c$ along with this ratio obtained from inclusive $p_T$ spectra in \pp collisions.  Data are compared to calculations from PYTHIA~\cite{Sjostrand:2006za} using the Perugia 2011 tunes~\cite{Skands:2010ak} and Tune A~\cite{Field:2005sa}.  This shows that, within the large uncertainties, there is no indication that the particle composition of jets is modified in \AplusA collisions, where $\Lambda$/$K^0_S$ reaches a maximum of 1.6~\cite{Agakishiev:2011ar}.
}
\label{Fig:bmratio}
\end{center}
\end{figure}

Theory predicts higher production of baryons and strange particles in jets fragmenting in the medium relative to jets fragmenting in the vacuum~\cite{Sapeta:2007ad}.  The only published study searching for modified particle composition in jets in heavy ion collisions is the $\Lambda$/$K^0_S$ ratio in the \ns \jlc of \dhcs in \Cu collisions at \sNN = 200 GeV by STAR~\cite{Abelev:2016dqe} shown in \Fref{Fig:bmratio}.  This measurement indicated that particle ratios in the \ns \jlc are comparable to the inclusive particle ratios in \pp collisions.  At high momenta, the inclusive particle ratios in \pp collisions are expected to be dominated by jet fragmentation and therefore are a good proxy for direct observation of the particle ratios in reconstructed jets.  PYTHIA studies show that the inclusive particle ratios in \pp collisions are approximately the same as the particle ratios in \dhcs with similar kinematic cuts; differences are well below the uncertainties on the experimental measurements.  The consistency between the $\Lambda$/$K^0_S$ ratio in the \jlc in \Cu collisions and the inclusive ratio in \pp collisions is therefore interpreted as evidence that the particle ratios in jets are the same in \AplusA collisions and \pp collisions, that at least the particle ratios are not modified.  In contrast, the inclusive $\Lambda$/$K^0_S$ reaches a maximum near 1.6~\cite{Agakishiev:2011ar}, a few times that in \pp collisions.  Preliminary measurements from both the STAR \dhcs~\cite{phdthesissuarez} and ALICE collaborations from both \dhcs~\cite{Veldhoen:2012ge} and reconstructed jets~\cite{Kucera:2015fni,Zimmermann:2015npa} support this conclusion.  However, experimental uncertainties are large and for studies in \dhcs, results are not available for the \as and the \ns is known to be surface biased.

\subsubsection{LeSub}\label{Sec:LeSub}
One of the new observables constructed in order to attempt to create well defined QCD observables is LeSub, defined as:
\begin{equation}
\textrm{LeSub} = {p}_{T}^{\textrm{lead,track}}-{p}_{T}^{\textrm{sublead,track}} 
\end{equation}
LeSub characterizes the hardest splitting, so it should be insensitive to background, however, it is not colinear safe and therefore cannot be calculated reliably in pQCD.  It agrees well with PYTHIA simulations of \pp collisions and is relatively insensitive to the PYTHIA tune~\cite{Cunqueiro:2015dmx}, which is not surprising as the hardest splittings in PYTHIA do not depend on the tune.
  LeSub calculated in PYTHIA agrees well with the data from \Pb~collisions for $R$ = 0.2 charged jets.  This indicates that the hardest splittings are likely unaffected by the medium.  Modifications may depend on the jet momentum, as the ALICE results are for relatively low momentum jets at the LHC.  The ALICE measurement is also for relatively small jets, which preferentially selects more collimated fragmentation patterns, but it indicates that observables that depend on the first splittings are insensitive to the medium.

\subsubsection{Jet Mass}\label{Sec:JetMass}

In a hard scattering the partons are produced off-shell, and the amount they are off-shell is the virtuality~\cite{Majumder:2014gda}.  When a jet showers in vacuum, at each splitting the virtuality is reduced and momentum is produced transverse to the original scattered parton's direction, until the partons are on-shell and thus hadronize.  For a vacuum jet, if the four vectors of all of the daughters from the original parton are combined, the mass calculated from the combination of the daughters would be precisely equal to the virtuality.  The virtuality of hard scattered parton is important as it is directly related to how broad the jet itself is, as it is directly related to how much momentum transverse to the jet axis the daughters can have.  

The mass of a jet might serve as a way to better characterize the state of the initial parton.  It is important to construct observables where the only difference between \pp~collisions compared to heavy ion collisions is due to the effects of jet quenching, and not the result of biases in the jet selection.  Jet mass may make a much closer comparison between heavy ion and \pp~observables by selecting more similar populations of parent partons than could be achieved by selecting differentially in transverse momentum alone.  Secondly, the measured jet mass itself could be affected by in-medium interactions as  the virtuality of the jet can increase for a given splitting due to the medium interaction, unlike in the vacuum case.

\begin{figure*}
\begin{center}
\rotatebox{0}{\resizebox{\textwidth}{!}{
	\includegraphics{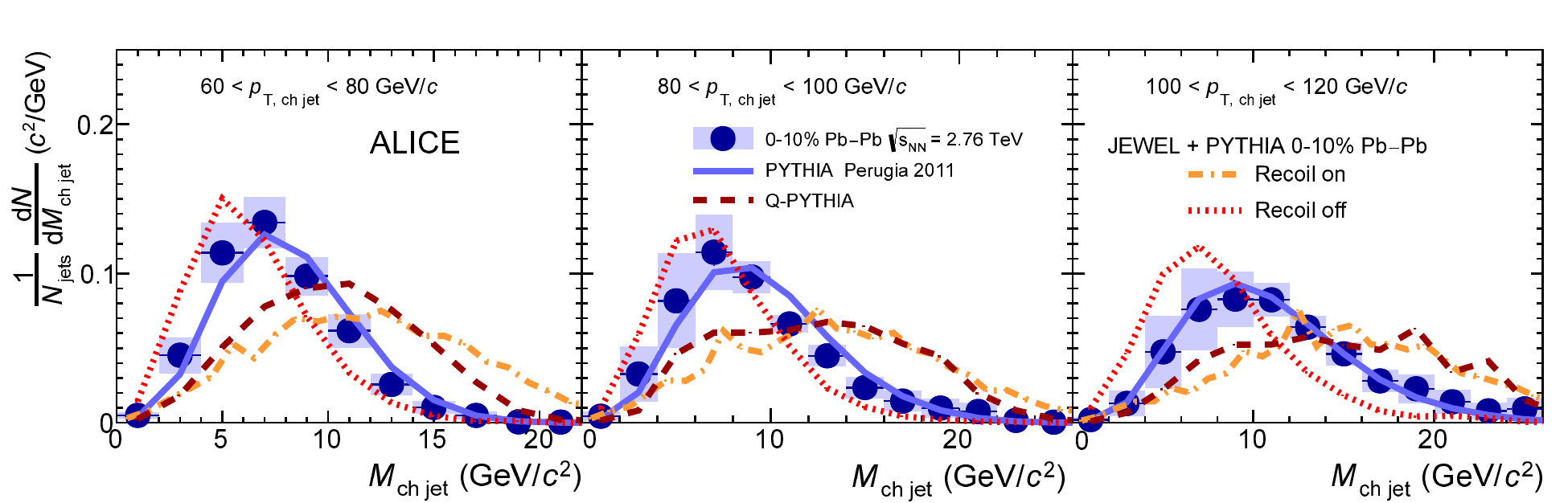}
}}\caption{
Figure from ALICE~\cite{Acharya:2017goa}.  Fully-corrected jet mass distribution for anti-\kT jets with R=0.4 in the 10\% most central \Pb collisions compared to PYTHIA~\cite{Sjostrand:2006za} with the Perugia 2011 tune~\cite{Skands:2010ak} and predictions from the jet quenching event generators JEWEL~\cite{Zapp:2013zya} and Q-PYTHIA~\cite{Armesto:2009fj}.  No difference is observed between PYTHIA and the data.  This shows that there is no modification of the jet mass within uncertainties.
 }\label{Fig:JetMass}
\end{center}
\end{figure*}

 \Fref{Fig:JetMass} shows the ALICE~\cite{Acharya:2017goa} jet mass measurement of charged jets for most central collisions.  No difference is observed between PYTHIA Perugia 2011 tune \cite{Skands:2010ak} and data from \Pb collisions in all jet $p_T$ bins indicating no apparent modification within uncertainties.  In addition to PYTHIA, these distributions were compared to three different quenching models, JEWEL \cite{Zapp:2013zya} with recoil on, JEWEL with recoil off, and Q-PYTHIA \cite{Armesto:2009fj}.  Both Q-PYTHIA and JEWEL with the recoil on produced jets with a larger mass distribution than in the data, whereas JEWEL with the recoil off gives a slightly lower value than the data. This implies that jet mass as a distribution in these energy and momentum ranges is rather insensitive to medium effects, as JEWEL and Q-PYTHIA both incorporate medium effects whereas PYTHIA describes vacuum jets. The agreement between PYTHIA and data could also indicate that the jets selected in this analysis were biased towards those that fragmented in a vacuum-like manner.  More differential measurements of jet mass are needed to determine the usefulness of jet mass variable.

\subsubsection{Dispersion}\label{Sec:Dispersion}
\begin{figure}
\begin{center}
\rotatebox{0}{\resizebox{8cm}{!}{
     \resizebox{8cm}{!}{\includegraphics{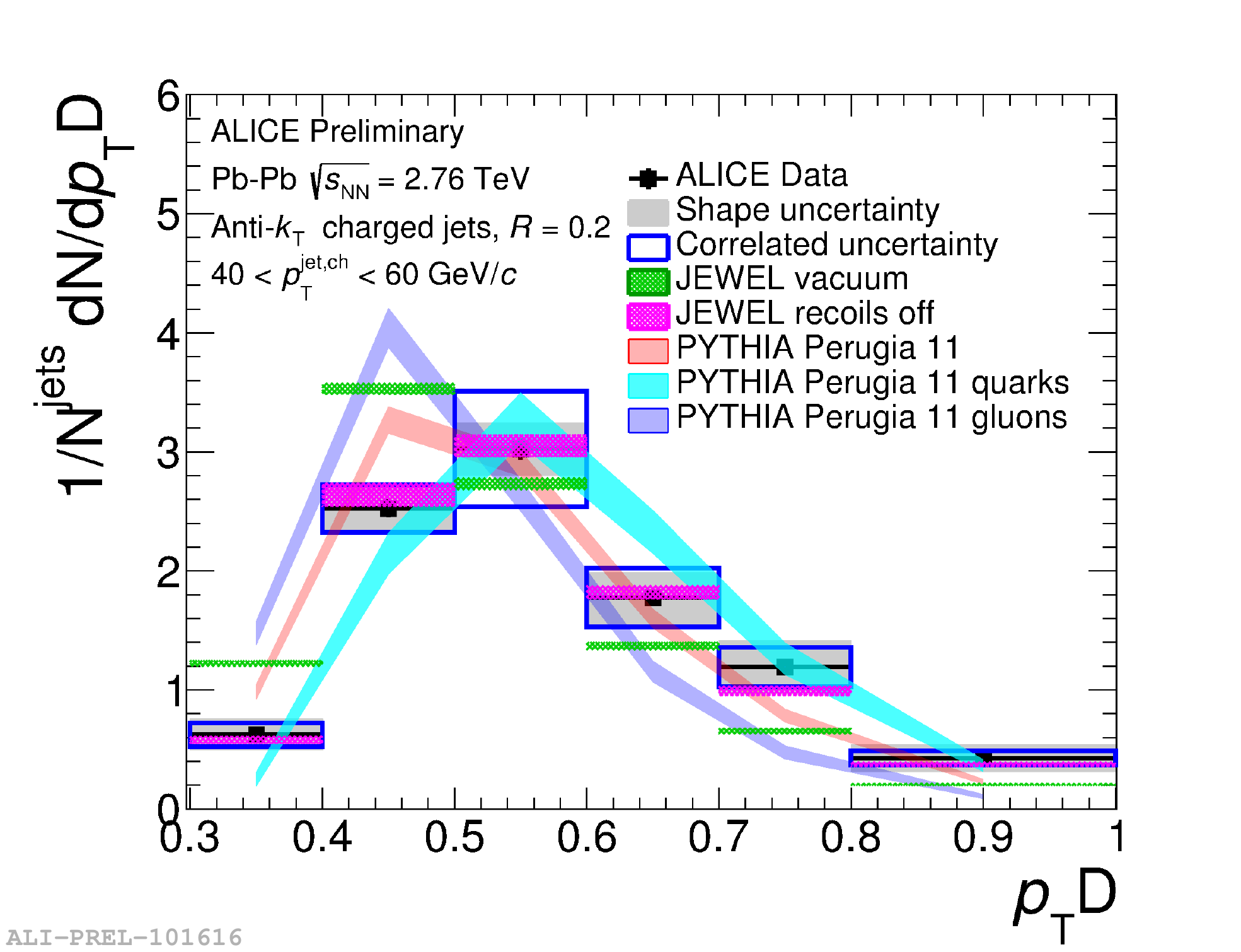}}
        }}\caption{        Figure from ALICE~\cite{Cunqueiro:2015dmx}.
       Unfolded $\rm p_T^D$ shape distribution in \Pb collisions for $R$=0.2 charged jets with momenta between 40 and 60 \GeV compared to PYTHIA simulations, to JEWEL calculations, and to q/g PYTHIA templates.  This shows that the dispersion is larger in \Pb collisions than in \pp collisions.  This may indicate either modifications or a quark bias.
        }
\label{Fig:ALICEpTD}
\end{center}
\end{figure}

 Since quark jets have harder fragmentation functions, they are more likely to produce jets with hard constituents that carry a significant fraction of the jet energy.  This can be studied with $\rm p_T^D=\sqrt{\Sigma_i p_{T,i}^2}/\Sigma_ip_{T,i}$. This observable was initially developed in order to distinguish between quark and gluon jets with quark jets yielding a larger mean $\rm p_T^D$~\cite{CMS:2013kfa}.
The ALICE experiment has measured $\rm p_T^D$ in \Pb~collisions, shown in Figure \ref{Fig:ALICEpTD}.  The data from \Pb~collisions for $R$ = 0.2 charged jets with transverse momentum between 40 and 60 GeV is compared to data from PYTHIA with the Perugia 11 tune.   In \Pb~collisions, the mean $\rm p_T^D$ was found to be larger compared to the PYTHIA reference, which had been validated by comparisons with \pp~data.  This may indicate either a selection bias towards quark jets or harder fragmenting jets.

\subsubsection{Girth}\label{Sec:Girth}

\begin{figure}
\begin{center}
\rotatebox{0}{\resizebox{8cm}{!}{
     \resizebox{8cm}{!}{\includegraphics{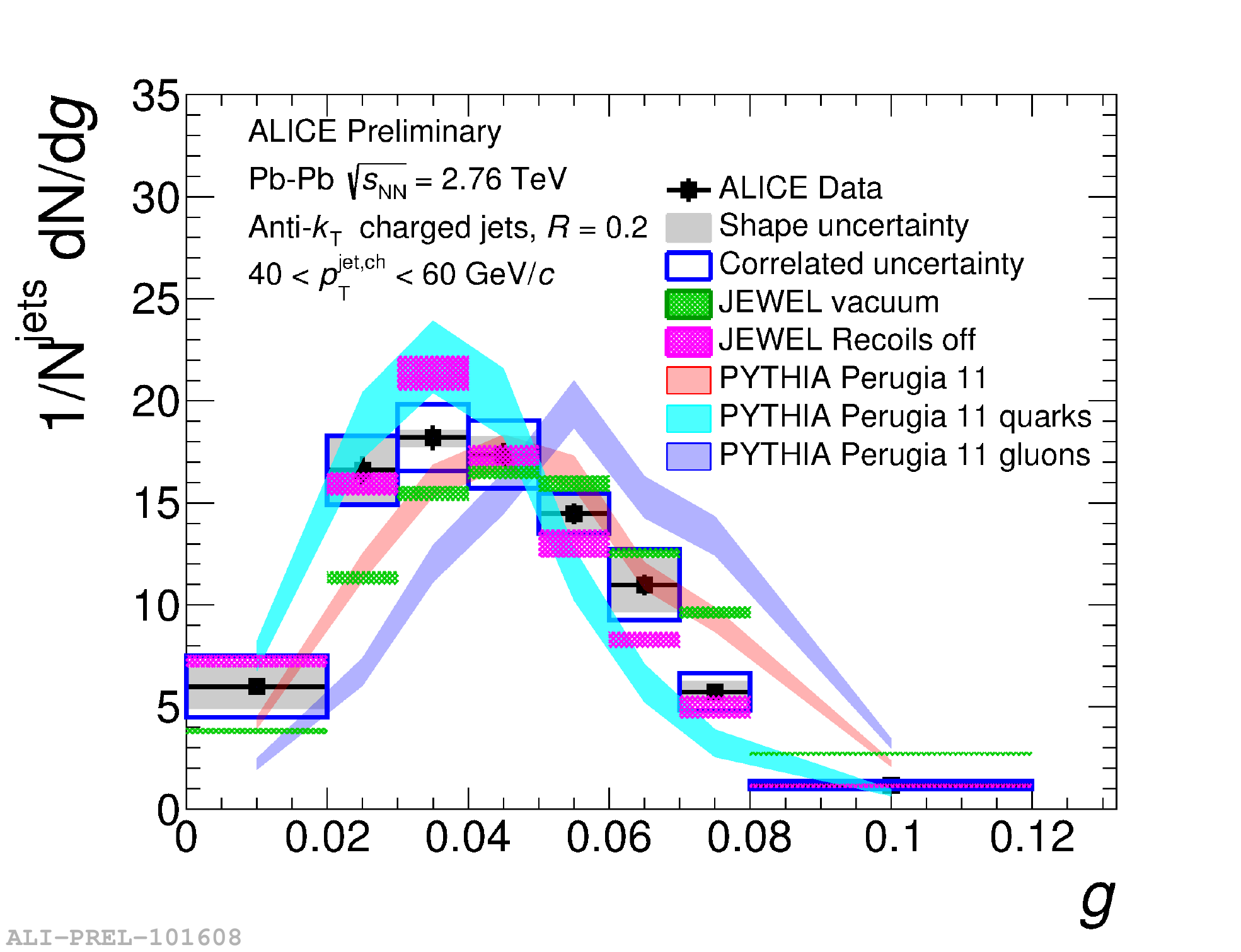}}
        }}\caption{        Figure from ALICE~\cite{Cunqueiro:2015dmx}.
        The girth $g$ for $R$=0.2 charged jets in \Pb collisions with jet $p_{T}^{ch}$ between 40 and 60 \GeV compared to a PYTHIA simulations, to JEWEL calculations, and to q/g PYTHIA templates.  This shows that jets are somewhat more collimated in \Pb collisions than in \pp collisions.  This may indicate a quark bias in surviving jets in \Pb collisions.}
\label{Fig:Girth1}
\end{center}
\end{figure}

The jet girth is another new observable describing the shape of a jet.  
The jet girth, $g$, is the \pT~weighted width of the jet
\begin{equation}
 g = \sum_{i}\cfrac{{p}_{T}^{i}}{{p}_{T}^{jet}}|{r}_{i}|,
\end{equation}
where $r_{i}$ is the angular distance between particle $i$ and the jet axis. If jets are broadened by the medium, we would expect that $g$ would be increased, and the converse would be that if jets were collimated than $g$ would be reduced.  
While the distributions overlap, the gluon jets are broader and have a higher average $g$ than quark jets.  The ALICE experiment has shown that distributions of $g$ in \pp collisions agree well with PYTHIA distributions, indicating that it is a reasonable probe and that PYTHIA can be used as a reference.  In \Pb~collisions, the ALICE experiment found that $g$ is slightly shifted towards smaller values compared to the PYTHIA reference for $R$ = 0.2 charged jets~\cite{Cunqueiro:2015dmx}, although the significance of this shift is unclear.  This indicates that the core may appear to be more collimated in \Pb~collisions than \pp~collisions.  Measurements are compared to JEWEL and PYTHIA calculations in \Fref{Fig:Girth1}.  JEWEL includes partonic energy loss and predicts little modification of the girth in heavy ion collisions.  PYTHIA calculations include inclusive jets, quark jets, and gluon jets.  The data are closest to PYTHIA predictions for quark jets.  This may be due to bias towards quarks in surviving jets in \Pb collisions.

One of the unanswered questions regarding jets in heavy ion collisions is whether jets start to fragment while they are in the medium, or whether they simply lose energy to the medium and then fragment similar to fragmentation in vacuum after reaching the surface.  If the latter is true, jet quenching would be described as a shift in parton \pT followed by vacuum fragmentation, which would mean that jets shapes in \Pb collisions would be consistent with jet shapes in \pp~collisions.  If $g$ is shifted, this would favor fragmentation in the medium and if it is not, it would favor vacuum fragmentation.  These observations are qualitatively consistent with the measurements of $\rm p_T^D$ discussed in \Sref{Sec:jetshape} and the jet shape discussed in \Sref{Sec:jetshape}.

\subsubsection{Grooming}\label{Sec:Grooming}

\begin{figure*}
\begin{center}
\rotatebox{0}{\resizebox{\textwidth}{!}{
	\includegraphics{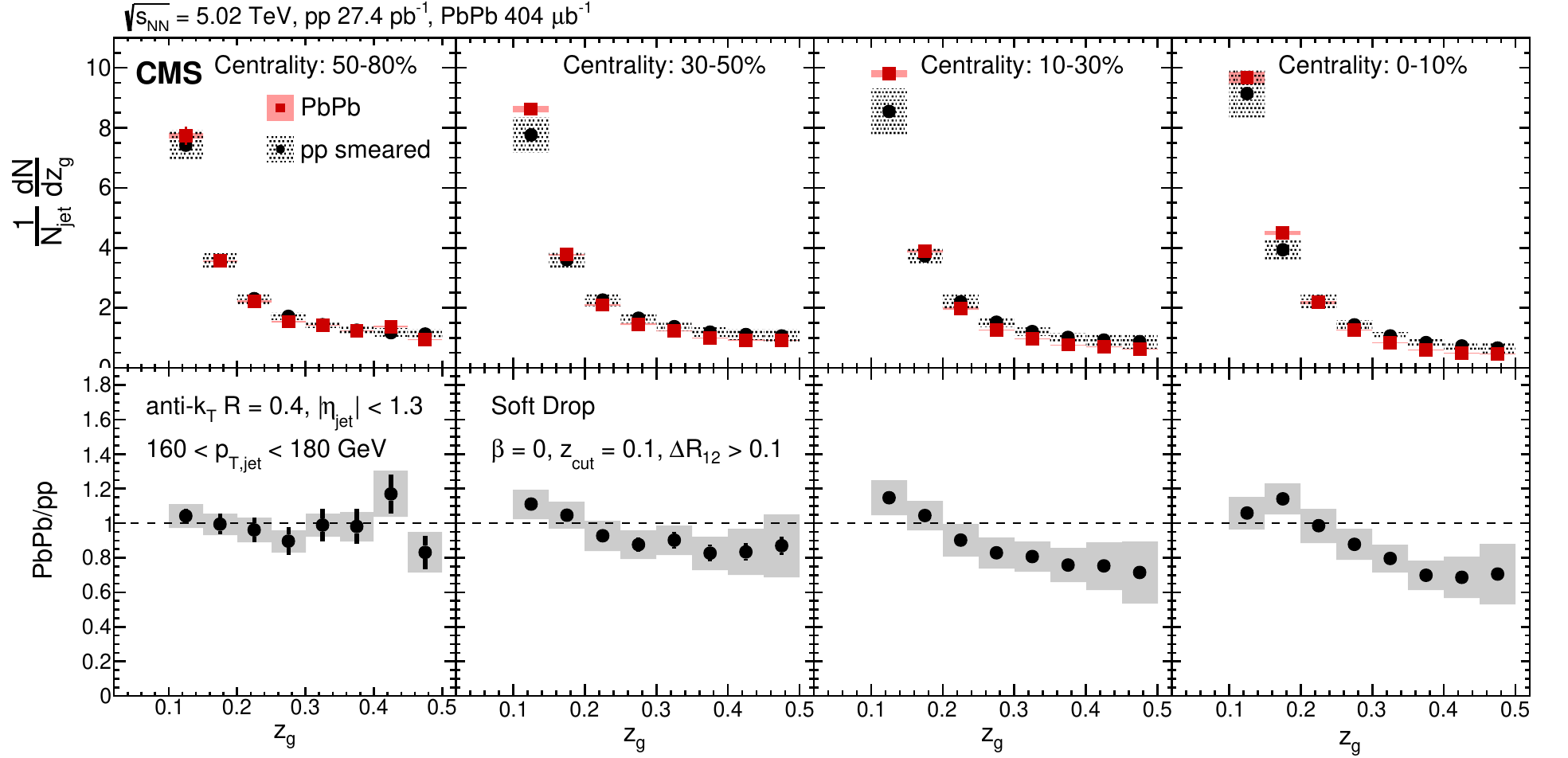}
}}\caption{
Figure from CMS~\cite{CMS:Splitting}.  Ratio of the splitting function $z_g = p_{T2}/( p_{T1} + p_{T2})$ in \Pb and \pp collisions with the jet energy resolution smeared to match that in \Pb for various centrality selections and 160 $<p_T^{jet}<$ 180 GeV. This shows that the splitting function is modified in central \Pb collisions compared to \pp collisions, which may indicate either a difference in the structure of jets in the two systems or an impact of the background.
 }\label{Fig:CMSZg}
\end{center}
\end{figure*}

Jet grooming algorithms~\cite{pruning1,pruning2,pruning3,pruning4} attempt to remove soft radiation from the leading partonic components of the jet, isolating the larger scale structure.  The motivation for algorithms such as jet grooming was to develop observables which can be calculated with perturbative QCD, and which are relatively insensitive to the details of the soft background.  This allows us to determine whether the medium affects the jet formation process from the hard process through hadronization, or whether the parton loses energy to the medium with fragmentation only affected at much later stages.  It is important to realize that the answers to these questions will depend on the jet energy and momentum, so there will not be a single definitive answer.  Jet grooming allows separation of effects of the length scale from effects of the hardness of the interaction.
Essentially this will allow us to see whether we are scattering off of point-like particles in the medium or scattering off of something with structure.  However, to properly apply this class of algorithms to the data, a precision detector is needed.

The jet grooming algorithm takes the constituents of a jet, and recursively declusters the jet's branching history and discards the resulting subjets until the transverse momenta, ${p}_{T,1}, {p}_{T,2}$, of the current pair fulfills the soft drop condition~\cite{Larkoski:2014wba}:
\begin{equation}
\cfrac{\textrm{min}({p}_{T,1},{p}_{T,2})}{{p}_{T,1}+{p}_{T,2}} > {z}_{\textrm{cut}} {\theta}^{\beta}
\end{equation}
\noindent where $\theta$ is an additional measure of the relative angular distance between the two sub-jets and ${z}_{\textrm{cut}}$ and ${\theta}^{\beta}$ are parameters which can select how strict the soft drop condition is.  For the heavy-ion analyses conducted so far, $\beta$ has been set to zero and $z_{\textrm cut}$ has been set to 0.1.

A measurement of the first splitting of a parton in heavy ion collisions is performed by the CMS collaboration in \Pb collisions at \sNN = 5 TeV.  The splitting function is defined as  $z_g = p_{T2}/( p_{T1} + p_{T2})$ with $p_{T2}$ indicating the transverse momentum of the least energetic subjet and $p_{T1}$ the transverse momentum of the most energetic subjet, applied to those jets that passed the soft drop condition outlined above.  \Fref{Fig:CMSZg} shows the ratio of $z_g$ in \Pb to that in \pp from CMS for several centrality intervals for jets within the transverse momentum range of 160--180 GeV/c~\cite{CMS:Splitting}. While the measured $z_g$ distribution in peripheral \Pb collisions is in agreement with the expected \pp measurement within uncertainties, a difference becomes apparent in the more central collisions. This observation indicates that the splitting into two branches becomes increasingly more unbalanced for more central collisions for the jets within the transverse momentum range of 160--180 GeV/c. A similar preliminary measurement by STAR observes no modification in $z_g$~\cite{KauderQM2017}.  The apparent modifications seen by CMS were proposed to be due to a restriction to subjets with a minimum separation between the two hardest subjets $R_{12}>0.1$~\cite{GuilhermeQM2017}.  This indicates that there may be modifications of $z_g$ limited to certain classes of jets but not observed globally.  This dependence of modifications on jets may be a result of interactions with the medium~\cite{Milhano:2017nzm}.  While grooming and measurements of the jet substructure are promising, we emphasize the need for a greater understanding of the impact of the large combinatorial background and the bias of kinematic cuts on $z_g$.

\subsubsection{Subjettiness}\label{Sec:Nsubjettiness}
The observable $\tau_{N}$ is a measure of how many hard cores there are in a jet.  This was initially developed to tag jets from Higgs decays in high energy \pp collisions.   A jet from a single parton usually has one hard core, but a hard splitting or a bremsstrahlung gluon would lead to an additional hard core within the jet.  An increase in the fraction of jets with two hard cores could therefore be evidence of gluon bremsstrahlung.

The jet is reclustered into $N$ subjets, and the following calculation is performed over each track in the jet:
\begin{equation}
{\tau}_{N} = \cfrac{\sum_{i=1}^{M}(p_{T}^{i}~min(\Delta {R}_{1,i},\Delta {R}_{2,i},....\Delta {R}_{N,i})} {R_0 \sum_{i=1}^{N} p_{T}^{i}}
\end{equation}
where $\Delta {R}_{N,i}$ is the distance in $\eta-\phi$ between the $i$th track and the axis of the Nth subjet and the original jet has resolution parameter $R_0$.   In the case that all particles are aligned exactly with one of the subjets' axes, $\tau_{N}$ will equal zero.  In the case where there are more than $N$ hard cores, a substantial fraction of tracks will be far from the nearest subjet axis, however, all tracks must have $min(\Delta {R}_{1,i},\Delta {R}_{2,i},....\Delta {R}_{N,i}) \leq R_0$ because they are contained within the original jet.  The maximum value of ${\tau}_{N}$ is therefore one, the case when all jet constituents are at the maximum distance from the nearest subjet axis.

\begin{figure}
\begin{center}
\rotatebox{0}{\resizebox{8.0cm}{!}{
	\includegraphics{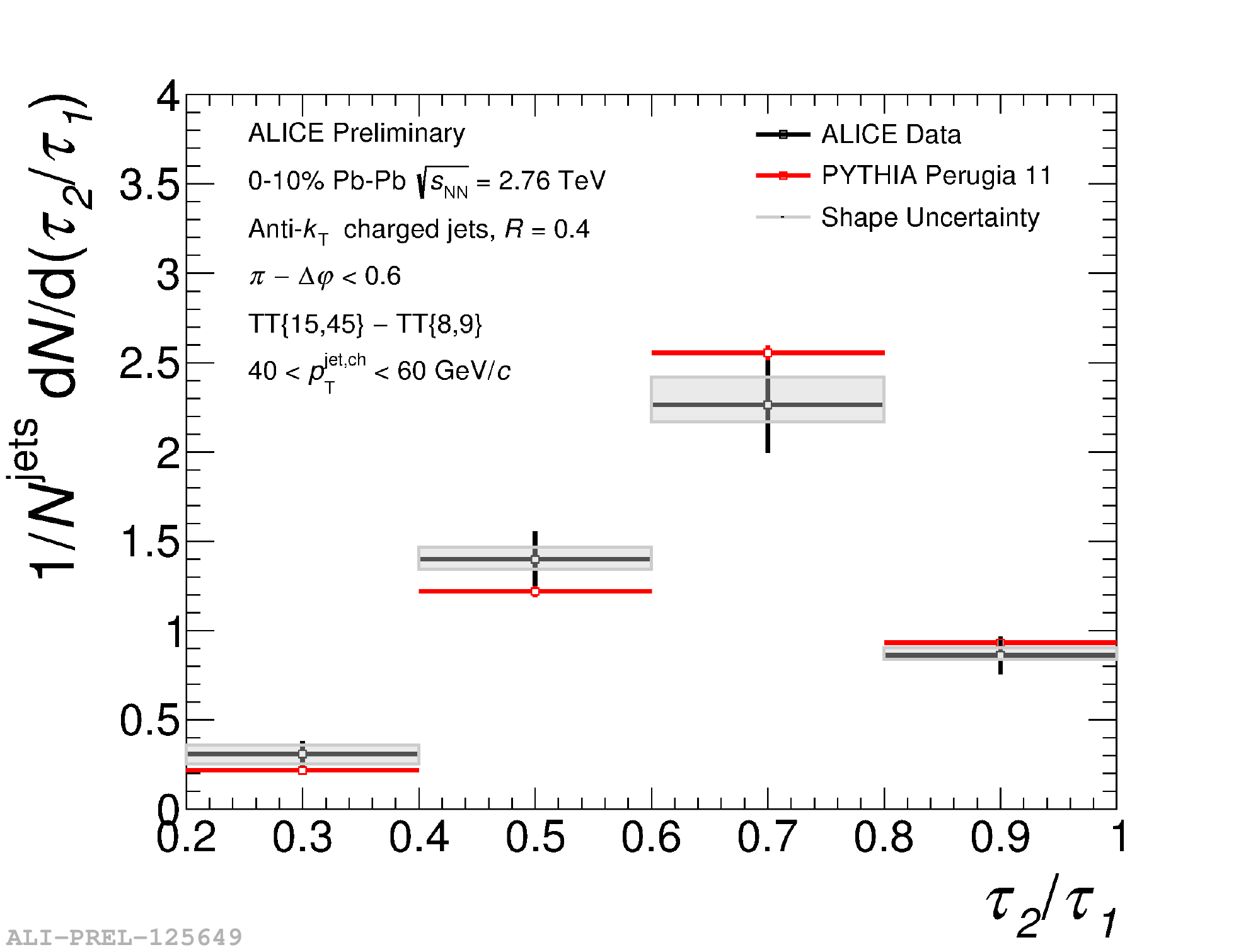}
}}
\caption{
Figure from~\cite{Zardoshti:2017yiy}.
$\it{\tau_{2}}/\it{\tau_{1}}$ fully corrected recoil R=0.4 jet shape in 0-10\% \Pb collisions at 40 $\leq p_{t,jet}^{ch} < 60$ \GeV.  This shows that, at least for this kinematic selection, the subjettiness is not modified.  The trigger tracks are 8--9 \GeV for the background dominated region and 15--45 \GeV for the signal dominated region.
}
\label{Fig:Subjettiness}
\end{center}
\end{figure}

Jets that have a low value of ${\tau}_{N}$ are therefore more likely to have $N$ or fewer well defined cores in their sub-structure, whereas jets with a high value are more likely to contain at least $N$+1 cores.  A shift in the distribution of ${\tau}_{N}$ in a jet population towards lower values can indicate fewer subjets while a shift to higher ${\tau}_{N}$ can indicate more subjets.  The observable  ${\tau}_{2}/{\tau}_{1}$ was constructed by the ALICE experiment~\cite{Zardoshti:2017yiy}.  Similar to the approach in~\cite{Adam:2015doa,Adamczyk:2017yhe}, background was subtracted using the coincidence between a soft trigger hadron, which should have only a weak correlation with jet production, and a high momentum trigger hadron, and can be seen in Figure \ref{Fig:Subjettiness}.  
A jet where this ratio is close to zero most likely has two hard cores.  This observable is relatively insensitive to the fluctuations in the background, as it would have to carry a significant fraction of the jet momentum to be modified.  The ALICE result shows that the structure of the jets was unmodified for $R$ = 0.4 charged jets with 40 $\leq p_{t,Jet}^{ch} < 60$ \GeV compared to PYTHIA calculations.  This implies that medium interactions do not lead to extra cores within the jet, at least for selection of jets in this measurement.  As for many jet observables, this observable may be difficult to interpret for low momentum jets in a heavy ion environment.

\subsubsection{Summary of experimental evidence for medium modification of jets}
The broadening and softening of jets due to interactions with the medium is demonstrated clearly by several mature observables which measure the average properties of jets.  This includes fragmentation functions measured with both jets and bosons, widths of \dhcs, \jhcs, and measurements of the jet shape.  On average, no change in the particle composition of jets in heavy ion collisions as compared to \pp collisions is observed.  There are some indications from \dhcs that quark and gluon jets do not interact with the medium in the same way.  These observables generally preferentially select quark jets over gluon jets, even in \pp collisions.  Some of the observables have a strong survivor bias due to the kinematic cuts that are applied in order to reduce the combinatorial background.

As our understanding of partonic energy loss has improved, the community has sought more differential observables.  This is motivated in part by an increased understanding of the importance of fluctuations -- while the average properties of jets are smooth, individual jets are lumpy, and by a desire construct well defined QCD observables.  These new observables give us access to different properties of jets, such as allowing distinction between quark and gluon jets, and therefore may be more sensitive to the properties of the medium.  Since the exploration of these observables is in its early stages, it is unclear whether we fully understand the impact of the background or kinematic cuts applied to the analyses.  It is therefore unclear in practice how much additional information these observables can provide about the medium, without applying the observables to Monte Carlo events with different jet quenching models.  We encourage cautious optimism and more detailed studies of these observables.

For future studies to maximize our understanding of the medium by the Jetscape collaboration using a Bayesian analysis, we propose first to produce comparisons between \dhcs, \jhcs, and $\gamma$-hadron correlations to insure that the models have properly accounted for the path length dependence, initial state effects and the basics of fragmentation and hadronization.  We do not list \RAA here as it is likely that this observable will be used to tune some aspects of the model, as it has been used in the past.  For the most promising jet quenching models, we would propose that these studies would be followed by comparisons of observables that depend more heavily on the details of the fragmentation, but are still based on the average distribution such as jet shapes, fragmentation functions, and particle composition.  Finally, it would be useful to see the comparison of ${z}_{g}$ to models.  We urge that initial investigations of the latter happen early so that the background effect can be quantified.

We note that the same analysis techniques and selection criteria must be used for analyses of the experiment and of the models in order for the comparisons to be valid.  This is particularly true for studies using reconstructed jets where experimental criteria to remove the effects of the background can bias the sample of jets used in construction of the observables.  We omit \Aj from consideration because nearly any reasonable model gives a reasonable value, thus it is not particularly differential.  We also omit heavy flavor jets because current data do not give much insight into modifications of fragmentation, and it is not clear whether it will be possible experimentally to measure jets with a low enough \pT that the mass difference between heavy and light quarks is relevant.  Inclusion of new observables into these studies may increase the precision with which medium properties can be constrained, but it is critical to replicate the exact analysis techniques.  

In order to compare experimental data, or to compare experimental data with theory, not only is it necessary for the analyses to be conducted the same way as it is stated above, but they should be on the same footing.  Thus comparing unfolded results to uncorrected results it not useful.  In general, we urge extreme caution in interpreting uncorrected results, especially for observables created with reconstructed jets.   Since it is unclear how much the process of unfolding may bias the results, an important check would be to compare the raw results with the folded theory.  However, this requires complete documentation of the raw results and the response matrix on the experimental side, and requires a complete treatment of the initial state, background, and hadronization on the theory side.  This comparison, which we could think of as something like a closure test, would still require that the same jet finding algorithms with the same kinematic elections are applied to the model.

 \subsection{Influence of the jet on the medium}\label{Sec:ResultsMedium}

The preceding sections have demonstrated that hard partons lose energy to the medium, most likely through gluon bremsstrahlung and collisional energy loss.  Often an emitted gluon will remain correlated with the parent parton so that the fragments of both partons are spatially correlated over relatively short ranges ($R = \sqrt{\Delta\phi^2+\Delta\eta^2}\lesssim 0.5$).  Hadrons produced from the gluon may fall inside or outside the jet cone of the parent parton, depending on the jet resolution parameter.  Whether or not this energy is then reconstructed experimentally as part of the jet depends on the resolution parameter and the reconstruction algorithm.   For sufficiently large resolution parameters, the ``lost'' energy will still fall within the jet cone, so that the total energy clustered into the jet would remain the same.  ``Jet quenching'' is then manifest as a softening and broadening of the structure of the jet.  The evidence for these effects was discussed in the previous section.

If, however, a parton loses energy and that energy interacts with or becomes equilibrated in the medium, it may no longer have short range spatial correlations with the parent parton.  This energy would then be distributed at distances far from the jet cone.  Alternately, the energy may have very different spatial correlations with the parent parton so that it no longer looks like a jet formed in a vacuum, and a jet finding algorithm may no longer group that energy with the jet that contains most of the energy of its parent parton.  Evidence for these effects is difficult to find, both because of the large and fluctuating background contribution from the underlying event, and because it is unclear how this energy would be different from the underlying event.  We discuss both the existing evidence that there may be some energy which reaches equilibrium with the medium, and the ridge and the Mach cone, which are now understood to be features of the medium rather than indications of interactions of hard partons with the medium.  We also discuss searches for direct evidence of Moli\`{e}re scattering off of partons in the medium.

\subsubsection{Evidence for out-of-cone radiation}

\begin{figure*}
\begin{center}

\rotatebox{0}{\resizebox{8cm}{!}{
        \includegraphics{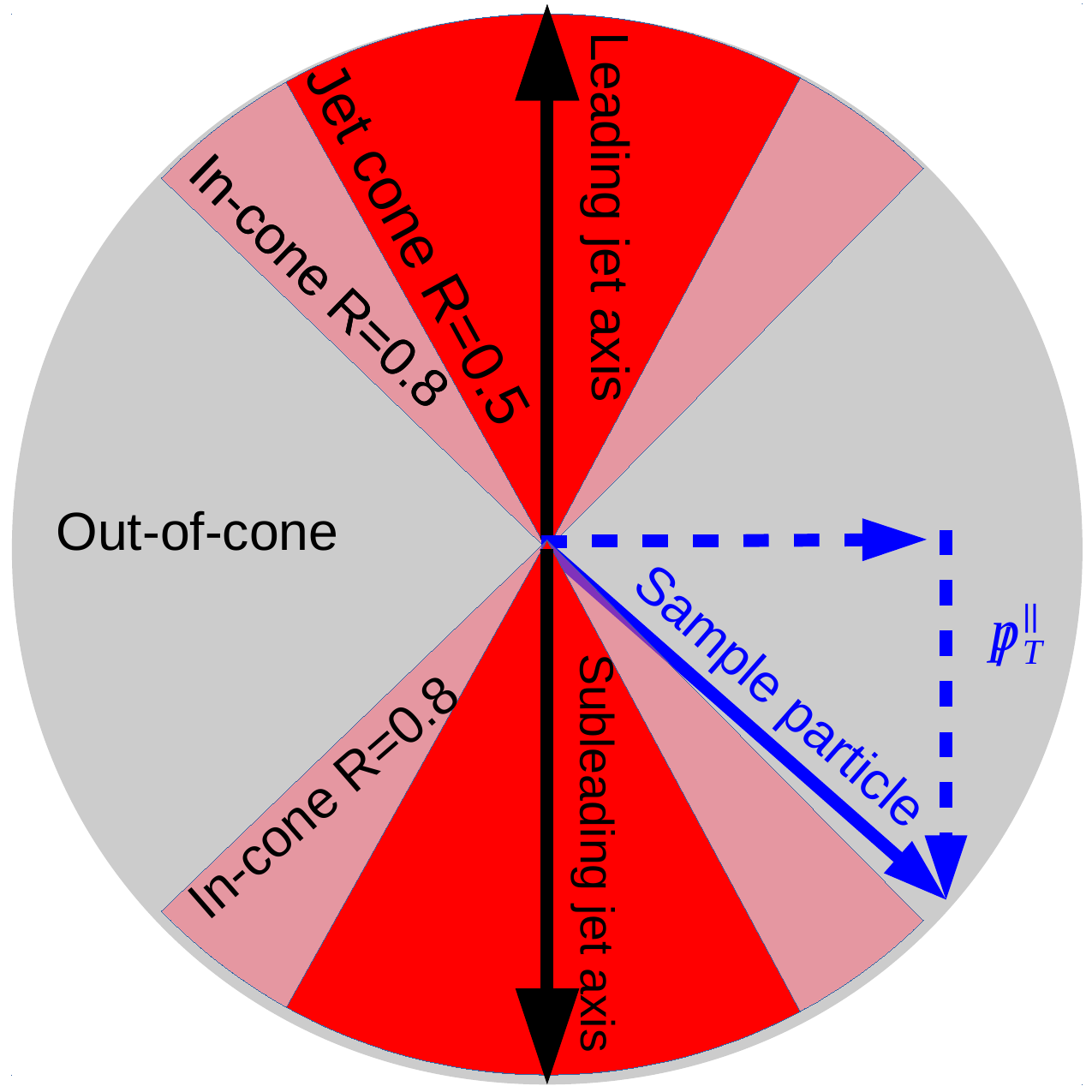}
}}\caption{
Schematic diagram showing the definitions used in \Fref{Fig:CMSInOutCone}.
}\label{Fig:CMSInOutConeSchematic}
\end{center}
\end{figure*}

\begin{figure*}
\begin{center}

\rotatebox{0}{\resizebox{13cm}{!}{
        \includegraphics{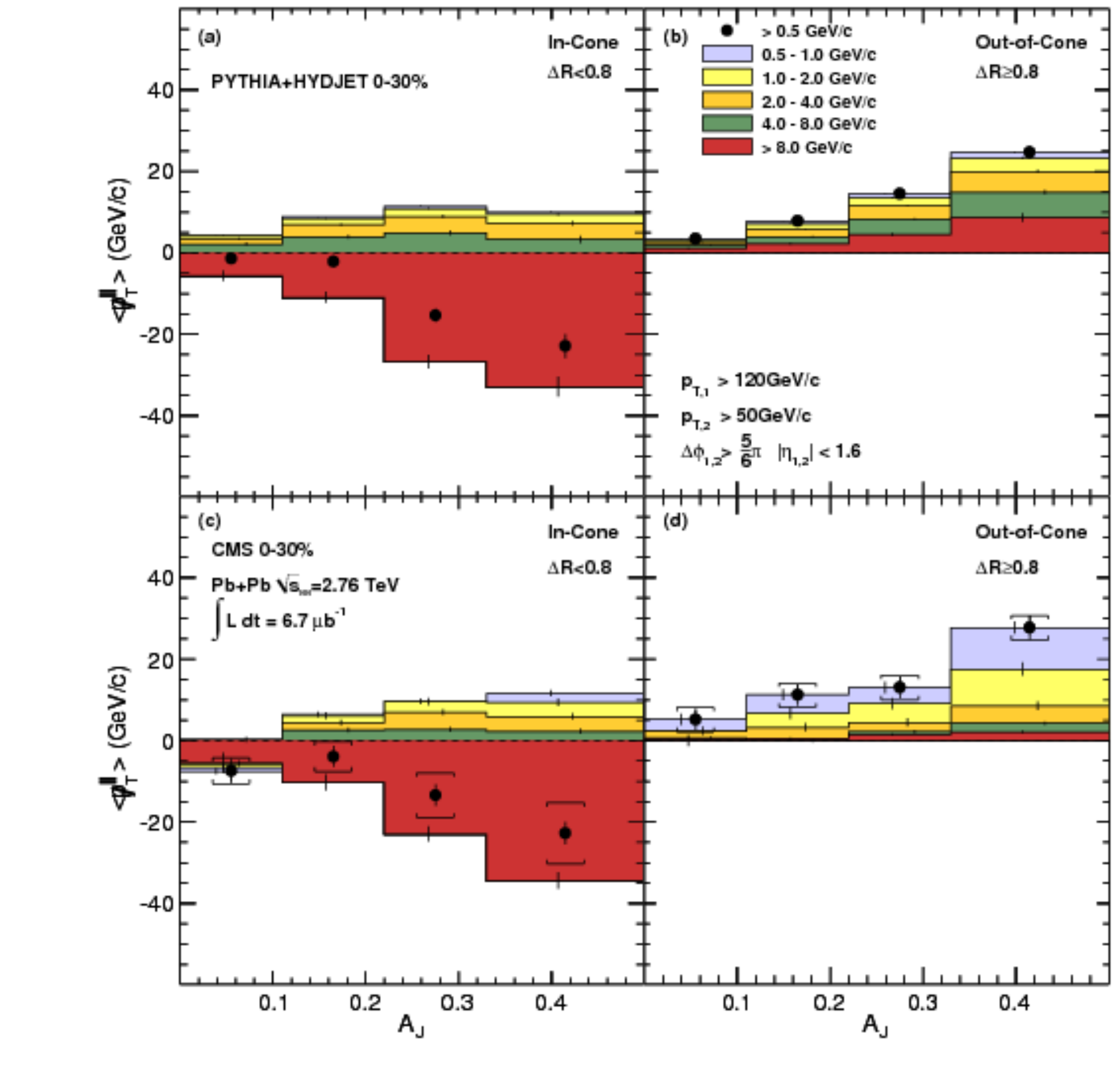}
}}\caption{
Figure from CMS~\cite{Chatrchyan:2011sx}. Average missing transverse momentum for tracks with \pT~$>0.5$ GeV/c, projected onto the leading jet axis is shown in solid circles. The average missing \pT~values are shown as a function of dijet asymmetry \Aj~for 0--30\% centrality, inside a cone of $\Delta R < 0.8$ of one of the leading or subleading jet cones on the left, and outside ($\Delta R > 0.8$) the leading and subleading jet cones on the right. The solid circles, vertical bars and brackets represent the statistical and systematic uncertainties, respectively. For the individual \pT ranges, the statistical uncertainties are shown as vertical bars.  This shows that missing momentum is found outside of the jet cone, indicating that the lost energy may have equilibrated with the medium.
}\label{Fig:CMSInOutCone}
\end{center}
\end{figure*}

The dijet asymmetry measurements demonstrate momentum imbalance for dijets in central heavy ion collisions, implying energy loss, but do not describe where that energy goes.  
To investigate this, CMS looked at the distribution of momentum parallel to the axis of a high momentum leading jet in three regions~\cite{Chatrchyan:2011sx}, shown schematically in \Fref{Fig:CMSInOutConeSchematic}.  The jet reconstruction used in this analysis was an iterative cone algorithm with a modification to subtract the soft underlying event on an event-by-event basis, the details of which can be found in \cite{Kodolova:2007hd}.  Each jet was selected with a radius $R = 0.5$  around a seed of minimum transverse energy of 1 GeV.   Since energy can be deposited outside $R > 0.5$ even in the absence of medium effects and medium effects are expected to broaden the jet, the momenta of all particles within in a slightly larger region, $R < 0.8$, were summed, regardless of whether or not the particles were jet constituents or subtracted as background.  This region is called in-cone and the region $R > 0.8$ is called out-of-cone.

CMS investigated these different regions of the events with a measurement of the projection of the
\pT of reconstructed charged tracks onto the leading jet axis. For each event, this 
projection was calculated as 
\begin{equation}
\displaystyle{\not} p_{\mathrm{T}}^{\parallel} = 
\sum_{\rm i}{ -p_{\mathrm{T}}^{\rm i}\cos{(\phi_{\rm i}-\phi_{\rm Leading\ Jet})}},
\end{equation}
\noindent where the sum is over all tracks with \pT $> 0.5$~\GeV. These results were 
then averaged over events to obtain $\langle \displaystyle{\not} p_{\mathrm{T}}^{\parallel} \rangle$. This momentum imbalance in-cone and out-of-cone as a function of \Aj, shown as black points in \Fref{Fig:CMSInOutCone}.  The momentum parallel to the jet axis in-cone is large, but should be balanced by the partner jet 180$^{\circ}$ away in the absence of medium effects.  A large \Aj indicates substantial energy loss for the \as jet, while a small \Aj indicates little interaction with the medium.  This shows that the total momentum in the event is indeed balanced.  For small \Aj, the $\langle \displaystyle{\not} p_{\mathrm{T}}^{\parallel} \rangle$ in the in-cone and out-of-cone regions is within zero as expected for balanced jets.  For large \Aj, the momentum in-cone is non-zero, balanced by the momentum out-of-cone.  These events were compared to PYTHIA+HYDJET simulations in order to understand which effects were simply due to the presence of a fluctuating background and which were due to jet quenching effects.  In both the central \Pb~data and the Monte Carlo, an imbalance in jet \Aj~also indicated an imbalance in the \pT~of particles within the cone of $R = 0.8$ about either the leading or subleading jet axes.  To investigate further, CMS added up the momentum contained by particles in different momentum regions.  The imbalance in the direction of the leading jet is dominated by particles with \pT $> 8$ \GeV, but is partially balanced in the subleading direction by particles with momenta below 8 \GeV.  The distributions look very similar in both the data and the Monte Carlo for the in-cone particle distribution.  The out-of-cone distributions indicated a slightly different story.  For both the data and the Monte Carlo, the missing momentum was balanced by additional, lower momentum particles, in the subleading jet direction.  The difference is that in the \Pb~data, the balance was achieved by very low momentum particles, between 0.5 and 1 \GeV.  In the Monte Carlo, the balance was achieved by higher momentum particles, mainly above 4 \GeV, which indicates a different physics mechanism.  In the Monte Carlo, the results could be due to semi-hard initial- or final-state radiation, such as three jet events.

The missing transverse momentum analysis was recently extended by examining the multiplicity, angular, and \pT~spectra of the particles using different techniques.  As above, these results were characterized as a function of the \Pb~collision centrality and \Aj~\cite{Khachatryan:2015lha}.  This extended the results to quite some distance from the jet axes, up to a $\Delta R$ of 1.8.  The angular pattern of the energy flow in \Pb~events was very similar to that seen in \pp~collisions, especially when the resolution parameter is small.  This indicates that the leading jet could be getting narrower, and/or the subleading jet is getting broader due to quenching effects.  For a given range in \Aj,  the in-cone imbalance in \pT~in \Pb~collisions is found to be balanced by relatively low transverse momentum out-of-cone particles with 0.5$<p_{T}<$2 \GeV.  This was quantitatively different than in \pp collisions where most of the momentum balance comes from particles with \pT between 2$<p_{T}<$8 \GeV.   This could indicate a softening of the radiation responsible for the \pT~imbalance of dijets in the medium formed in \Pb collisions. In addition, a larger
multiplicity of associated particles is seen in \Pb than in \pp collisions.  In every case, the difference between \pp~and \Pb~observations increased for more central \Pb~collisions.

However, some caution should be used in interpreting the result as these measurements make assumptions about the background, and require certain jet kinematics, which may limit how robust the conclusions are.  It is unlikely that the medium would focus the leading jet so that it would be more collimated, for instance, but that a selection bias causes narrower jets to be selected in \Pb~collisions for a given choice in $R$ and jet kinematics.  Additionally, as with any analysis that attempts to disentangle the effects of the medium on the jet with the jet on the medium, the ambiguity in what is considered part of the medium and what is considered part of the jet can also complicate the interpretation of this result.  While the results demonstrate that there is a difference in the missing momentum in \Pb and \pp collisions, in order to identify the mechanism responsible, the data would need to be compared to a Monte Carlo model that incorporates jet quenching, and preserves momentum and energy conservation between the jet and medium.

\subsubsection{Searches for Moli\`{e}re scattering}
The measurement of jets correlated with hard hadrons in~\cite{Adam:2015doa} was also used to look for broadening of the correlation function between a high momentum hadron and jets.  Such broadening could result from Moli\`{e}re scattering of hard partons off other partons in the medium, coherent effects from the scattering of a wave off of several scatterers.  No such broadening is observed, although the measurement is dominated by the statistical uncertainties.  Similarly, STAR observes no evidence for Moli\`{e}re scattering~\cite{Adamczyk:2017yhe}.  We note that this would mainly be sensitive to whether or not the jets are deflected rather than whether or not jets are broadened.

\subsubsection{The rise and fall of the Mach cone and the ridge}\label{Sec:MachCone}
Several theoretical models proposed that a hard parton traversing the medium would lose energy similar to the loss of energy by a supersonic object traveling through the atmosphere~\cite{CasalderreySolana:2004qm,Ruppert:2005uz,Renk:2005si}.  The energy in this wave forms a conical structure about the object called a Mach cone.
Early \dhcs studies observed a displaced peak in the \as~\cite{Adler:2005ee,Adare:2006nr,Adare:2007vu,Aggarwal:2010rf}.  Three-particle correlation studies observed that this feature was consistent with expectations from a Mach cone~\cite{Abelev:2008ac}.  
Studies indicated that its spectrum was softer than that of the \jlc on the \ns~\cite{Adare:2007vu} and its composition similar to the bulk~\cite{Afanasiev:2007wi}, as might be expected from a shock wave from a parton moving faster than the speed of light in the medium.  Curiously, the Mach cone was present only at low momenta~\cite{Adare:2008ae,Aggarwal:2010rf}, whereas some theoretical predictions indicated that a true Mach cone would be more significant at higher momenta~\cite{Betz:2008wy}.

At the same time, studies of the \ns indicated that there was a feature correlated with the trigger particle in azimuth but not in pseudorapidity~\cite{Abelev:2009af,Alver:2009id}, dubbed the ridge.  The ridge was also observed to be softer than the \jlc~\cite{Abelev:2009af} and to have a particle composition similar to the bulk~\cite{phdthesissuarez,Bielcikova:2008ad}.  Several of the proposed mechanisms for the production of the ridge involved interactions between the hard parton and the medium, including collisional energy loss~\cite{Wong:2007pz,Wong:2008yh} and recombination of the hard parton with a parton in the medium~\cite{Hwa:2008jt,Hwa:2008um,Chiu:2008ht}.

However, the observation of odd \vn in heavy ion collisions~\cite{ALICE:2011ab,Adamczyk:2013waa,Adare:2011tg} indicated that the Mach cone and the ridge may be an artifact of erroneous background subtraction.  Since the ridge was defined as the component correlated with the trigger in azimuth but not in pseudorapidity, it is now understood to be entirely due to \vthree.  Initial \dhc studies after the observation of odd \vn are either inconclusive about the presence or absence of shape modifications on the \as~\cite{Adare:2012qi} or indicate that the shape modification persists~\cite{Agakishiev:2014ada}.  A reanalysis of STAR \dhcs~\cite{Agakishiev:2010ur,Agakishiev:2014ada} using a new method for background subtraction~\cite{Sharma:2015qra} found that the Mach cone structure is not present~\cite{Nattrass:2016cln}.  This new analysis indicates that jets are broadened and softened~\cite{Nattrass:2016cln}, as observed in studies of reconstructed jets~\cite{Aad:2014wha,Chatrchyan:2014ava}.

While the ridge is currently understood to be due to \vthree in heavy ion collisions, a similar structure has also been observed in high multiplicity \pp collisions~\cite{Khachatryan:2010gv,Aaboud:2016yar}.  There are some hypotheses that this might indicate that a medium is formed in violent \pp collisions~\cite{Khachatryan:2016txc}, although there are other hypotheses such as production due to gluon saturation~\cite{Ozonder:2016xqn} or string percolation~\cite{Andres:2016ahv}.  Whatever the production mechanism for the ridge in \pp collisions, there is currently no evidence that it is related to or correlated with jet production in either \pp or heavy ion collisions.

\subsubsection{Summary of experimental evidence for modification of the medium by jets}
Measurements of the impact of jets on the medium are difficult because of the large combinatorial background.  The background may distort reconstructed jets and requiring the presence of a jet may bias the event selection.  Because the energy contained within the background is large compared to the energy of the jet, even slight deviations of the background from the assumptions of the structure of the background used to subtract its effect could skew results.  A confirmation of the CMS result indicating that the lost energy is at least partially equilibrated with the medium will require more detailed theoretical studies, preferably using Monte Carlo models so that the analysis techniques can be applied to data.  The misidentification of the ridge and the Mach cone as arising due to partonic interactions with the medium highlights the perils of an incomplete understanding of the background.
 
\subsection{Summary of experimental results}
\Sref{Sec:ResultsColdNuclearMatter} reviews studies of cold nuclear matter effects, indicating that currently it does not appear that there are substantial cold nuclear matter effects modifying jets at mid-rapidity and that therefore effects observed thus far on jets in A+A collisions are primarily due to interactions of the hard parton with the medium.  We note, however, that our understanding of cold nuclear matter effects is evolving rapidly and recommend that each observable is measured in both cold and hot nuclear matter in order to disentangle effects from hot and cold nuclear matter.
\Sref{Sec:EnergyLoss} shows that there is ample evidence for partonic energy loss in the QGP.  Nearly every measurement demonstrates that high momentum hadrons are suppressed relative to expections from \pp and \pPb collisions in the absence of quenching.  \Sref{Sec:ResultsFragmentation} reviews the evidence that these partonic interactions with the medium result in more lower momentum particles and particles at larger angles relative to the parent parton, as expected from both gluon bremsstrahlung and collisional energy loss. \Tref{Tab:FragmentationObservables} summarizes physics observations, selection biases and ability to constrain the initial kinematics for the measured observables.  \Sref{Sec:ResultsMedium} discusses the evidence that at least some of this energy may be fully equilibrated with the medium and no longer distinguishable from the background.

For future studies to maximize our understanding of the medium, most observables can be incorporated into a Bayesian analysis.  We encourage exploration of comparisons of new observables to describe the jet structure.  However, we caution that many observables are sensitive to kinematic selections and analysis techniques so that a replication of these techniques is required for the measurements to be comparable to theory. 
 \section{Discussion and the path forward}\label{Sec:Discussion}

In the last several years, we have seen a dramatic increase in the number of experimentally accessible jet observables for heavy-ion collisions.  During the early days of RHIC, measurements were primarily limited to \RAA and \dhcs, and reconstructed jets were measured only relatively recently.  Since the start of the LHC, measurements of reconstructed jets have become routine, fragmentation functions have been measured directly, and the field is investigating and developing more sophisticated observables in order to quantify partonic energy loss and its effects on the QGP.  The constraint of \qhat, the energy loss squared per fm of medium traversed, using \RAA measurements by the JET collaboration is remarkable.  However, studies of jets in heavy ion collisions largely remain phenomenological and observational.  This is probably the correct approach at this point in the development of the field, but a quantitative understanding of partonic energy loss in the QGP requires a concerted effort by both theorists and experimentalists to both make measurements which can be compared to models and use those measurements to constrain or exclude those models.  

Below we lay out several of the stpdf we think are necessary to reach this quantitative understanding of partonic energy loss.  We think that it is critical to quantitatively understand the impact of measurement techniques on jet observables in order to make meaningful comparisons to theory.  We encourage the developments in new observables but urge caution -- new observables may not have as many benefits as they first appear to when their biases and sensitivities to the medium are better understood.  Many experimental and theoretical developments pave the way towards a better quantitative understanding of partonic energy loss.  However, we think that the field will not fully benefit from these without discussions targeted at a better understanding of and consistency between theory and experiment and evaluating the full suite of observables considering all their biases.  One of the dangers we face is that  many observables are created by experimentalists, which often yields observables that are easy to measure such as \Aj, but that are not particularly differential with respect to constraining jet quenching models.

\subsection{Understand bias}
As we discussed in \Sref{Sec:ExpMethods}, all jet measurements in heavy ion collisions are biased towards a particular subset of the population of jets produced in these collisions.  The existence of such biases is transparent for many measurements, such as surface bias in measurements of \dhcs at RHIC.  However, for other observables, such as those relating to reconstructed jets, these biases are not always adequately discussed in the interpretation of the results.  As the comparison between ALICE, ATLAS, and CMS jet \RAA at low jet momenta shows, requiring a hard jet core in order to suppress background and reduce combinatorial jets leads to a strong bias which cannot be ignored.  The main biases that pertain to jets in heavy ion collisions are: fragmentation, collision geometry, kinematic and parton species bias.  The fragmentation bias can be simply illustrated by the jet \RAA measurement. Requiring a particular value of the resolution parameter, a particular constituent cut, or even the particular trigger detector used by the experiment selects a particular shower structure for the jet.  The geometry bias is commonly discussed as a surface bias, since the effect of the medium increases with the path length causing more hard partons come from the surface of the QGP.  The kinematic bias is somewhat related to the fragmentation bias as the fragmentation depends on the kinematics of the parton, but the energy loss in the medium means that jets of given kinematics do not come from the same selection of initial parton kinematics in vacuum and in heavy ion collisions.  The parton species bias results as the gluons couple more strongly with the medium, and thus are expected to be more modified.  This can be summarized by stating that nearly every technique favors measurement of more quark jets over gluon jets, is biased towards high z fragments, and is biased towards jets which have lost less energy in the medium.

While some measurements may claim to be bias free because they deal with the background effects in a manner which makes comparisons with theoretical models more straightforward, they still contain biases, usually towards jets which interacted less with the medium and therefore have lost less energy.  For example, for the hadron-jet coincidence measurements, it is correct to state that the away side jet does not have a fragmentation bias since the hadron trigger is not part of its shower. However, this does not mean that this measurement is completely unbiased since the trigger hadron may select jets that have traveled through less medium or interacted less with the medium.  In addition, the very act of using a jet finding algorithm introduces a bias (particularly toward quark jets) that is challenging to calculate.  Given the large combinatorial background, such biases are most likely unavoidable. 

We propose that these biases should be treated as tools through jet geometry engineering rather than a handicap.  These experimental biases should also be made transparent to the theory community.  Frequently the  techniques which impose these biases are buried in the experimental method section, with no or little mention of the impact of these biases on the results in the discussion.  Theorists should not neglect the discussion of the experimental techniques, and experimentalists should make a greater effort to highlight potential impacts of the techniques to suppress and subtract the background on the measurement.

\subsection{Make quantitative comparisons to theory}
With the explosion of experimentally accessible observables, much of the focus has been on making as many measurements as possible with less consideration of whether such observables are calculable, or capable of distinguishing between different energy loss models.  Even without direct comparisons to theory, these studies have been fruitful because they contribute to a phenomenological understanding of the impact of the medium on jets and vice versa.  While we still feel that such exploratory studies are valuable, the long term goal of the field is to measure the properties of the QGP quantitatively, making theoretical comparisons essential.  Some of the dearth of comparisons between measurements and models is due to the relative simplicity of the models and their inability to include hadronization.  

The field requires another systematic attempt to constrain the properties of the medium from jet measurements.  The Jetscape collaboration has formed in order to incorporate theoretical calculations of partonic energy loss into Monte Carlo simulations, which can then be used to directly calculate observables using the same techniques used for the measurements.  This will then be followed up by a Bayesian analysis similar to previous work~\cite{Novak:2013bqa,Bernhard:2016tnd} but incorporating measurements of jets.  This is essential, both to improve our theoretical understanding and to provide Monte Carlo models which can be used for more reliable experimental corrections.  In our opinion, it should be possible to incorporate most observables into these measurements.  However, we urge careful consideration of all experimental techniques and kinematic selections in order to ensure an accurate comparison between data and theory.  The experimental collaborations should cooperate with the Jetscape collaboration to ensure that response matrices detailing the performance of the detectors for different observables are available.

\subsection{More differential measurements}
The choices of what to measure, how to measure it and how to both define and treat the background are key to our quantitative understanding of the medium.  There have been substantial improvements in the ability to measure jets in heavy ion collisions in recent years, such as the available kinematic reach due to accelerator and detector technology improvements.  Additionally, our quantitative understanding of the effect of the background in many observables has also significantly improved.  Given the continuous improvement in technology and analysis techniques, it is vital that the some of the better understood observables such as \RAA~and \IAA~are repeated with higher precision. 
Theoretical models should be able to simultaneously predict these precisely measured jet observables with different spectral shapes and path length dependencies. While this is necessary it is not sufficient to validate a theoretical model.  Given that these will also depend on the collision energy, comparisons between RHIC and the LHC would be valuable, but again only when all biases are carefully considered. 
Now that the era of high statistics and precision detectors is here, the field is currently exploring several new observables to attempt to identify the best observables to constrain the properties of the medium.  Older observables, such as \RAA, were built with the mindset that the final state jet reflects the kinematics of its parent parton, and the change in these kinematics due to interactions with the medium would be reflected in the change in the jet distributions.  One of the lessons learned is that the majority of the modification of the fragmentation occurs at a relatively low \pT~compared to the momentum of the jet.  However, jet finding algorithms were specifically designed in order to not be sensitive to the details of the soft physics, which means that the very thing we are trying to measure and quantify is obscured by jet finder.  The new observables are based on the structure of the jet, rather than on its kinematics alone.  Specifically, they recognize that a hard parton could split into two hard daughters.  If this splitting occurs in the medium, not only can the splitting itself be modified by the presence of the medium, but each of the daughters could lose energy to the medium independently.  This would be actually be rather difficult to see in an ensemble structure measurement such as the jet fragmentation function, which yields a very symmetric picture of a jet about its axis, and so requires the specific structures within the jet to be quantified.  While these new observables hold a lot of promise in terms of our understanding, caution must also be used in interpreting them until precisely how the background removal process or the detector effects will play a role in these measurements is carefully studied.

The investigations into these different observables are very important, since we have likely not identified the observables most sensitive to the properties of the medium.  We cannot forget that we want to quantify the temperature dependence of the jet transport coefficients, as well as determine the size of the medium objects the jets are scattering off of.  While these are global and fundamental descriptors of a medium, the fact that the process by which we make these measures is statistical means that the development of quantitative Monte Carlo simulations is key.  Not only will they allow calculations of jet quenching models to be compared with the same initial states, hadronization schemes, etc, but they also could make the calculations of even more complicated observables feasible.

However, the sensitivity of simple observables should not be underestimated as with every set of new observables there are new mistakes to be made, and we can be reasonably sure that we understand the biases inherent in these simple observables.  While it is not likely that comparison between \RAA and theories will constrain the properties of the medium substantially better than the JET collaboration's calculation of \qhat, calculations of $\gamma$-hadron, dihadron, and jet hadron correlations are feasible with the development of realistic Monte Carlo models.  The relative simplicity of these observables makes them promising for subsequent attempts to constrain \qhat and other transport coefficients, especially since we now have a fairly precise quantitative experimental understanding of the background.  This may be a good initial focus for systematic comparisons between theory and experiment.  
Interpreting a complicated result with a simple model that misses a lot of physics is a misuse of that model, and can lead to incorrect assumptions.

We caution against overconfidence, and encourage scrutiny and skepticism of measurement techniques and all observables.  For each observable, an attempt needs to be made to quantify its biases, and determine which dominate.  Observables should be measured in the same kinematic region and, if possible, with the same resolution parameters in order to ensure consistency between experiments.  If initial studies of a particular observable reveal that it is either not particularly sensitive to the properties of the medium, or that it is too sensitive to experimental technique, we should stop measuring that observable.  We urge caution when using complicated background subtraction and suppression techniques, which may be difficult to reproduce in models and requires Monte Carlo simulations that accurately model both the hard process that has produced the jet and the soft background.  Given that the response of the detector to the background is different from experiment to experiment, complicated subtraction processes may make direct comparisons across experiments and energies difficult.

We also caution against the overuse and blind use of unfolding.  Unfolding is a powerful technique which is undoubtedly necessary for many measurements.  It also has the potential to impose biases by shifting measurements towards the Monte Carlo used to calculate the response matrix, and obfuscating the impact of detector effects and analysis techniques.  When unfolding is necessary, it should be done carefully in order to make sure all effects are understood and that the result is robust.  Since most effects are included in the response matrix rather than corrected for separately, it can be difficult to understand the impact of different effects, such as track reconstruction efficiency and energy resolution.  Unfolding is not necessarily superior to careful studies of detector effects and corrections, and attempts to minimize their impact on the observables chosen.  Given the relative simplicity of folding a result, for all observables we should perform a theory-experiment closure test where the theoretical results are folded and compared to the raw data. Since the robustness of a particular measurement depends on the unfolding corrections, the details of the unfolding method should be also transparent to both experimental and theoretical communities. 

Of course making more differential measurements is aided by better detectors.  The LHC detectors use advanced detector technology, and are designed for jet measurements.  However, the current RHIC detectors were not optimized for jet measurements, which has limited the types of jet observables at these lower energies.  Precise measurements of jets over a wide range of energies is necessary to truly understand partonic energy loss.  The proposed sPHENIX detector will greatly aid these measurements by utilizing some of the advanced detector technology that has been developed since the design of the original RHIC experiments \cite{Adare:2015kwa}.  The high rate and hermetic detector will improve the results by reducing detector uncertainties and increasing the kinematic reach so that a true comparison between RHIC and LHC can be made. In particular, upgrades at both RHIC and LHC will make precise measurements of heavy-flavor tagged jets and boson-tagged jets, which constrain the initial kinematics of the hard scattering, possible. 

\subsection{An agreement on the treatment of background in heavy ion collisions}
The issues we listed above are complicated and require substantive, ongoing discussions between theorists and experimentalists.  A start in this direction can be found in the Lisbon Accord where the community agreed to use Rivet \cite{Buckley:2010ar}, a C++ library which provides a framework and tools for calculating observables at particle level developed for particle physics.   Rivet allowed event generator models and experimental observables to be validated.  Agreeing on a framework that all physicists can use is an important first step, however it is not sufficient.  It would not prevent a comparison of two observables with different jet selection criteria, or a comparison of a theoretical model with a different treatment or definition of the background than a similar experimental observable.  The problems we face are similar to those faced by the particle physics community as they learned how to study and utilize jets, to make them one of the best tools we have for understanding the Standard Model.  An agreement on the treatment of the background in heavy ion collisions experimentally and theoretically is required as it is part of the definition of the observable.  Theorists and experimentalists need to understand each other's techniques and find common ground, to define observables that experimentalists can measure and theorists can calculate.  We need to recognize that observables based on pQCD calculations are needed if we are to work towards a text-book formulation of jet quenching, and what we learn about QCD from studying the strongly coupled QGP.  However, observables that are impossible to measure are not useful, nor is it useful to measure observables that are impossible to calculate or are insensitive to the properties of the medium.  We propose a targeted workshop to address these issues in heavy ion collisions with the goal of an agreement similar to the Snowmass Accord.  Ideally we would agree on a series of jet algorithms, including selection criteria, that all experiments can measure, and a background strategy that can be employed both in experiment and theory.
 \section{Acknowledgements}
We thank Will Witt for productive discussions about unfolding.  We thank  Redmer Bertens, Jana Biel\v{c}ikova, Leticia Cunqueiro Mendez,  Kate Jones, Kolja Kauder, Abhijit Majumder, Jaki Noronha-Hostler, Thomas Papenbrock,  Dennis Perepelitsa, J\"orn Putschke, Soren Sorensen, Peter Steinberg,  and Giorgio Torrieri for useful advice on the manuscript and useful discussions.  We thank Raghav Elayavalli Kunnawalkam, Boris Hippolyte, Kurt Jung,  Igor Lokhtin, Thomas Ullrich, and the experimental collaborations for permission to reproduce their figures for this work.    This material is based upon work supported by the Division of Nuclear Physics of the U.S. Department of Energy under Grant No. DE-FG02-96ER40982 and by the National Science Foundation under Grant Nos. 1352081 and 1614474.

\end{document}